\documentclass[11pt]{article}
\pdfoutput=1 

\usepackage{jheppub_softqed1} 
\def\@preprint{MOO}

\usepackage[T1]{fontenc} 

\usepackage[usenames,dvipsnames]{color}

\usepackage{graphicx}
\usepackage{bm}
\usepackage{amsfonts}
\usepackage{booktabs}

\usepackage{tikz}
\newcommand{\Plus}{{\mathord{\tikz\draw[line width=0.2ex, x=1ex, y=1ex] (0.5,0) -- (0.5,1)(0,0.5) -- (1,0.5);}}}

\usepackage{xcolor}

\usepackage{dsfont}

\def\b{{\bm b}}
\def\x{{\bm x}}
\def\p{{\bm p}}
\def\q{{\bm q}}
\def\k{{\bm k}}

\def\B{{\bm B}}

\def\P{{\bm P}}

\def\eps{\epsilon}
\def\beps{{\bm\epsilon}}
\def\vareps{\varepsilon}

\def\Nc{N_{\rm c}}
\def\Nf{N_{\rm f}}

\def\alphas{\alpha_{\rm s}}

\def\Re{\operatorname{Re}}
\def\Im{\operatorname{Im}}

\def\sh{\operatorname{sh}}

\def\csch{\operatorname{csch}}
\def\grad{{\bm\nabla}}

\def\yfrak{{\mathfrak y}}
\def\yfrakE{\yfrak_\ssE}

\def\Time{\mathbb{T}}

\def\tform{t_{\rm form}}

\def\gammaE{\gamma_{\rm\scriptscriptstyle E}}

\def\qhat{\hat q}

\def\NLO{{\rm NLO}}

\def\xe{x_e}
\def\E{{\rm E}}
\def\Ebar{{\bar\E}}
\def\ssE{{\scriptscriptstyle{\E}}}
\def\ssEbar{{\scriptscriptstyle{\Ebar}}}
\def\xE{x_\ssE}
\def\xEbar{x_\ssEbar}

\def\net{{\rm net}}

\def\uee{{\underline{e\to e}}}

\def\form{{\rm form}}

\def\BH{{\rm BH}}
\def\quoteBH{{\rm ``BH"}}
\def\LPM{{\rm LPM}}
\def\LPMplus{{{\rm LPM}\Plus}}
\def\Elpma{E_{\rm LPM}}
\def\Elpm{\hat E_{\rm LPM}}
\def\kgamma{k_\gamma}
\def\bmin{b_{\rm min}}
\def\el{{\rm el}}
\def\scatt{{\rm scatt}}
\def\QCD{{\rm QCD}}
\def\brem{{\rm brem}}
\def\br{0}
\def\pair{{\rm pair}}
\def\pr{{\rm pr}}
\def\Mo{{\bar M}_0}
\def\Mpr{M_\pr}
\def\Omegapr{\Omega_\pr}
\def\Time{\mathbb{T}}
\def\vacloop{{\rm vac\text{-}loop}}

\def\dlangle{\langle\!\langle}
\def\drangle{\rangle\!\rangle}
\def\bigdlangle{\big\langle\!\big\langle}
\def\bigdrangle{\big\rangle\!\big\rangle}
\def\Bigdlangle{\Big\langle\!\!\Big\langle}
\def\Bigdrangle{\Big\rangle\!\!\Big\rangle}

\def\bxi{{\bm\xi}}
\def\bpi{{\bm\pi}}

\newcounter{savefootnote}
\newcommand{\astfootnote}[1]{%
  \setcounter{savefootnote}{\value{footnote}}%
  \setcounter{footnote}{0}%
  \let\oldthefootnote=\thefootnote
  \renewcommand{\thefootnote}{\fnsymbol{footnote}}%
  \footnote{#1}%
  \let\thefootnote=\oldthefootnote%
  \setcounter{footnote}{\value{savefootnote}}%
}

\begin {document}



\title{Revisiting extremely high energy QED bremsstrahlung in matter:
       large modifications to the LPM effect}

\author[a]{Peter Arnold,}
\author[a]{Joshua Bautista,}
\author[b,c]{Omar Elgedawy,}
\author[d]{Shahin Iqbal}


\affiliation[a]{Department of Physics, University of Virginia,
  P.O.\ Box 400714, 
  Charlottesville, VA 22904, U.S.A.}
\affiliation[b]{
  State Key Laboratory of Nuclear Physics and
  Technology, Institute of Quantum Matter, South China Normal
  University, Guangzhou 510006, China}
\affiliation[c]{
  Guangdong Basic Research Center of Excellence for
  Structure and Fundamental Interactions of Matter, Guangdong
  Provincial Key Laboratory of Nuclear Science, Guangzhou
  510006, China}
\affiliation[d]{National Centre for Physics,
  Shahdra Valley Road,
  Islamabad, 45320 Pakistan}

\emailAdd{parnold@virginia.edu}
\emailAdd{rzp9zf@virginia.edu}
\emailAdd{oelgedawy@scnu.edu.cn}
\emailAdd{shahin@ncp.edu.pk}

\begin {abstract}%
{%
  Very high energy electrons initiate electromagnetic showers in
  ordinary matter that branch and multiply
  through bremsstrahlung and pair production.
  At extremely high energies, the quantum mechanical duration of these
  processes becomes longer than the mean free time to
  elastically scatter from the medium, which leads to a very significant
  suppression of bremsstrahlung (and pair production) known as the
  Landau-Pomeranchuk-Migdal (LPM) effect.  We revisit the LPM effect for
  bremsstrahlung of energy $k_\gamma$ from an electron of energy $E$.
  We find that there are very large corrections to the LPM bremsstrahlung rate
  for certain regions of $(\kgamma,E)$ due to quantum overlap of
  bremsstrahlung and subsequent pair production.
  This possibility was first raised in the 1960s, when it was argued
  qualitatively that pair production would significantly
  decrease the bremsstrahlung rate in those regions of $(\kgamma,E)$
  compared to the already-suppressed LPM bremsstrahlung rate.
  We find the opposite --- quantum
  overlap of bremsstrahlung with pair production
  significantly \textit{increases} the bremsstrahlung rate compared to the
  LPM calculation --- and we
  verify our qualitative arguments with an analytic calculation of the effect.
}%
\end {abstract}

\maketitle
\thispagestyle {empty}

\newpage


\section{Introduction and Preview of Results}
\label{sec:intro}

\subsection {Introduction}

\begin {figure}[t]
\begin {center}
  \includegraphics[scale=0.7]{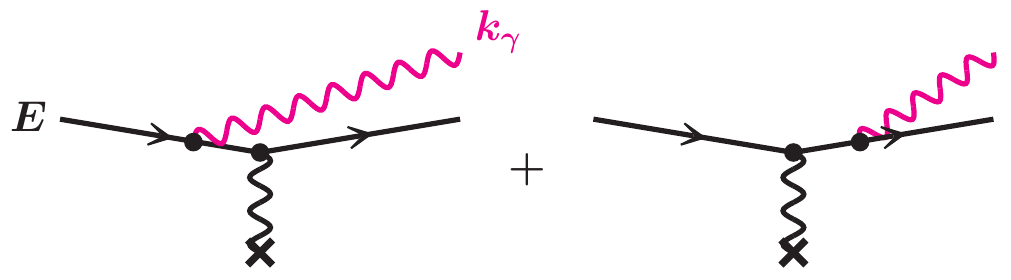}
  \caption{
     \label{fig:BHdiags}
     Relevant Feynman diagrams for the Bethe-Heitler limit of
     high-energy bremsstrahlung in the
     Coulomb field of an atomic nucleus (in the absence of a medium).
  }
\end {center}
\end {figure}

Roughly 75 years after the completion%
\footnote{
  which we count as the work of
  Tomanaga, Schwinger and Feynman recognized by the 1965 Nobel Prize in Physics.
}
of the relativistic theory of quantum electrodynamics,
it is delightful to still be able to discuss qualitatively new
features of purely-QED processes at very high energy.
In this paper, we revisit the theory of bremsstrahlung from
extremely high-energy electrons (or positrons) passing through ordinary matter,
which is the dominant mechanism of $e^\pm$ energy loss at high energy.%
\footnote{
  See, for example, section 34.4.4 of the
  2024 Review of Particle Physics \cite{RPP2024}.
}
The QED calculation of the bremsstrahlung rate in the electric field of
a single atomic nucleus, represented (anachronistically)
by the Feynman diagrams of
fig.\ \ref{fig:BHdiags}, was carried out by Bethe and Heitler in 1934
\cite{BH}.
In 1953, Landau and Pomeranchuk pointed out that the quantum duration
of bremsstrahlung (known as the formation or coherence time)
grows with electron energy
and eventually becomes larger than the mean free time
between elastic scatterings from the medium \cite{LP1,LP2}.%
\footnote{
  The papers of Landau and Pomeranchuk \cite{LP1,LP2} are also available in
  English translation \cite{LPenglish}.
}
Accounting for
multiple collisions during the bremsstrahlung process, they calculated
the bremsstrahlung rate in the case where
the high-energy photon is relatively soft
compared to the initial high-energy electron
($\kgamma{\ll}E$ in our notation), using classical radiation
arguments and so ignoring the back-reaction of the electron.
They found that, at sufficiently high energy, the bremsstrahlung
rate was suppressed compared to the Bethe-Heitler (BH) rate.
Parametrically, in our notation (and ignoring a mild logarithmic dependence
on energy),
\begin {equation}
   \left[ \frac{d\Gamma}{d\kgamma} \right]_{\rm LP}
   \sim
   \sqrt{ \frac{\kgamma \Elpma}{E^2} }
   \times
   \left[ \frac{d\Gamma}{d\kgamma} \right]_\BH ,
\label {eq:LPrate}
\end {equation}
where $\Elpma$ is an energy scale determined by properties of the medium
and is,
for example, about $\Elpma = 2.5$ TeV for Gold and $\Elpma = 234$ PeV
for air.%
\footnote{
\label{foot:uggerhoj}
  Our numbers here are taken from
  Table 1 of ref.\ \cite{SpencerReview}, which lists $\Elpma$ for a
  selection of different materials.
  Throughout this paper, we consider only amorphous materials and not
  crystalline ones (though the initial electron would have to be closely aligned
  with one of the major crystallographic directions for that to be
  an issue).
  A review of the interactions of
  relativistic particles with crystalline matter
  may be found in ref.\ \cite{LPMcrystal}.
}
A few years after Landau and Pomeranchuk, Migdal carried out the first
full quantum mechanical calculation of the effect \cite{Migdal},
which could treat the hard-bremsstrahlung case $\kgamma{\sim}E$
as well.
Indications of the suppression, known as the
Landau-Pomeranchuk-Migdal (LPM) effect, were observed in various experiments
over the ensuing decades, culminating in a detailed examination in the
1990s by
the dedicated experiment E-146 at SLAC \cite{SLAC1,SLAC2}.
(See Klein \cite{SpencerReview} for a comprehensive review.%
\footnote{
  Readers of Migdal's paper \cite{Migdal} and Klein's review
  \cite{SpencerReview} should be alert that those references
  use units for electric charge
  where the fine structure constant is $\alpha = e^2/\hbar c$ instead of
  $\alpha = e^2/4\pi\varepsilon_0\hbar c$.
  (Throughout this paper, we will use natural units and so take
  $\varepsilon_0\hbar c = 1$.)
}%
)

In 1964, Galitsky and Gurevich \cite{Galitsky} realized that there are
regions of $(\kgamma,E)$ where the formation time for bremsstrahlung
is not only larger than the mean free time between elastic scatterings
but is \textit{also} larger than the typical time for the
resulting bremsstrahlung photon to pair produce $(\gamma\to e^+ e^-)$
in the medium.  In this case, they gave a qualitative argument, accompanied by
estimates, that the quantum mechanical overlap
of the bremsstrahlung process with subsequent pair production
further suppresses the bremsstrahlung rate to be significantly less
than the already-suppressed LPM behavior (\ref{eq:LPrate}).%
\footnote{
  A helpful summary of Galitsky and Gurevich's analysis can be found
  in section II.D of Klein's review \cite{SpencerReview}.
}

In this paper, we argue that Galitsky and Gurevich were correct that
pair production can in some cases have a significant effect on the
bremsstrahlung rate but that they misidentified the direction of the effect.
We find that, for the region of $(\kgamma,E)$ where pair
production is relevant, pair production instead \textit{disrupts} the
mechanism of LPM suppression and so leads to a bremsstrahlung rate
significantly \textit{larger}
than the LPM behavior (\ref{eq:LPrate}).  We will first make a qualitative
argument, with parametric estimates.  Then we verify our
qualitative argument by doing a full calculation of the rate,
including the effects of possible pair production.
Our calculation uses methods similar to Migdal \cite{Migdal},
but generalized here to include the effects of pair production
during bremsstrahlung.

To keep the discussion of the full calculation
manageable, in this first paper
we will carefully choose our battleground
--- the particular region of $(\kgamma,E)$ ---
to keep the calculation simple.


\subsection {Assumptions and Preview of Results}

Our first simplification for this paper will be to focus on the case
of extremely high energy bremsstrahlung with
\begin {subequations}
\label {eq:Econdition}
\begin {equation}
   E \ge \kgamma \gg \Elpma \qquad \mbox{(this paper)},
\label{eq:Econdition1}
\end {equation}
with the qualification that we also avoid getting extremely close
to the limit $\kgamma{\to}E$ by assuming%
\footnote{
  One may easily eliminate the particular
  constraint (\ref{eq:Econdition2})
  by using Migdal's full formula (\textit{including} electron mass) for the
  LPM rate as $\kgamma\to E$ in our (\ref{eq:qualLPM}) below.
  Since we are not going to treat the electron mass for overlap effects in this
  paper (which are important for $E \lesssim \Elpma$), we chose
  to keep Migdal's rate simple in our presentation by imposing
  (\ref{eq:Econdition2}) so that we could ignore the electron
  mass everywhere in region (\ref{eq:Econdition}).
}
\begin {equation}
   1 - \frac{\kgamma}{E} \gg \frac{\Elpma}{E} .
\label {eq:Econdition2}
\end {equation}
\end {subequations}
Note that the right-hand side of (\ref{eq:Econdition2}) is very small
because of (\ref{eq:Econdition1}).
In the region of $(\kgamma,E)$ defined by (\ref{eq:Econdition}),
it turns out that the mass $m$ of the high-energy electron may be
ignored, and the medium-induced mass for the photon (known in
this context as the ``dielectric effect'') may also be ignored.
We leave to future work \cite{softqed2} the treatment of the massive case,
which will allow analysis of the LPM effect at lower energies.

In the region (\ref{eq:Econdition}), we will first give qualitative
arguments in section \ref{sec:qualitative} that quantum overlap of
bremsstrahlung and subsequent pair production becomes important when
\begin {equation}
  \kgamma \lesssim \alpha E \qquad\quad \mbox{(overlap effects)}.
\label {eq:ovlapimportant}
\end {equation}
For the remainder of this paper, we will refer to bremsstrahlung as
\begin {align*}
   \mbox{soft:}      &\quad \kgamma \ll E , \\
   \mbox{very soft:} &\quad \kgamma \ll \alpha E .
\end {align*}
Remember that we are analyzing the region (\ref{eq:Econdition}), and so
even our ``very soft'' photons are extremely high energy
($\kgamma \gg \Elpma$).
We will argue parametrically that, within region (\ref{eq:Econdition}),
the net differential rate for the electron to lose energy $\kgamma$ by
photon emission (inclusive of whether or not the subsequent pair production
by that photon overlaps the bremsstrahlung process itself) is
\begin {subequations}
\label {eq:qual}
\begin {align}
   \frac{d\Gamma}{dk_\gamma}
   \simeq \left[ \frac{d\Gamma}{dk_\gamma} \right]_\LPM
   \sim
   \sqrt{ \frac{\kgamma \Elpma}{(E{-}\kgamma)E} }
     \times
     \left[ \frac{d\Gamma}{d\kgamma} \right]_\BH
   & \mbox{ for $\alpha E \ll \kgamma \le E$ ~~
             {\footnotesize [subject to (\ref{eq:Econdition})]}, }
\label {eq:qualLPM}
\\
   \frac{d\Gamma}{dk_\gamma}
   \sim
   \sqrt{ \frac{\alpha^2\Elpma}{\kgamma} }
     \times
     \left[ \frac{d\Gamma}{d\kgamma} \right]_\BH
   \gg \left[ \frac{d\Gamma}{dk_\gamma} \right]_\LPM
   & \mbox{ for $\Elpma \lesssim \kgamma \ll \alpha E$. }
\label{eq:qualLPM+}
\end {align}
\end {subequations}
The parametric form for the LPM rate shown in (\ref{eq:qualLPM}) differs
slightly from (\ref{eq:LPrate}) because here we do not assume the
soft photon limit.  We will review this formula later.

\begin {figure}[t]
\begin {center}
  \includegraphics[scale=0.9]{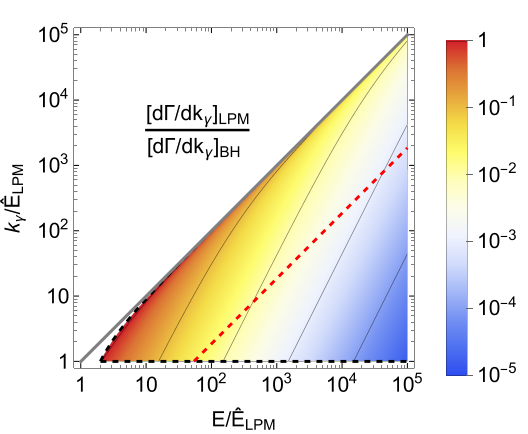}
  \hspace{0.1in}
  \includegraphics[scale=0.9]{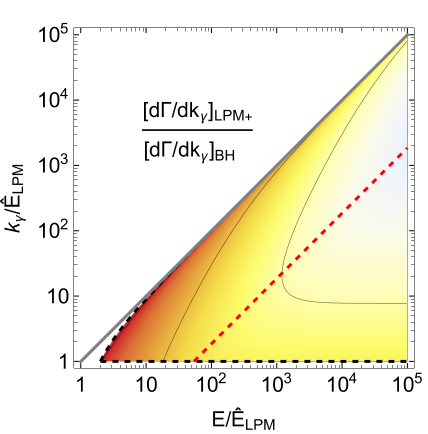}
  \caption{
     \label{fig:overBH}
     (left) Log-log-log contour plot of the ratio $\LPM/\BH$
     of the original
     LPM rate (\ref{eq:LPM})
     to the Bethe-Heitler rate (\ref{eq:BH1}) vs.\ $E/\Elpm$ and
     $\kgamma/\Elpm$.
     (right) The same, but now for $\protect\LPMplus/\BH$, which includes
     the effects of overlapping pair
     production of the bremsstrahlung photon according to (\ref{eq:LPMplus}).
     The mass of the electron has been ignored in these calculations, which
     means that the results break down close to or outside of the boundary
     depicted by the dashed black lines [see condition (\ref{eq:Econdition})].
     The precise meaning of the
     red dashed line in both plots is explained later in the caption of
     fig.\ \ref{fig:LPM+overLPM}; here it is intended to guide the eye
     to the transition where the $\protect\LPMplus$ rate starts to deviate from
     ordinary $\LPM$ at $k_\gamma \sim \alpha E$.
     See appendix \ref{app:bmin} for some minor
     qualifications concerning ignoring
     scale dependence (\ref{eq:qhat}) of $\qhat$ when taking the ratios
     $\LPM/\BH$ and $\protect\LPMplus/\BH$.
  }
\end {center}
\end {figure}

\begin {figure}[t]
\begin {center}
  \includegraphics[scale=0.9]{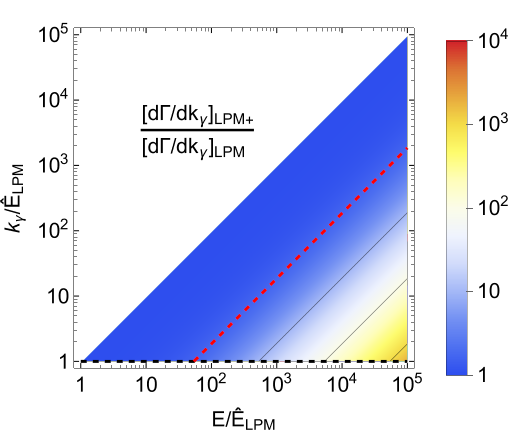}
  \hspace{0.1in}
  \includegraphics[scale=0.75]{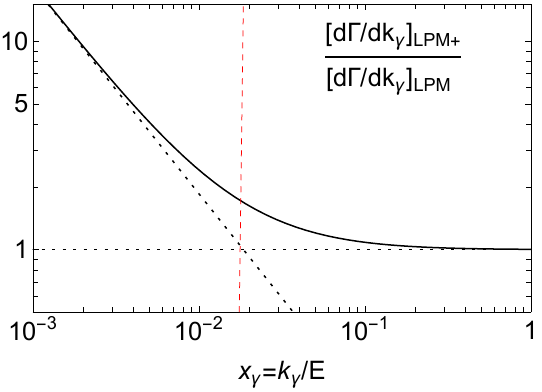}
  \caption{
     \label{fig:LPM+overLPM}
     (left) Log-log-log contour plot of the $\protect\LPMplus/\LPM$ ratio
     of the $\protect\LPMplus$ rate (which includes overlapping pair production)
     to the LPM rate (which does not) vs.\
     $E/\Elpm$ and $\kgamma/\Elpm$.
     (right) The same ratio plotted on
     a log-log plot as $\protect\LPMplus/\LPM$ vs.\ $x_\gamma = \kgamma/E$.
     The sloped dotted line shows the $x_\gamma \ll \Nf\alpha$ result
     limit (\ref{eq:previewLPMplus}) [for $\Nf{=}1$ and $\alpha{\simeq}1/137$].
     The red dashed line here and in fig.\ \ref{fig:overBH} marks the value
     of $x_\gamma$ where these two limiting formulas are equal.
     See appendix \ref{app:bmin} for some minor
     qualifications concerning ignoring
     scale dependence (\ref{eq:qhat}) of $\qhat$ when taking the ratio
     $\protect\LPMplus/\LPM$.
  }
\end {center}
\end {figure}

To make (\ref{eq:qual}) more concrete, we now segue from
previewing results of qualitative arguments to previewing results
from a full calculation of the rate
(with a caveat to be explained shortly).
Figure \ref{fig:overBH} compares the ordinary LPM suppression factor
(the ratio of the LPM rate to the Bethe-Heitler rate)
to a calculation that also includes the effects of bremsstrahlung overlapping
with pair production, which we will refer to as the ``$\LPMplus$'' effect
in order to give it a simple acronym.
Comparison of the left and right plots shows a transition from (i) the
$\LPMplus/\BH$ ratio matching the LPM rate for $k_\gamma \gg \alpha E$ to (ii)
$\LPMplus/\BH \gg \LPM/\BH$  (i.e.\ much less suppression) for
$k_\gamma \ll \alpha E$.
To more directly display the enhancement, fig.\ \ref{fig:LPM+overLPM}
shows the ratio $\LPMplus/\LPM$ of the
$\LPMplus$ rate to the ordinary $\LPM$ rate.
On the left of fig.\ \ref{fig:LPM+overLPM}, we
show this ratio as a contour plot for easy reference back
to fig.\ \ref{fig:overBH}.
But the $\LPMplus/\LPM$ ratio depends only on $\kgamma/E$ in
region (\ref{eq:Econdition1})
and so may be more simply and accurately plotted as on the right
of fig.\ \ref{fig:LPM+overLPM}.

A minor caveat concerning
figs.\ \ref{fig:overBH} and \ref{fig:LPM+overLPM}.
Though our qualitative arguments will not depend on this, we
will find it convenient to initially perform our
\textit{explicit} calculations in the
large-$\Nf$ limit of QED in order to reduce the number of
diagrams that must be evaluated.  That is,
we formally pretend that there are
$\Nf{\gg}1$ flavors of electron with negligible mass.
That said, we will explain at the end of this paper why we believe that
our large-$\Nf$ results are nonetheless reliable for small $\Nf$
(e.g.\ $\Nf{=}1$) as well.

To place the number $\Nf$ of flavors in context,
we should clarify that the constant $\Elpma$ turns out to
depend on the fourth power of the mass of the high-energy lepton,
and that the examples $\Elpma = 2.5$ TeV for Gold and $\Elpma = 234$ PeV
for air (at standard temperature and pressure)
were for bremsstrahlung by high-energy electrons, as opposed to muons.
So, in the real world, $\Nf = 1$ for the situation we consider in
(\ref{eq:Econdition}) unless one pushes to energies that are
roughly $(m_\mu/m_e)^4 \sim 10^9$ times higher than the values
quoted for $\Elpma$.


\subsubsection*{\boldmath$\Elpm$, $\Elpma$, and $\qhat$}
\label {sec:Elpm}

We find it convenient to define the $\Elpm$ appearing in
figs.\ \ref{fig:overBH} and \ref{fig:LPM+overLPM} with slightly
different normalization than the $\Elpma$
\cite{RPP2024,SpencerReview} whose values were quoted earlier.

$\Elpma$ is usually defined in terms of the radiation length $X_0$
in the medium, which is the distance over which a
high-energy
particle (but not so high that the LPM effect is relevant)
loses $1/e$ of its energy.  We won't be focused on the details of
the exact value of $X_0$ for different materials,
but it will be relevant that at leading-log order
\begin {equation}
   X_0 \approx \left[ \frac{4 n Z^2 \alpha^3}{m^2} \ln(a_Z m) \right]^{-1}
\label {eq:X0ll}
\end {equation}
for a high-energy lepton of mass $m$ traveling through
a medium composed of a single (non-light) element with atomic number $Z$,
number density $n$, and the relevant screening distance of the Coulomb fields
of the nuclei given by the Thomas-Fermi atomic radius
$a_Z \sim Z^{-1/3} a_{\rm Bohr} \sim Z^{-1/3}/\alpha m_e$.  Above, the
argument of the logarithm $\ln(a_Z m_e)$ arises because there is a range
of impact parameters, from $a_Z$ down to the Compton wavelength $1/m$
of the lepton, that contribute equally to the Bethe-Heitler radiation
rate, and the argument of the logarithm is (parametrically) the ratio
of those two scales.  A common approximation \cite{BH}%
\footnote{
  For more accuracy, and for mixtures of elements,
  see the review in section 34.4.2
  of ref.\ \cite{RPP2024}, which is for $m = m_e$.
  Note that, in formulas, the $X_0$ there 
  corresponds to what we call $X_0$ here times the mass density of the medium.
  Our (\ref{eq:X0}), which is also used in the Klein review
  \cite{SpencerReview}, corresponds to keeping just the $L_{\rm rad}$ term of
  eq.\ (34.25) of ref.\ \cite{RPP2024}.
}
which goes beyond leading-log
order by adjusting the coefficient inside the logarithm is
\begin {equation}
   X_0 \simeq \left[ \frac{4 n Z^2 \alpha^3}{m^2} \ln(184\, Z^{-1/3} m/m_e)
                     \right]^{-1}
\label {eq:X0}
\end {equation}
for non-light atoms.  $\Elpma$ represents the energy scale above which
the mass $m = m_e$ of the electron starts to become ignorable in calculations
of the LPM rate for democratic splittings, by which we mean the case
where neither daughter of a single splitting process (whether bremsstrahlung
$e{\to}e\gamma$ or pair production $\gamma{\to}e^+ e^-$) has significantly
less energy than the other.  Parametrically, $\Elpma$ turns out to be
of order $\alpha m^2 X_0$.  Its exact normalization at the level of
factors of 2 is a matter of convention, but
refs.\ \cite{SpencerReview,RPP2024} adopt the convention%
\footnote{
  As noted in ref.\ \cite{RPP2024}, this definition differs by a factor
  of two from the definition used in ref.\ \cite{SLAC1}. 
}
\begin {equation}
   \Elpma \equiv \frac{\alpha m^2 X_0}{4\pi} \,.
\label {eq:Elpma}
\end {equation}
Note that (\ref{eq:X0}) and (\ref{eq:Elpma}) imply that $\Elpma$ scales
with mass as $m^4$ (up to a logarithm), as mentioned earlier.

The reason for reviewing this is in part because we are interested
in a region of energies where the mass $m$ of the high-energy lepton
is negligible, and so it is somewhat circuitous to describe rates in
terms of variables like $X_0$ that depend on $m$, especially given
that $X_0$ is no longer the actual radiation length once the LPM effect
becomes significant.  Another reason is that we will be adopting
techniques previously used to study overlapping formation times for
the case of very high energy partons traversing a quark-gluon plasma,
where the convention is to present formulas in terms of a different
characteristic of the medium known as the transverse momentum diffusion
parameter (sometimes called the jet quenching parameter)
$\qhat$, which is independent of the mass $m$.  The variable $\qhat$
also incorporates changes to the argument of the Bethe-Heitler
logarithm [$\ln(a_Z m_e)$ in our case] that arise as the LPM effect
becomes more and more significant.

Physically, $\qhat$ is defined in terms of the typical transverse
momentum kick $\Delta p_\perp$ that the high-energy charged particle picks
up in a time $\Delta t$ traversing the medium, in the limit that
$\Delta t$ is large compared to the
mean-free time between elastic collisions with the medium.
Since the transverse momentum transfers $\delta\p_\perp$
from each elastic collision are random, the total $\Delta p_\perp$ grows
like a random walk, so that
$(\Delta p_\perp)^2 \propto \Delta t$.
The (medium-dependent) proportionality constant
is called $\qhat$:
\begin {equation}
  (\Delta p_\perp)^2_{\rm typical} \simeq \qhat \, \Delta t .
\label {eq:qhatdef}
\end {equation}
For the type of medium considered here, $\qhat$ is given
at leading-log order by%
\footnote{
  \label{foot:qhat}
  See, for example, appendix C.4 of ref.\ \cite{qedNf}.
  The notation $\qhat_{\rm eff}(b^{-1})$ there corresponds, for
  $b \ll 1/m_e$, to the notation
  $\qhat(b)$ in (\ref{eq:qhat}) here.
}
\begin {equation}
   \qhat(\bmin) \approx 8\pi n Z^2 \alpha^2 \ln\Bigl( \frac{a_Z}{\bmin} \Bigr)
\label {eq:qhat}
\end {equation}
where $\bmin$ represents the
minimum relevant impact parameter for individual elastic collisions
with the medium.  This makes $\qhat$ logarithmically
sensitive to the relevant time scale $\Delta t$ (which for LPM rate
calculations will be the formation time) because collisions with
extremely small $b$ are only relevant if they are likely
to occur within the time $\Delta t$.
In our problem, the relevant values of
$\bmin(\Delta t)$ can vary from
the Compton wavelength $1/m$ of the high-energy particle down to the
nuclear radius $R_A$:%
\footnote{
  Impact parameters small compared to $R_A$ lose the factor
  of $Z^2$ in rate coming from interacting coherently with the entire nuclear
  charge.
}
\begin {equation}
   \frac{1}{m} \gtrsim \bmin \gtrsim R_A .
\end {equation}
A formula for $\bmin$, and its
relation to Migdal's notation \cite{Migdal}, is given in
appendix \ref{app:bmin}.
Because of an accident of nature in the size of $a_Z$, $m_e^{-1}$, and
$R_A$, the limits of the range
$\qhat(m_e^{-1}) \lesssim \qhat(\bmin) \lesssim \qhat(R_A)$ differ by
approximately a factor of two,
\begin {equation}
   \qhat(R_A) \simeq 2\kern1pt\qhat(m_e^{-1})
\label {eq:accident}
\end {equation}
for $m = m_e$.

At leading-log order, the relation between $X_0$ (\ref{eq:X0ll}) and
$\qhat$ (\ref{eq:qhat}) is simply
\begin {equation}
   X_0 \approx \frac{2\pi m^2}{\alpha\kern1pt\qhat(m^{-1})} \,,
\label {eq:X0vsqhat}
\end {equation}
and so (\ref{eq:Elpma}) becomes
\begin {equation}
   \Elpma \approx \frac{m^4}{2\kern1pt\qhat(m^{-1})} \,.
\label {eq:Elpmaqhat}
\end {equation}

Since our derivations will all be in terms of $\qhat$, we find it
more convenient to define our $\Elpm$ as simply
\begin {equation}
  \Elpm \equiv \frac{m^4}{\qhat(\bmin)} \,,
\label {eq:Elpm}
\end {equation}
which differs from (\ref{eq:Elpmaqhat}) by
(i) a factor of 2 and (ii) a mild logarithmic dependence on $(\kgamma,E)$.

When it came to solving for the LPM rate, Migdal was able to find
analytic answers by treating the logarithm in (\ref{eq:X0}) as large,
and working to leading-log order.  We will do the same here.
In what follows, we will stop writing ``$\approx$'' for this
particular approximation but will instead use equal signs.
The approximation turns out to be equivalent to treating the distribution
of $\Delta p_\perp$ from multiple collisions with the medium as
(by the central limit theorem and its generalizations) approximately
Gaussian, with width given by (\ref{eq:qhatdef}), ignoring
power-law tails to the distribution.%
\footnote{
  For a discussion of how to systematically include the sub-leading
  corrections from the power-law tails, order by order, see
  the discussion of the Improved Optical Expansion (IOE) in
  refs.\ \cite{IOE1,IOE2,IOE3}, for example.
}
This is the \textit{only}
large-logarithm approximation we will make in this paper; we
do not ignore any other effects that are sub-leading in some logarithm.%
\footnote{
  See, for example, eq.\ (\ref{eq:previewLPMplus}).
}


\subsubsection*{Rate formulas}

We wrap up our preview by summarizing the explicit formulas
used to generate figs.\ \ref{fig:overBH} and \ref{fig:LPM+overLPM}.

Here and in the rest of the paper, we will often specify the photon
energy by its energy fraction relative to the original electron,
\begin {equation}
   x_\gamma \equiv \frac{\kgamma}{E} \,.
\end {equation}
With this notation, the Bethe-Heitler rate formula, written in terms of
$\qhat(m^{-1})$, is
\begin {equation}
  \left[ \frac{d\Gamma}{dx_\gamma} \right]_\BH \simeq 
  \frac{\alpha \qhat}{6\pi m^2} \,
  \bigl[ 2 P_{e\to\gamma}(x_\gamma) + x_\gamma \bigr] ,
\label{eq:BH1}
\end {equation}
where $P_{e\to\gamma}(x_\gamma)$ is the unregulated
Dokshitzer-Gribov-Lipatov-Altarelli-Parisi (DGLAP) splitting function
\begin {equation}
  P_{e\to\gamma}(x_\gamma) = \frac{1 + (1{-}x_\gamma)^2}{x_\gamma} \,.
\label {eq:Peg}
\end {equation}
Throughout this paper, we will simplify our writing of formulas
by not explicitly indicating the appropriate scale $\bmin$ of each
$\qhat = \qhat(\bmin)$. See appendix \ref{app:bmin} for clarification.

In the deep LPM regime, where the mass $m$ is ignorable, the standard
LPM rate \cite{Migdal} is
\begin {equation}
  \left[ \frac{d\Gamma}{dx_\gamma} \right]_\LPM \simeq 
  \frac{\alpha}{2\pi} \, P_{e\to\gamma}(x_\gamma)
     \sqrt{ \frac{x_\gamma\qhat}{(1{-}x_\gamma)E} } \,.
\label{eq:LPM}
\end {equation}

The new analytic result derived in this paper
is for the $\LPMplus$ rate, which includes the effects
of overlapping pair production.  We find
\begin {subequations}
\label {eq:LPMplus}
\begin {equation}
  \left[ \frac{d\Gamma}{dx_\gamma} \right]_{\LPMplus}
  =
  \left[ \frac{d\Gamma}{dx_\gamma} \right]_\LPM
  \left\{
    1 + \frac{\Nf\alpha}{2 x_\gamma}  \, f_\Plus(x_\gamma)
  \right\} ,
\label {eq:LPMplus1}
\end {equation}
where
\begin {equation}
  f_\Plus(x_\gamma) \simeq
  -\tfrac34 \left[
    \psi\bigl(
      1{+}\tfrac{3\Nf\alpha}{16x_\gamma}
    \bigr)
    + \ln(\pi x_\gamma)
  \right]
  + \tfrac18
  + \tfrac54 \ln 2
  \,.
\label {eq:fplus}
\end {equation}
\end {subequations}
Above, $\psi$ is the digamma function $\psi(z) \equiv \Gamma'(z)/\Gamma(z)$.
Formally, the large-$\Nf$ limit of QED
corresponds to taking ${\Nf\to\infty}$ while
keeping $\Nf\alpha$ finite.  In detail,
we will make some arguments in our derivation of (\ref{eq:fplus})
easier to justify by specifically
considering the case $\Nf\alpha \ll 1 \ll \Nf$.

In the limit $x_\gamma \gg \Nf\alpha$, the
$\Nf\alpha f_\Plus(x_\gamma)/2x_\gamma$ term in (\ref{eq:LPMplus1}) is suppressed,
and so the $\LPMplus$ rate becomes simply the ordinary
$\LPM$ rate (\ref{eq:LPM}), up to small corrections, as seen in
fig.\ \ref{fig:LPM+overLPM}.
In the opposite limit,
the $\Nf\alpha f_\Plus(x_\gamma)/2x_\gamma$ term in (\ref{eq:LPMplus1})
is dominant, with
\begin {equation}
  f_\Plus(x_\gamma) \simeq
    \tfrac34 \ln \bigl( \tfrac{32}{3\pi\Nf\alpha} \bigr)
    + \tfrac18 + \tfrac12 \ln2
  \qquad \mbox{for $x_\gamma \ll \Nf\alpha$} .
\end {equation}
Correspondingly,
\begin {equation}
  \left[ \frac{d\Gamma}{dx_\gamma} \right]_{\LPMplus}
  \simeq
  \frac{\Nf\alpha^2}{2\pi x_\gamma^{3/2}}
  \sqrt{ \frac{\qhat}{E} } \,
  \Bigl[
    \tfrac34 \ln \bigl( \tfrac{32}{3\pi\Nf\alpha} \bigr)
    + \tfrac18 + \tfrac12 \ln2
  \Bigr]
  \qquad \mbox{for $x_\gamma \ll \Nf\alpha$} ,
\label {eq:previewLPMplus}
\end {equation}
Parametrically, this limit matches the qualitative behavior
previewed earlier in (\ref{eq:qualLPM+}) for $\Nf{=}1$,
except for the additional logarithmic dependence on $\alpha$.


\subsection {Outline}

The next section gives qualitative arguments for our claims and
contrasts them with the qualitative arguments of Galitsky and Gurevich.
Section \ref{sec:LPM} reviews the calculation of the ordinary
LPM effect in the formalism that we will use for our later
calculations.  Section \ref{sec:NLO} applies that formalism to
analytically compute the effect of bremsstrahlung overlapping with
pair production in large-$\Nf$ QED for soft (but not yet very soft) photons
($\Nf\alpha \ll x_\gamma \ll 1$).  That analytic result matches
numerical extrapolation of the same limit from more general
($\Nf\alpha \ll x_\gamma \le 1$) numerical results of ref.\ \cite{qedNfenergy}.
Section \ref{sec:mainresult} then extends the calculation to \textit{also}
treat very-soft bremsstrahlung ($x_\gamma \lesssim \Nf\alpha$) and derives
the main analytic result (\ref{eq:LPMplus}) of this paper.
We then argue in section \ref{sec:smallNf}
that the large-$\Nf$ approximation that we used to
simplify the set of diagrams to analyze is not actually necessary in the
soft-photon limit $x_\gamma{\ll}1$, and we believe that our final result
(\ref{eq:LPMplus}) should apply equally well to any $\Nf$, including $\Nf{=}1$.

In section \ref{sec:log}, we give a physical interpretation of the
logarithm in the $\LPMplus$ rate (\ref{eq:previewLPMplus}) as arising,
at leading-log order, from combining the
ordinary LPM pair production rate ($\gamma\to e^-e^+$) with
the Weizs\"acker-Williams probability (like QCD parton
distribution functions) of finding the photon $\gamma$
in the initial electron.

We end with a brief conclusion in section \ref{sec:conclusion}.


\section{Qualitative Arguments}
\label {sec:qualitative}

\subsection{Review of the ordinary LPM effect}

In order to clearly explain the qualitative differences between our results
and those of Galitsky and Gurevich, it will be important to first review
a simple qualitative picture of the original LPM effect in bremsstrahlung
(without yet considering the effect of overlapping pair production)
and to review how LPM suppression of the Bethe-Heitler rate depends
on the formation time (and how formation time grows with energy).
Like Landau and Pomeranchuk, in this section
we will focus on
bremsstrahlung with $x_\gamma \ll 1$ so that we can ignore back-reaction
on the electron.  But our formulas in this section will only be
parametric and so apply more generally to $x_\gamma \lesssim 1$,
provided that $1{-}x_\gamma$ is not small.


\subsubsection{A useful qualitative picture of the LPM effect}
\label{sec:LPMintro}

\begin {figure}[t]
\begin {center}
  \begin{picture}(380,420)(0,0)
    \put(18,5){\includegraphics[scale=0.6]{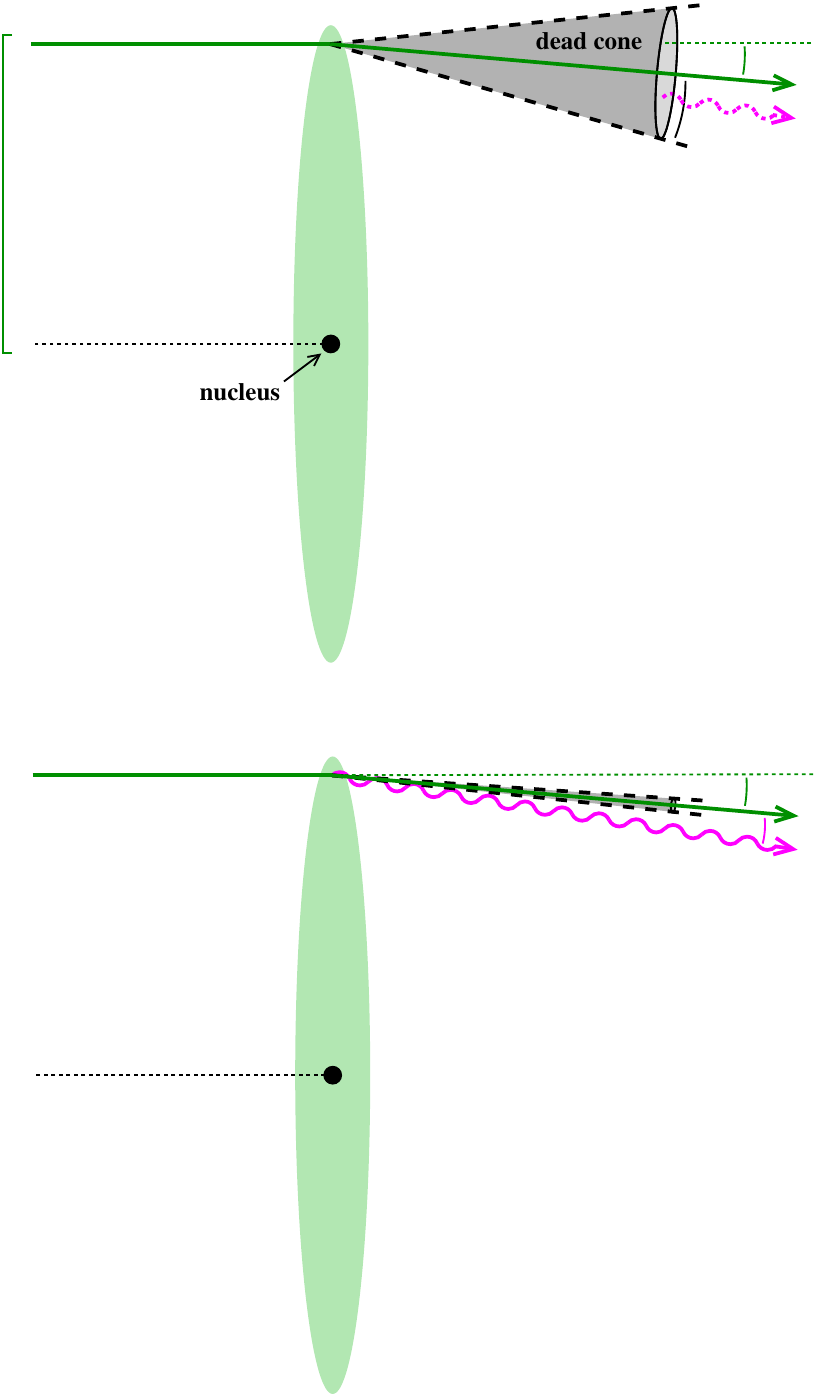}}
    \put(5,390){ \color{PineGreen}
       \rotatebox{-90}{\footnotesize
          \textbf{sets scale for \boldmath$\sigma_{\scriptscriptstyle\rm el}$}} }
    \put(28,398){ \color{PineGreen}
       \footnotesize \boldmath$b \sim a_Z$ }
    \put(215,369){ \rotatebox{-10}{\footnotesize
                     \boldmath$\theta_m \sim m/E$} }
    \put(248,168){ \rotatebox{-5}{\footnotesize
                     \boldmath$\theta_m \ll \color{PineGreen} \theta_\scatt$} }
    \put(235,388){ \color{PineGreen}
       \rotatebox{0}{\footnotesize
                     \boldmath$\theta_\scatt \sim 1/bE \sim 1/a_ZE$} }
    \put(235,176){ \color{PineGreen}
       \rotatebox{0}{\footnotesize
                     \boldmath$\theta_\scatt$} }
    \put(248,158){ \color{magenta}
       \rotatebox{-5}{\footnotesize
           \boldmath$\theta_{\gamma e} \color{black}
                     \sim \color{PineGreen} \theta_\scatt$} }
    \put(170,340){\boldmath$\boxed{m=m_e}$}
    \put(170,135){\boldmath$\boxed{m\ll \alpha m_e}$}
    \put(163,118){(sec.\ \protect\ref{sec:LPMintro} only)}
  \put(0,0){.}
  \put(0,420){.}
  \put(380,0){.}
  \put(380,420){.}
  \end{picture}
  \caption{
    \label{fig:deadcone1}
    A cartoon depicting the typical case (impact parameter $b \sim a_Z$) of a
    high-energy charged lepton of mass $m$ scattering from the Coulomb field
    of the atom's nucleus, along with the possibility that a bremsstrahlung
    photon is produced by the scattering.
    The top figure depicts
    the case $m = m_e$, where radiation from most scatterings will
    be suppressed by the dead-cone ($\theta \lesssim \theta_m \sim m/E$).
    The bottom figure depicts an imaginary world where the high-energy
    charged lepton has mass $m \ll \alpha m_e$ and the dead-cone is negligible.
    In vacuum, the collinear logarithm associated with
    $\theta_m \lesssim \theta_\gamma \ll \theta_\scatt$ is
    ignored for the purpose of this figure.
    (Also for the sake of keeping the figure simple,
    the drawing shows the dead-cone associated
    with final-state but not initial-state radiation.)
  }
\end {center}
\end {figure}
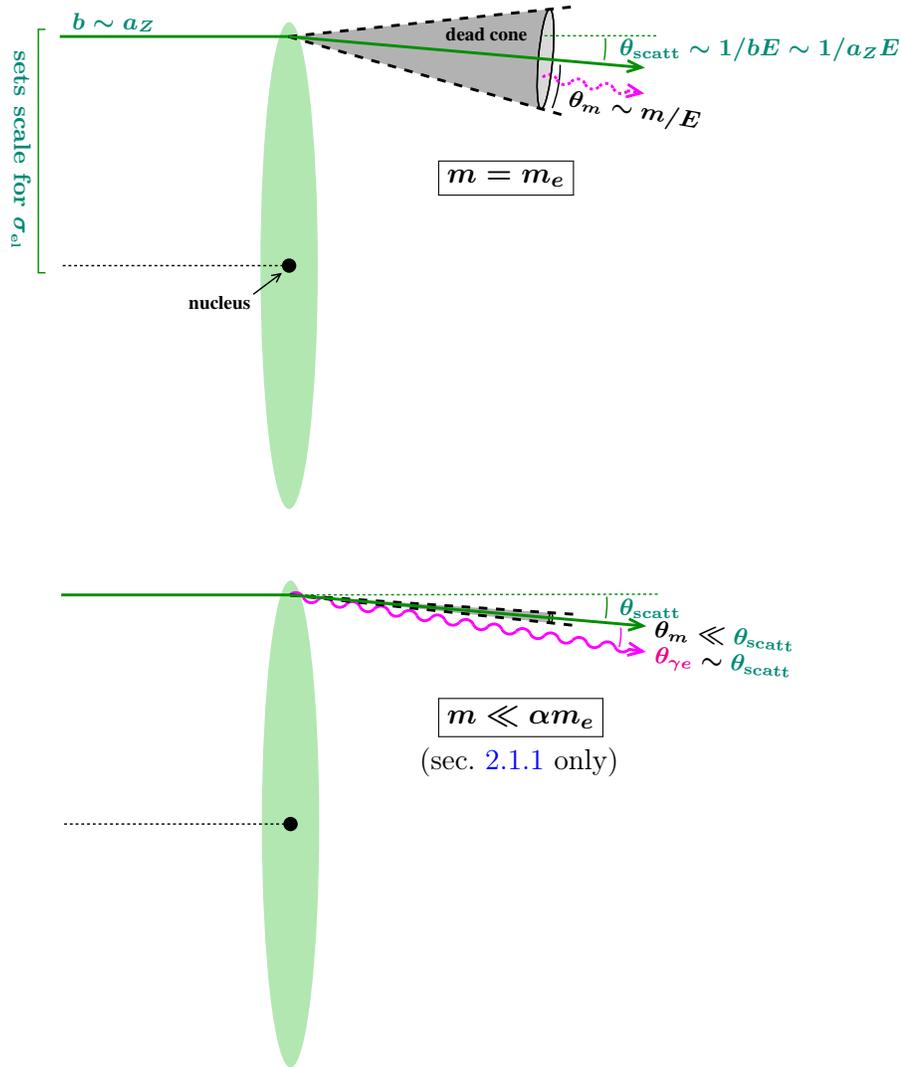

Since the high-energy particle mass does not matter in the deep LPM
regime, it will be simplest to illustrate the principles we want to
highlight by treating the high-energy lepton as massless.
However, the Bethe-Heitler rate (which we wish to compare to the LPM
rate to discuss the amount of LPM suppression) depends strongly on
$m$, as in (\ref{eq:BH1}).  This is a consequence of the dead-cone effect,
which is that radiation from a high-energy
massive particle is suppressed at angles small compared to
$\theta_m \equiv m/E$, which is large compared to the typical angle
of scattering $\theta_{\rm scatt} \sim 1/(a_Z E)$ of a high-energy electron
from an atom, as depicted in fig.\ \ref{fig:deadcone1} (top).
This slightly complicates
a back-of-the-envelope qualitative comparison of the physics of
the LPM effect in the deep LPM regime (where the dead-cone effect turns
out to be unimportant) and
the Bethe-Heitler limit.
So it will be useful to start by imagining an alternate universe
with a species of very light charged lepton with mass $m$ somewhat smaller than
$\alpha m_e$, so that the dead-cone angle $\theta_m$ is small compared to
the typical angle $\theta_{\rm scatt}$ for high-energy scattering of
that light lepton from real-world atoms made with ordinary electrons,%
\footnote{
  Alternatively, one could instead keep closer to the real world
  but replace
  the medium by an ultra-relativistic QED plasma.  In that case,
  the Debye electric screening length is order $a \sim 1/eT$ and
  the effective electron mass is $m_{\rm eff} \sim eT$ (due to
  interactions with the medium), so that $\theta_m \sim \theta_{\rm scatt}$.
  Since we are only making parametric estimates, that's as good
  as having $\theta_m \ll \theta_{\rm scatt}$.
}
as depicted in fig.\ \ref{fig:deadcone1} (bottom).
When we're done, we'll explain the slight
modification needed to conform with the real world of $m{=}m_e$.

Let $\Gamma_\el$ be the rate of scattering of a high-energy charged lepton
from the medium, and $\tau_\el \sim 1/\Gamma_\el$ be the corresponding
mean free path between such scatterings.
In our fictional world
where the dead-cone effect is never important, then, in the absence of
the LPM effect, each such scattering would
be accompanied by a probability of order
$d({\rm Prob})/dx_\gamma \sim \alpha P_{e\to\gamma}(x_\gamma) \sim \alpha/x_\gamma$
for radiating a photon (times a collinear logarithm we will not keep track of
for these parametric arguments).
The $\alpha P_{e\to\gamma}(x_\gamma)$ is associated with adding the
photon emission vertex to the scattering diagram.
Altogether, the corresponding ``Bethe-Heitler'' rate in this situation is
then%
\footnote{
  \label{foot:tauel}
  It won't be important for our argument but, for the
  sake of concreteness,
  \[
     \Gamma_\el
     \sim \sigma_\el n
     \sim (Z\alpha)^2 \sigma_{\rm geom} n
     \sim (Z\alpha)^2 a_Z^2 n,
  \]
  where $\sigma_{\rm geom} \sim \pi a_Z^2$ is the geometric cross-section
  of the atom, and so
  $\tau_\el \sim [(Z\alpha)^2 a_Z^2 n]^{-1}$.
  This means that (\ref{eq:BHfake}) is parametrically
  similar to the real Bethe-Heitler rate (\ref{eq:BH1})
  with $1/m^2$ replaced by $a_Z^2$
  [and using (\ref{eq:qhat}) for $\qhat$].
}
\begin {equation}
  \left[ \frac{d\Gamma}{dx_\gamma} \right]_\quoteBH
  \sim \alpha P_{e\to\gamma}(x_\gamma) \times \Gamma_\el
  \sim \frac{\alpha P_{e\to\gamma}(x_\gamma)}{\tau_\el}
  \sim \frac{\alpha}{x_\gamma \tau_\el}
  \qquad
  \mbox{(for $m \ll a_{\rm Bohr}^{-1} = \alpha m_e$)}
\label {eq:BHfake}
\end {equation}
(up to a collinear log).

\begin {figure}[t]
\begin {center}
  \includegraphics[scale=0.5]{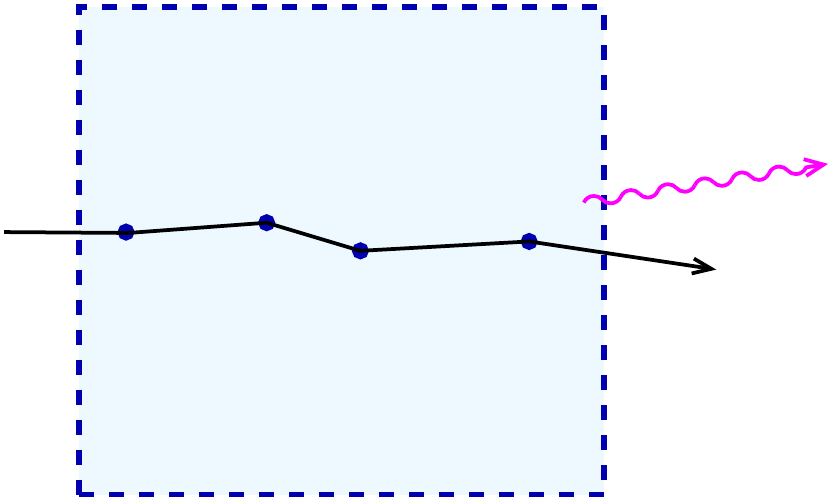}
  \caption{
    \label{fig:explainLPM1}
    A high-energy electron undergoing multiple typical (small-angle)
    collisions in a section of the medium (represented by the dashed
    blue square), with the possibility of a high-energy bremsstrahlung
    photon resulting from those collisions.
  }
\end {center}
\end {figure}

\begin {figure}[t]
\begin {center}
  \includegraphics[scale=0.5]{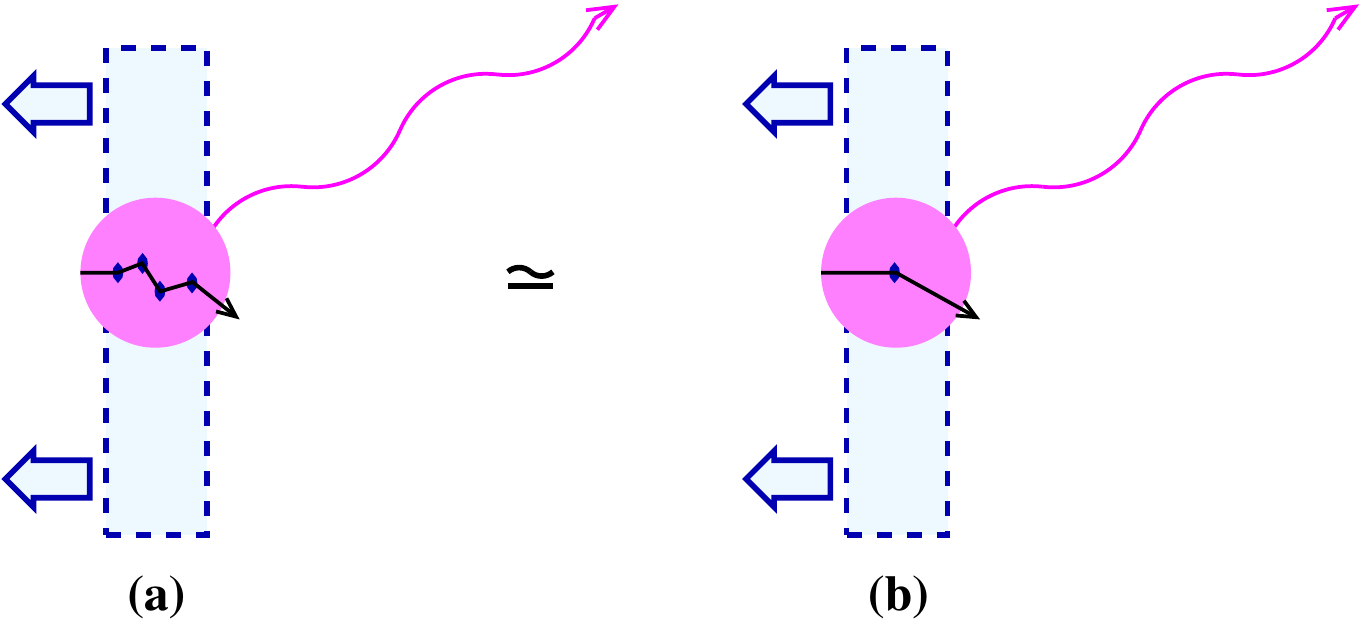}
  \caption{
    \label{fig:explainLPM2}
    (a) Like fig.\ \ref{fig:explainLPM1} but now highly boosted to the point
    of view of an observer who was moving in the direction of the original
    electron.  The pink circle represents details of the collision on scales
    smaller than the bremsstrahlung photon wavelength, which cannot be
    resolved.  The photon cannot tell the difference between this situation
    and (b) the same picture except that there was only one collision with
    the medium.
  }
\end {center}
\end {figure}

In order to address the LPM effect, consider
several elastic collisions in a row, as shown in
fig.\ \ref{fig:explainLPM1} in the rest frame of the medium.
(The figure shows just one selected section of the medium; imagine that there
are more collisions both before and after.)
Deflection angles are almost always very small at high energy, and so
the bremsstrahlung radiation is nearly collinear.
Now imagine boosting to a frame that is moving to the right at
close to the speed of light --- say, the center-of-momentum frame
of the two daughters $e\gamma$ of the bremsstrahlung $e{\to}e\gamma$.
In that boosted frame,
as depicted in fig.\ \ref{fig:explainLPM2}a,
the medium is extremely Lorentz contracted, the
electron and bremsstrahlung photon have extremely less energy,
and so the photon has extremely longer wavelength than in the
original frame of fig.\ \ref{fig:explainLPM1}.
That means that
the Lorentz-expanded photon wavelength may be large compared
to the Lorentz-contracted elastic mean free path, as we've crudely depicted.

\begin {figure}[t]
\begin {center}
  \includegraphics[scale=0.48]{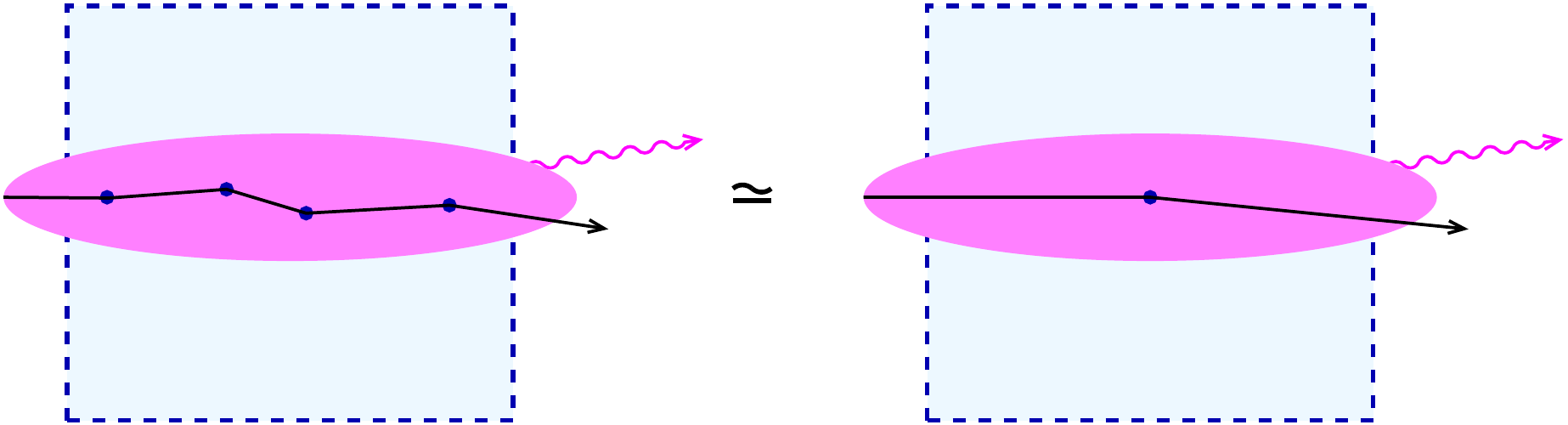}
  \caption{
    \label{fig:explainLPM3}
    The equivalence of fig.\ \ref{fig:explainLPM2}, now depicted back
    in the frame of fig.\ \ref{fig:explainLPM1} (the rest frame of
    the medium).  The stretched-out ellipse indicates the shape of the
    region that cannot be resolved by the high-energy, nearly collinear
    bremsstrahlung photon.
  }
\end {center}
\end {figure}

Since a single photon cannot effectively resolve details smaller than its
wavelength, the bremsstrahlung photon in the boosted frame
cannot know whether it was produced as the result of multiple elastic
collisions
with the medium (fig.\ \ref{fig:explainLPM2}a) or a single collision
with the medium (fig.\ \ref{fig:explainLPM2}b), and so the total
probability for photon production from the multiple collisions
of fig.\ \ref{fig:explainLPM2}a must be approximately the same
as the probability for photon production from a single collision,
which we denote by the ``$\simeq$'' sign in the figure.%
\footnote{
\label{foot:mclerran}
  We are unsure of the history of this qualitative approach to arguing
  for the LPM effect by switching frames.  The argument presented
  here is our own elaboration of a brief comment made to one of us
  by Larry McLerran circa 1987.  McLerran tells us that he heard it
  from J. D. Bjorken.  So far, we have not managed to track
  it back further.
}
Boosting
back to the rest frame of the medium, the unresolved region of
fig.\ \ref{fig:explainLPM2} becomes the long, stretched-out region
of fig.\ \ref{fig:explainLPM3}, whose length corresponds to the
``formation time'' $t_\form$ for the bremsstrahlung photon.  The LPM effect
is that multiple collisions with the medium do not give independent
chances for bremsstrahlung but instead give only one chance.
So, instead of the probability for bremsstrahlung being of order
$\alpha\,P_{e\to\gamma}(x_\gamma)$ per elastic collision, it is
instead $\alpha\,P_{e\to\gamma}(x_\gamma)$ per formation time, in the
case where the formation time exceeds the mean free path $\tau_\el$.
Correspondingly, the rate (\ref{eq:BHfake}) is replaced by
\begin {equation}
  \left[ \frac{d\Gamma}{dx_\gamma} \right]_\LPM
  \sim \frac{\alpha P_{e\to\gamma}(x_\gamma)}{t_\form}
  \sim \frac{\alpha}{x_\gamma t_\form}
  \qquad
  \mbox{(for $t_{\rm form} \gg \tau_\el$)} .
\label {eq:LPMqual}
\end {equation}
In terms of our ``Bethe-Heitler'' rate (\ref{eq:BHfake}), the
LPM suppression effect is
\begin {equation}
  \left[ \frac{d\Gamma}{dx_\gamma} \right]_\LPM
  \sim \frac{\tau_\el}{t_\form} \left[ \frac{d\Gamma}{dx_\gamma} \right]_\quoteBH
  \qquad
  \mbox{(for $t_{\rm form} \gg \tau_\el$)} .
\label {eq:LPMvBHqual}
\end {equation}
The suppression factor is the inverse of the number $t_\form/\tau_\el$ 
of elastic scatterings
that the bremsstrahlung photon cannot resolve.

When we later incorporate the effects of overlapping pair
production, our identification of the LPM suppression factor as
order $\tau_\el/t_\form$ will be the most significant difference
between our qualitative arguments and those of Galitsky and Gurevich,
even though our different approaches, and different formulas for
the LPM suppression factor, give the same qualitative results
(those of Landau and Pomeranchuk)
when overlapping pair production is \textit{not} considered.

In this paper, we will refer to the case of significant LPM suppression,
where the formation time encompasses many elastic scatterings, as the
``deep LPM'' or ``multiple scattering'' regime.


\subsubsection {Real world revisited}
\label {sec:realworld}

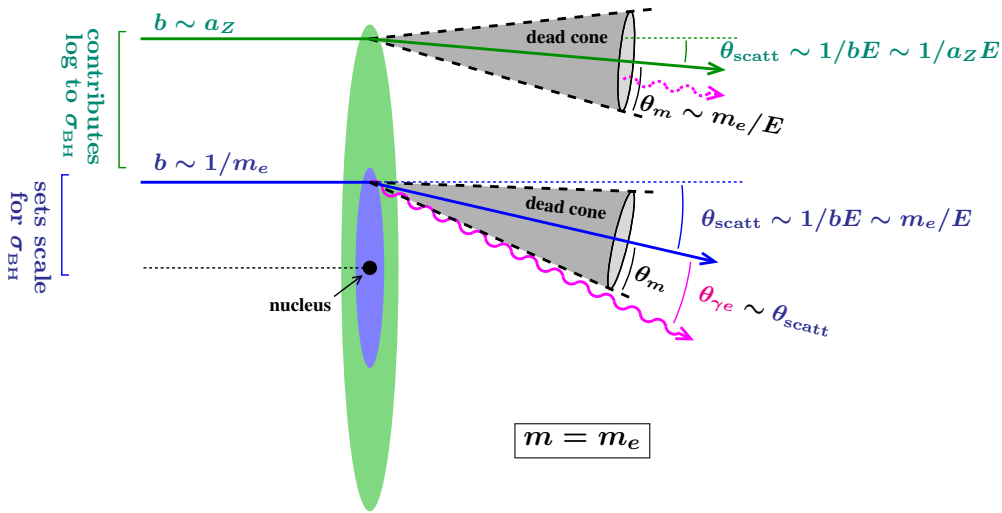
\begin {figure}[t]
\begin {center}
  \begin{picture}(380,200)(0,0)
    \put(28,5){\includegraphics[scale=0.6]{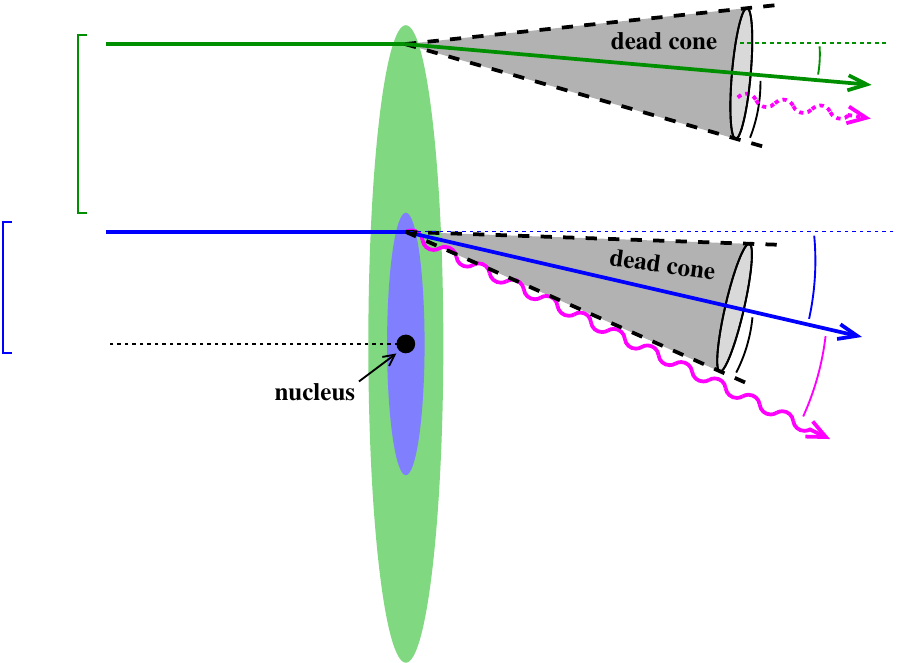}}
    \put(35,188){ \color{PineGreen}
       \rotatebox{-90}{\footnotesize \textbf{contributes}} }
    \put(25,186){ \color{PineGreen}
       \rotatebox{-90}{\footnotesize
            \textbf{log to \boldmath$\sigma_{\scriptscriptstyle\rm BH}$}} }
    \put(15,130){ \color{Blue}
       \rotatebox{-90}{\footnotesize \textbf{sets scale}} }
    \put(5,126){ \color{Blue}
       \rotatebox{-90}{\footnotesize
            \textbf{for \boldmath$\sigma_{\scriptscriptstyle\rm BH}$}} }
    \put(60,187){ \color{PineGreen}
       \footnotesize \boldmath$b \sim a_Z$ }
    \put(60,134){ \color{Blue}
       \footnotesize \boldmath$b \sim 1/m_e$ }
    \put(243,157){ \rotatebox{-10}{\footnotesize
                     \boldmath$\theta_m \sim m_e/E$} }
    \put(241,93){ \rotatebox{-20}{\footnotesize \boldmath$\theta_m$} }
    \put(274,176.5){ \color{PineGreen}
       \rotatebox{0}{\footnotesize
                     \boldmath$\theta_\scatt \sim 1/bE \sim 1/a_ZE$} }
    \put(267,113){ \color{Blue}
       \rotatebox{0}{\footnotesize
                      \boldmath$\theta_\scatt \sim 1/bE \sim m_e/E$} }
    \put(265,85){ \color{magenta}
       \rotatebox{-15}{\footnotesize
           \boldmath$\theta_{\gamma e} \color{black} \sim
                     \color{Blue} \theta_\scatt$} }
    \put(200,30){\boldmath$\boxed{m=m_e}$}
%
  \end{picture}
  \caption{
    \label{fig:deadcone2}
    Like fig.\ \ref{fig:deadcone1} (top) for impact parameter $b \sim a_Z$,
    but here compared to smaller impact parameters $b\sim 1/m_e$ that are
    not significantly dead-cone suppressed.  The green region depicts the
    geometric cross-sectional area of the atom (viewed at an angle), whereas
    the blue indicates the parametrically smaller geometric cross-sectional
    area (of radius ${\sim}m^{-1}$)
    where radiation is not suppressed by the dead-cone effect.
  }
\end {center}
\end {figure}

That said, we should now take a moment to return to the real world
and briefly explain how the preceding should be adapted.
The most probable impact parameter $b$ for scattering is the size
of the atom ($b \sim a_Z$) because that's the biggest target.
However, in the context of the Bethe-Heitler calculation,
this produces small momentum transfers $\delta p_\perp \sim 1/a_Z$ and
so small deflections $\theta_{\rm scatt}$, which in our fictional world
of a very light mass lepton would produce radiation at angles
$\theta_\gamma \lesssim \theta_{\rm scatt}$.  In the real world, these
angles are deeply inside the dead cone, and so, even though elastic
scattering is dominated by $b \sim a_Z$, the probability of bremsstrahlung
for each such scattering is deeply suppressed by the dead-cone effect.
To avoid the dead-cone effect, we need $\theta_{\rm scatt} \gtrsim \theta_m$
and so $b \lesssim 1/m$, with $m=m_e$ for high-energy electrons.
Of those impact parameters, scattering with $b \sim 1/m$
is the most likely, and so this is the impact parameter that determines
the overall size of the real-world Bethe-Heitler rate (\ref{eq:BH1}),
as depicted in fig.\ \ref{fig:deadcone2}.
There is a log enhancement in the BH rate
because, though radiation
from scattering with large impact parameter is more dead-cone suppressed,
the chance of such scattering is larger, and the two effects cancel, leading
to a log enhancement $\ln(a_Z m)$ of the ratio of scales from
$b \sim 1/m$ up to $b \sim a_Z$.

The take-away is that, if not worrying about keeping track of
logarithms in qualitative estimates, then the
dead-cone effect may be ignored provided that,
for the purpose of the Bethe-Heitler radiation rate,
we consider the only relevant elastic scattering to be
$b \sim 1/m$.  Our previous qualitative argument then goes through,
except that
$\tau_\el$ should be re-interpreted as the mean free path $\tau_{\el,m}$
for elastic scattering with transverse momentum transfers $\gtrsim m$.%
\footnote{
\label {foot:tauelm}
  The only change to footnote \ref{foot:tauel} is that then
  $\sigma_{\rm geom} \sim \pi/m^2$ in the real world,
  and so the re-interpreted mean free path becomes
  $\tau_{\el,m} \sim [(Z\alpha)^2 m^{-2} n]^{-1}$.
  This makes
  (\ref{eq:BHfake}) now match the real-world Bethe-Heitler rate,
  other than the factor of $\ln(a_Z m)$ that we ignored in our
  parametric argument above that $\tau_\el$ can
  be re-interpreted as $\tau_{\el,m}$.
}

For soft bremsstrahlung, the condition $\tform \gg \tau_\el$ that arose in
our qualitative discussion of the LPM effect, now re-interpreted as
$\tform \gg \tau_{\el,m}$, turns out to be equivalent%
\footnote{
  For example, from footnote \ref{foot:tauelm} and eq.\ (\ref{eq:qhat}),
  $\tau_{el,m} \sim m^2/\qhat$.  Using (\ref{eq:tform}) in
  $\tform \gg \tau_{el,m}$, along with
  $\qhat \sim m^4/\Elpma$ from (\ref{eq:Elpmaqhat}),
  then yields (\ref{eq:bremcondition}).
}
(up to factors of the logarithm ignored above)
to the condition identified by Landau and Pomeranchuk \cite{LP1,LP2}:
\begin {equation}
  k_\gamma \ll \frac{E^2}{\Elpma}
  \label {eq:bremcondition}
\end {equation}
This is always satisfied in the region (\ref{eq:Econdition}) of extremely
high energy $E \gg \Elpma$ that we consider in this paper.


\subsubsection{The LPM formation time and rate}

It will be useful to quickly
review the formation time for the ordinary LPM effect, which we now
discuss in
the context of figs.\ \ref{fig:explainLPM1}--\ref{fig:explainLPM3}.
In fig.\ \ref{fig:explainLPM2}, the bremsstrahlung is insensitive to
details of the collisions that fit within a wavelength.
We can write this condition
as roughly $|\k_\gamma\cdot\Delta\x| \lesssim 1$ in that frame.%
\footnote{
   The directional dependence of the condition
   $|\k_\gamma\cdot\Delta\x| \lesssim 1$ in this frame
   is a slightly more general statement about resolution failure
   than the non-relativistic rule of thumb that
   ``a photon cannot resolve features smaller than its wavelength.''
   Note also that, in the relativistic case,
   these non-relativistic statements do not capture the
   region of poor resolution (e.g.\ the bremsstrahlung
   formation length in fig.\ \ref{fig:explainLPM3} is much longer than
   the wavelength of the high-energy bremsstrahlung photon).  We employ them
   just to invoke the non-relativistic physics intuition of, for example,
   the resolution of conventional optical microscopes,
   and to use our earlier qualitative arguments in section \ref{sec:LPMintro} as
   motivation for (\ref{eq:resolution}) as
   the relativistic generalization.
}
That could just as well be written covariantly as
\begin {equation}
  |k^\mu \Delta x_\mu| \lesssim 1 ,
\label {eq:resolution}
\end {equation}
and in this form will be true in any frame, including the
rest frame of the medium as in fig.\ \ref{fig:explainLPM3}.
The formation time (or equivalently length) will then be given
in that frame as
\begin {equation}
  1 \sim |k^\mu \Delta x_\mu| \simeq \kgamma \tform (1 - \cos\theta_{\gamma e})
  \sim \kgamma \tform \theta_{\gamma e}^2 ,
\label {eq:tform1}
\end {equation}
and so
\begin {equation}
  \tform \sim \frac{1}{\kgamma \theta_{\gamma e}^2} \,,
\label {eq:tform2}
\end {equation}
where $\theta_{\gamma e}$ is the angle between the photon and final electron.
In QED, $\theta_{\gamma e}$ is of order the scattering angle $\theta_\scatt$
of the electron%
\footnote{
  If not for medium effects, then we should also consider
  $\theta_{\gamma e} \ll \theta_\scatt$, associated with
  a collinear logarithm in the bremsstrahlung rate.
  However, such collinear logarithms
  are not present in the ordinary LPM effect because smaller $\theta_{\gamma e}$
  leads to yet-longer formation times, whose contribution to
  bremsstrahlung is therefore yet-more LPM suppressed.
  Put another way, unlike in vacuum,
  the electron and photon cannot remain close to collinear
  forever because continued scattering of the electron with the medium
  will increase $\theta_{\gamma e}$ over time.
}
during the formation time.  This in turn is
\begin {equation}
  \theta_{\gamma e} \sim \theta_\scatt \sim \frac{\Delta p_\perp}{p_z}
  \simeq \frac{(\qhat t_\form)^{1/2}}{E}
\label {eq:thetascatt}
\end {equation}
by using (\ref{eq:qhatdef}) for the last equality.  Plugging this into
(\ref{eq:tform2}), and then solving for
$t_\form$,
\begin {equation}
  t_\form^\LPM \sim \sqrt{ \frac{E^2}{\qhat\kgamma} }
         \sim \sqrt{ \frac{E}{x_\gamma\qhat} } \,.
\label {eq:tform}
\end {equation}

The LPM bremsstrahlung rate is then given by (\ref{eq:LPMqual}),
\begin {equation}
  \left[ \frac{d\Gamma}{dx_\gamma} \right]_\LPM
  \sim \frac{\alpha}{x_\gamma t_\form^\LPM}
  \sim \frac{\alpha}{x_\gamma} \sqrt{ \frac{x_\gamma \qhat}{E} }
  \qquad
  \mbox{(for $t_{\rm form} \gg \tau_\el$)} .
\label {eq:LPMqual2}
\end {equation}
This is parametrically the same as the previously quoted result
(\ref{eq:LPM}), assuming that $x_\gamma$ is not so close to 1 that
$1{-}x_\gamma$ is small (which has been our assumption throughout
this qualitative discussion).


\subsection{The effect of pair production}
\label{sec:disruption}

To explain the effect of pair production, we will use an analogy
to QCD made by some of us in ref.\ \cite{qcdNf}.
It is well known in the study of the QCD LPM effect%
\footnote{
  The treatment of the LPM effect in QCD was originally worked out by
  Baier, Dokshitzer, Mueller, Peigne, and Schiff
  \cite{BDMPS1,BDMPS2,BDMPS3,BDMS}
  and by Zakharov \cite{Zakharov1,Zakharov2,Zakharov3}
  (BDMPS-Z).
}
that soft gluon
bremsstrahlung ($q \to qg$)
differs qualitatively from soft photon bremsstrahlung
because gluons carry color charge but photons are neutral.
For this reason, a bremsstrahlung gluon scatters from a QCD medium
roughly as easily as a quark does, whereas the scattering of a bremsstrahlung
photon in a QED medium is negligible (at high energy) compared to that
of an electron.  Since soft particles are easier to deflect with a given
transverse momentum kick $\Delta p_\perp$ than hard particles are,
the angle $\theta_{gq}$ between the soft bremsstrahlung gluon and a quark
will be much more affected by the soft gluon's interaction with the medium
than by the hard quark's interaction with the medium.
So the QED estimate (\ref{eq:thetascatt}) for
$\theta_{\gamma e}$ is replaced by%
\footnote{
  $\qhat$ for gluons is different from $\qhat$ for quarks because they
  have different color charge.  The $\qhat$ in (\ref{eq:scattQCD}) is
  the gluon one, but this makes no difference in parametric
  estimates.
}
\begin {equation}
  \theta_{gq} \sim \frac{\Delta p_\perp}{k_g}
  \simeq \frac{(\qhat t_\form)^{1/2}}{k_g} \,,
\label {eq:scattQCD}
\end {equation}
where the denominator is the soft bremsstrahlung
gluon energy $k_g$ instead of the initial quark energy $E$.
Using this angle [instead of (\ref{eq:thetascatt})]
in (\ref{eq:tform2}), and then solving for
$t_\form$, gives
\begin {equation}
  t_\form^\QCD \sim \sqrt{ \frac{k_g}{\qhat} }
         \sim \sqrt{ \frac{x_g E}{\qhat} }
\end {equation}
in contrast to (\ref{eq:tform}).
The soft gluon bremsstrahlung rate given by (\ref{eq:LPMqual}) is then
\begin {equation}
  \left[ \frac{d\Gamma}{dx_g} \right]_\LPM^\QCD
  \sim \frac{\alpha}{x_g t^\QCD_\form}
  \sim \frac{\alpha}{x_g} \sqrt{ \frac{\qhat}{x_g E} }
  \qquad
  \mbox{(for $t^\QCD_{\rm form} \gg \tau_\el$)}
\label {eq:LPMqualQCD}
\end {equation}
in contrast to (\ref{eq:LPMqual2}).
Unlike QED, where the formation time grows in the soft bremsstrahlung limit
$x_\gamma{\to}0$,
the QCD formation time \textit{shrinks}
in the corresponding limit $x_g{\to}0$.  A smaller formation time
means less LPM suppression, and so the QCD LPM bremsstrahlung
rate (\ref{eq:LPMqualQCD}) (proportional to $x_g^{-3/2}$)
grows substantially faster in the soft limit than does
the QED rate (\ref{eq:LPMqual2}) (proportional to $x_\gamma^{-1/2}$).

\begin {figure}[t]
\begin {center}
  \includegraphics[scale=0.8]{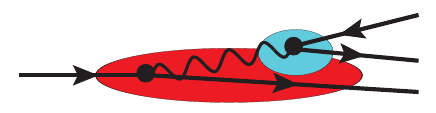}
  \caption{
    \label{fig:overlap}
    A depiction of bremsstrahlung (red formation region)
    overlapping with subsequent pair production (blue formation region).
  }
\end {center}
\end {figure}

The analogy to our QED problem is that, if soft bremsstrahlung
$e^\pm \to e^\pm\gamma$
overlaps subsequent pair production $\gamma\to e^+ e^-$, as depicted in
fig.\ \ref{fig:overlap}, then, while the bremsstrahlung is still
taking place, the soft photon has converted to a soft $e^+$ and soft $e^-$
that, like a bremsstrahlung gluon, can easily interact with and be deflected
by the medium.  This effect will shorten the overall formation time
for the bremsstrahlung process.
In particular, if the mean time $1/\Gamma_\pair$ for the photon to
convert to a pair is smaller than the LPM formation time (\ref{eq:tform}),
we can expect that the coherence of bremsstrahlung will be
disrupted so that the actual formation time will be $1/\Gamma_\pair$.
So the actual formation time behaves as
\begin {equation}
   t_\form^{\LPMplus} \sim
   \min\left( t_\form^\LPM \,,\, \frac{1}{\Gamma_\pair} \right) .
\label {eq:tformplus1}
\end {equation}
According to earlier qualitative arguments, a smaller formation
time means less LPM suppression and so a larger bremsstrahlung rate.
Specifically, (\ref{eq:LPMqual}) gives
\begin {equation}
  \left[ \frac{d\Gamma}{dx_\gamma} \right]_\LPMplus
  \sim \frac{\alpha}{x_\gamma t_\form^\LPMplus}
  \sim \frac{\alpha}{x_\gamma}
     \max\biggl( \sqrt{ \frac{x_\gamma \qhat}{E} } \,,\, \Gamma_\pair \biggr)
  \qquad
  \mbox{(for $t_{\rm form} \gg \tau_\el$)} .
\label {eq:LPMqualplus1}
\end {equation}

We now just need the pair production rate.  Landau, Pomeranchuk and Migdal
\cite{LP1,LP2,Migdal} showed that there is also
LPM suppression of pair production at
high enough energy.  The analog of the
bremsstrahlung LPM rate (\ref{eq:LPMqual}) is, for pair production,%
\footnote{
  The corresponding DGLAP splitting function
  $P_{\gamma\to e}(x) = \tfrac12 [ x^2 + (1{-}x)^2 ]$
  is $O(1)$ and so does not need to be explicitly accounted for in our
  parametric estimate (\ref{eq:pairqual1}).
}
\begin {equation}
  \Gamma_\pair^\LPM
  \sim \frac{\Nf\alpha}{\tform^\pair}
  \qquad
  \mbox{(for $\tform^\pair \gg \tau_\el$)} .
\label {eq:pairqual1}
\end {equation}
In this paper, we will only be interested in the total rate for
pair production,
which is dominated by democratic splitting.
The formation time for democratic pair production
turns out to be parametrically the same as for
democratic bremsstrahlung [(\ref{eq:tform}) with $x_\gamma{\sim}1$],
except that the initial particle energy is now that of the photon,
which we call $\kgamma$, instead of the energy $E$ we took for the
initial electron in bremsstrahlung.  So
\begin {equation}
   \tform^\pair \sim \sqrt{ \frac{\kgamma}{\qhat} }
   \qquad
   \mbox{(democratic pair production)}
\label {eq:tformpair}
\end {equation}
in (\ref{eq:pairqual1}), giving
\begin {equation}
  \Gamma_\pair^\LPM
  \sim \Nf\alpha \sqrt{ \frac{\qhat}{\kgamma} }
  \qquad
  \mbox{(for $\tform^\pair \gg \tau_{\el,m}$)} .
\label {eq:pairqual}
\end {equation}
For \textit{democratic} bremsstrahlung, the condition (\ref{eq:bremcondition})
for deep-LPM bremsstrahlung becomes $E \gg \Elpma$.  For the total 
pair rate, the condition is the same except that the initial electron
energy $E$ is replaced by the initial photon energy $k_\gamma$:
\begin {equation}
  k_\gamma \gg \Elpma .
  \label {eq:paircondition}
\end {equation}
This is the reason that, for the sake of simplicity,
we made $k_\gamma \gg \Elpma$
a condition on the region (\ref{eq:Econdition}) studied in this
paper.

Using (\ref{eq:pairqual}), the $\LPMplus$ rate (\ref{eq:LPMqualplus1})
becomes
\begin {equation}
  \left[ \frac{d\Gamma}{dx_\gamma} \right]_\LPMplus
  \sim
    \left[ \frac{d\Gamma}{dx_\gamma} \right]_\LPM
    \max\biggl( 1, \frac{\Nf\alpha}{x_\gamma} \biggr)
  \sim
     \max\biggl( 
       \frac{\alpha}{x_\gamma} \sqrt{ \frac{x_\gamma \qhat}{E} }
       \,,\, 
       \frac{\Nf\alpha^2}{x_\gamma} \sqrt{ \frac{\qhat}{x_\gamma E} }
     \biggr) .
\label {eq:LPMqualplus}
\end {equation}
This differs significantly from the ordinary LPM rate in the case
of very soft bremsstrahlung
($x_\gamma \ll \Nf\alpha$), as we previewed (without explicit $\Nf$
dependence) in (\ref{eq:ovlapimportant}).
Notice also that in this case the $\LPMplus$ rate behaves like the ordinary
QCD LPM bremsstrahlung rate (\ref{eq:LPMqualQCD}) except
multiplied by an extra factor of $\Nf\alpha$ associated with the
cost of producing the $e^+e^-$ pair from the bremsstrahlung photon.
The qualitative estimate (\ref{eq:LPMqualplus}) parametrically matches
the behavior of the more precise result previewed in (\ref{eq:previewLPMplus})
[except for the logarithmic factor $\ln(1/\alpha)$, which will be
discussed later].


\subsection{Comparison of qualitative arguments with Galitsky and Gurevich}

Galitsky and Gurevich \cite{Galitsky} also concluded that the bremsstrahlung
formation time could be shortened by pair production, just as in our
(\ref{eq:tformplus1}).  However, their version of our parametric relation
(\ref{eq:LPMvBHqual}) between the LPM
rate and the Bethe-Heitler rate was%
\footnote{
  In their paper, Galitsky and Gurevich write (\ref{eq:GGrelation}) above
  with an equal sign.  Even though their calculations are not exact,
  they attempt throughout to estimate overall factors (such as $2$ and
  $\pi$) as best they can.  Here, though, we just focus on the
  parametric behavior.
}
\begin {equation}
  \left[ \frac{d\Gamma}{dx_\gamma} \right]_\LPM
  \sim~ \frac{\tform}{\tform^\BH}
       \left[ \frac{d\Gamma}{dx_\gamma} \right]_\BH ,
\label {eq:GGrelation}
\end {equation}
with suppression factor
\begin {equation}
    \frac{\tform}{\tform^\BH} ~ \mbox{(Galitsky\&Gurevich)}
    \qquad \mbox{instead of} \qquad
    \frac{\tau_{\el,m}}{t_\form} ~ \mbox{(us)} .
\label {eq:Scompare}
\end {equation}
Here, $\tform^\BH$ is the formation time for Bethe-Heitler, which is%
\footnote{
  One way to get (\ref{eq:tformBH}) is to use (\ref{eq:tform2}) with
  $\theta_{\gamma e} \sim \theta_\scatt \sim \Delta p_\perp/E$ and then
  $\Delta p_\perp \sim m$ as in the discussion of section \ref{sec:realworld}.
}
\begin {equation}
  \tform^\BH \sim \frac{E^2}{\kgamma m^2} \,.
\label {eq:tformBH}
\end {equation}
Using this, both versions of (\ref{eq:Scompare}) parametrically
reproduce
the \textit{same} ordinary LPM brem\-sstrah\-lung suppression factor
$\sim \sqrt{\kgamma m^4/\qhat E^2} \sim \sqrt{\kgamma \Elpm/E^2}$
as in (\ref{eq:LPrate}).
But they suggest radically different behavior when overlapping
pair production becomes important and so reduces the LPM formation time
as in (\ref{eq:tformplus1}).  Galitsky and Gurevich's formula
for the LPM suppression factor $t_\form/t_\form^\BH$ suggests that the
rate is proportional to the formation time (all other things being
equal), and so a reduction in the formation time will
\textit{further suppress}
the bremsstrahlung rate compared to the ordinary LPM rate.
In contrast, our version $\tau_{\el,m}/t_\form$ is inversely proportional
to the formation time, and so a decrease in the formation time
will \textit{increase} the bremsstrahlung rate compared to the
ordinary LPM rate.

Naturally, we think our argument is the better one --- a decrease
in the formation time means fewer scatterings with the medium are blurred
together, resulting in less LPM suppression and a larger rate because of
the simple
physical picture of the LPM effect presented in section \ref{sec:LPMintro}.
Readers may remain justifiably unconvinced since qualitative estimates (ours
or others') can hide subtle errors of reasoning.
However, our qualitative conclusions will be verified by our detailed
calculations later.

In fact, there is already evidence for our point of view from a previous
large-$\Nf$ calculation \cite{qedNfenergy}, which motivated the present study.
That calculation studied the small corrections to
$[d\Gamma/dx_\gamma]_\LPM$ from overlap with pair production
under the formal assumption that the overlap effects were
\textit{small}.  The calculation did not assume that the photon was
soft, and it required numerical evaluation.  In the
soft photon limit ($x_\gamma \ll 1$), the result of ref.\ \cite{qedNfenergy}
was (in our language here)%
\footnote{
  The rate $[d\Gamma/dx_\gamma]_\LPMplus$ here to lose energy by
  $\kgamma = x_\gamma E$ was called the
  ``net rate $[d\Gamma/dx_e]_\uee$'' in ref.\ \cite{qedNfenergy},
  with $x_e = 1{-}x_\gamma$.  Eq.\ (\ref{eq:fplusold}) above is the
  $x_e{\to}1$ limit of eq.\ (3.10a) of ref.\ \cite{qedNfenergy}.
  That equation was actually an interpolation of numerical results computed
  for discrete values of $x_e$ and so has interpolation/extrapolation error.
  Ignoring that source of error, the $x_e{\to}1$ limit
  of that equation would give $0.56709$ for the last term
  in our eq.\ (\ref{eq:fplusold})
  above.  We have truncated to three digits ($0.567$) based on the size
  of interpolation error found in sec.\ 5.5 of ref.\ \cite{qedNfenergy}.
}
\begin {subequations}
\label {eq:LPMplusold}
\begin {equation}
  \left[ \frac{d\Gamma}{dx_\gamma} \right]_{\LPMplus}
  \simeq
  \left[ \frac{d\Gamma}{dx_\gamma} \right]_\LPM
  \left\{
    1 + \frac{\Nf\alpha}{2 x_\gamma}  \, f_\Plus(x_\gamma)
  \right\}
\label {eq:LPMplusold1}
\end {equation}
with
\begin {equation}
  f_\Plus(x_\gamma) \simeq
  \tfrac34 \ln\bigl( \tfrac{1}{x_\gamma} \bigr) + 0.567
  \,.
\label {eq:fplusold}
\end {equation}
\end {subequations}
Though that calculation was only valid in the limit that the correction
was small, and so valid only for $x_\gamma \gg \Nf\alpha$, the result
(\ref{eq:LPMplusold})
is \textit{larger} than the ordinary LPM rate.  So the ratio
of the $\LPMplus$ to $\LPM$ rate indeed grows as the photon becomes softer,
consistent with our earlier qualitative arguments.
But (\ref{eq:LPMplusold}) was extracted from the result of an incredibly
long, difficult, and subtle calculation \cite{2brem,seq,dimreg,qedNf},%
\footnote{
  The formalism of \cite{2brem,seq,dimreg} was developed in the context
  of large-$\Nc$ QCD, and \cite{qedNf} developed it further for
  the large-$\Nf$ QED case
  as a warm-up for continued QCD development \cite{qcd,qcdI}.
}
and one could worry about the possibility of a mistake somewhere along
the way.  Moreover, that calculation was invalid (the overlap correction
is not small) in precisely the range
$x_\gamma \lesssim \Nf\alpha$ where the overlap correction dominates.
In this paper, we present a simpler way to calculate the rate in
the soft photon limit, independent of most of the complexity of
refs.\ \cite{2brem,seq,dimreg,qedNf} (complexity needed for
the case of hard bremsstrahlung).  We are then able to generalize
that simpler method to study the case $x_\gamma \lesssim \Nf\alpha$ of interest.

In passing, we also find a simple analytic formula for the numerical
constant in (\ref{eq:fplusold}), providing a welcome
cross-check of earlier calculations.  That analytic result can be
extracted now from Taylor expanding our previewed result
(\ref{eq:fplus}) to first order in $\Nf\alpha/x_\gamma$, which gives
\begin {equation}
  f_\Plus(x_\gamma) \simeq
  \tfrac34 \ln\bigl( \tfrac{1}{x_\gamma} \bigr)
  + \left[ \tfrac34 \gammaE + \tfrac18 - \tfrac34\ln\pi + \tfrac54 \ln 2 \right]
  \qquad
  \mbox{(for $\Nf\alpha \ll x_\gamma \ll 1$)}
\label {eq:fplusAB}
\end {equation}
and matches fairly well the
previous, numerically-extracted result (\ref{eq:fplusold}).
Above, $\gammaE = 0.57721\cdots$ is the Euler-Mascheroni constant.

There is an additional, relatively unimportant difference
between our conclusions and those of Galitsky and Gurevich, which we
describe in appendix \ref{app:minor} for the sake of anyone making
a detailed comparison.


\section {Review: Calculating the ordinary LPM effect}
\label{sec:LPM}

Before proceeding to analytic calculations of overlapping
bremsstrahlung and pair production, we first
review the calculation of the ordinary LPM bremsstrahlung rate
(\ref{eq:LPM}), and also the LPM pair production rate, in order to
establish the formalism and notation that we will use.


\subsection {Basic formalism}
\label {sec:basic}

\begin {figure}[t]
\begin {center}
  \includegraphics[scale=0.6]{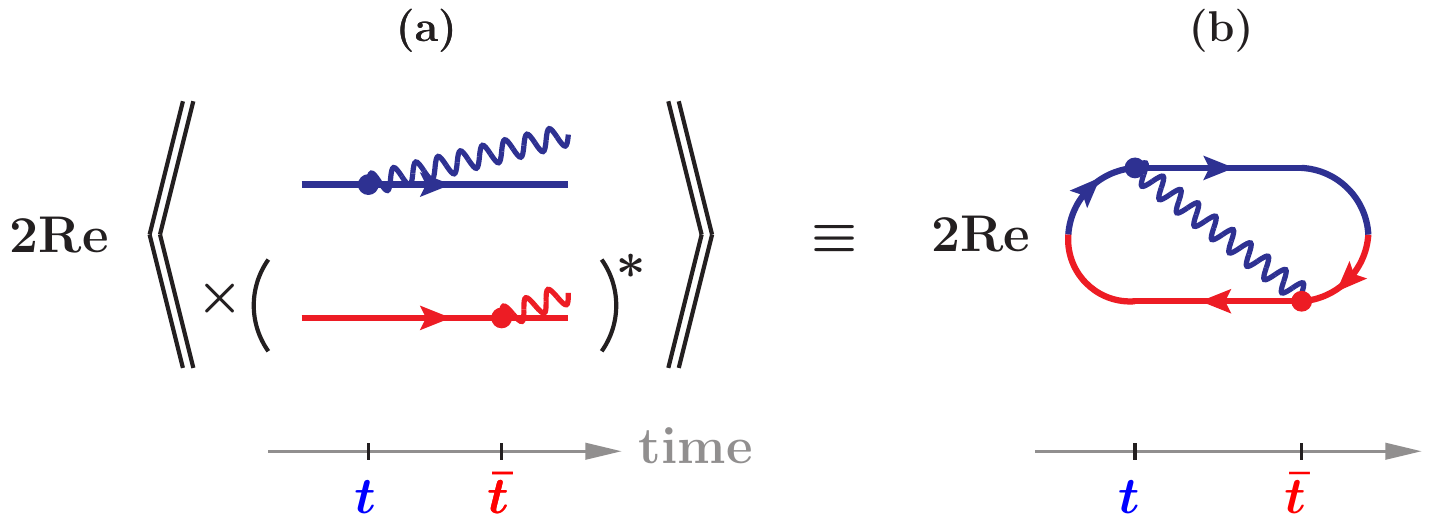}
  \caption{
    \label{fig:lpm}
    (a) A graphical representation of the LPM bremsstrahlung rate,
    consisting of (i) the amplitude (blue) times the conjugate amplitude
    (red) for time ordering $t < \bar t$, implicitly integrated over $t$
    and $\bar t$, (ii) $2\Re(\cdots)$ to add in
    the other time ordering $\bar t < t$, and (iii) averaging
    $\dlangle \cdots \drangle$ the rate
    over the randomness of the amorphous medium.
    All electron lines are implicitly summed over arbitrary numbers of
    elastic collisions with the medium.
    (b) depicts the short-hand graphical notation that we use for (a). 
  }
\end {center}
\end {figure}

The starting point is time-ordered perturbation theory for
bremsstrahlung, depicted by the top (blue) diagram in
fig.\ \ref{fig:lpm}a, which is integrated over all possible times
$t$ for the photon emission vertex.
Though not explicitly shown, each electron line in the diagram should
be understood to experience arbitrarily many elastic scatterings with
the medium.  In the figure, the amplitude (blue) is multiplied by
the conjugate amplitude (red), which is also integrated over the
time $\bar t$ of its photon emission vertex, but the diagram shown
in the picture refers to the contribution from the
explicit time ordering $t < \bar t$.  The operation $2\Re(\cdots)$
adds this diagram to its complex conjugate.  The latter
exchanges the amplitude
and conjugate amplitude, and so corresponds to the other time ordering
$t > \bar t$ of the original diagram.
The double angle brackets $\dlangle \cdots \drangle$ indicate
averaging of the rate over the randomness of the amorphous medium.

It is convenient to represent the time-ordered interference
of fig.\ \ref{fig:lpm}a more compactly
by the single time-ordered interference diagram drawn in
fig.\ \ref{fig:lpm}b, where again blue and red respectively
identify the contributions to the amplitude and conjugate amplitude.
Medium averaging $\dlangle \cdots \drangle$ of the rate is still taken
in the short-hand drawing of fig.\ \ref{fig:lpm}b, but we do not
write it explicitly.

Choose the $z$ axis to be closely aligned with the high-energy, nearly-collinear
bremsstrahlung process, so that $z \simeq t$ for all the high-energy
particles, and let $\b_i$ refer to the transverse positions of the
particles.

\begin {figure}[t]
\begin {center}
  \includegraphics[scale=0.6]{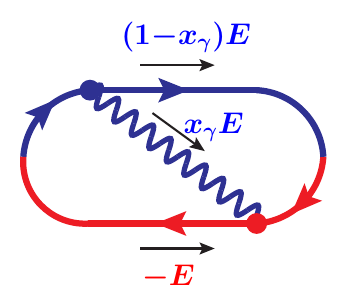}
  \caption{
    \label{fig:lpmflow}
    The approximate values (in the high-energy limit),
    of energies $\simeq$ longitudinal momenta $p_z$ of
    the particles in fig.\ \ref{fig:lpm}b in its formal re-interpretation as
    a three-particle system $e^-\gamma e^+$ evolving forward in time.
    Here, $E$ is the initial energy of the
    electron in the splitting process, and $x_\gamma$ is the
    energy fraction ($\simeq$ longitudinal momentum fraction) of the photon.
  }
\end {center}
\end {figure}

We now follow Zakharov \cite{Zakharov3} by conceptually re-interpreting
the diagram of fig.\ \ref{fig:lpm}b as three high-energy
particles $e^-\gamma e^+$ propagating forward in time from $t$ to $\bar t$.
In this picture, we will interpret the energy and momentum of all three of
these ``particles'' as flowing from left to right in the figure.
And so, in this interpretation,
the energy and $p_z$ of the ``positron'' $e^+$ are negative (from
conservation at the vertices), as shown in fig.\ \ref{fig:lpmflow}.
Though a specific configuration of atoms breaks translation invariance,
medium averaging restores it, and so total 3-momentum is conserved in
our three-particle interpretation of fig.\ \ref{fig:lpm}b:
\begin {equation}
   \p_{e^-} + \k_\gamma + \p_{e^+} = 0 ,
\label {eq:pconserved}
\end {equation}

Zakharov showed that the medium-averaged evolution of this ``3-particle
system'' can be described by a two-dimensional Schr\"odinger-like equation
that describes the transverse dynamics of the particles
with an effective Hamiltonian%
\footnote{
  One should not be disturbed that the effective Hamiltonian (\ref{eq:HeffV})
  is not Hermitian.  If it weren't for medium averaging, then the
  evolution of both the amplitude and conjugate amplitude contributions to
  fig.\ \ref{fig:lpm} would be unitary.  But an average of unitary operators
  need not be unitary.  So the effective medium-averaged evolution
  $e^{-i{\cal H}t}$ of our 3-particle system is not unitary,
  and ${\cal H}$ is not Hermitian.
}
\begin {subequations}
\label {eq:HeffV}
\begin {equation}
  {\cal H} =
  \frac{p_{\perp e^-}^2{+}m^2}{2(1{-}x_\gamma)E}
     + \frac{k_{\perp \gamma}^2}{2x_\gamma E} 
     - \frac{p_{\perp e^+}^2{+}m^2}{2E}
  + V(\b_{e^-} - \b_{e^+})
\label {eq:Heff}
\end {equation}
in our notation,
where the analog of the ``potential'' is
\begin {equation}
  V(\b) = -i \bigl[ \gamma_\el(0) - \gamma_\el(\b) \bigr]
  \qquad \mbox{with} \qquad
  \gamma_\el(\b) \equiv
  \int d^2q_\perp \frac{d\Gamma_\el}{d^2q_\perp} \, e^{i\q_\perp\cdot\b} ,
\label {eq:V}
\end {equation}
\end {subequations}
defined in terms of the differential rate
$d\Gamma_\el/d^2q_\perp = d\Gamma_\el/\pi d(q_\perp^2)$
for elastic scattering
of a high-energy electron from the medium with transverse momentum
transfer $\q_\perp$.

For the sake of readers unfamiliar with this formalism,
let us give a sense of what (\ref{eq:Heff}) represents.
Consider first the simpler case of free evolution of the particles
in vacuum.
In the high-energy limit,
the energy of a free particle with mass $m$ and
momentum $\p$ nearly-collinear
with the $z$ axis is
\begin {equation}
  \varepsilon_\p = \sqrt{|\p|^2 + m^2}
  \simeq  p_z + \frac{p_\perp^2{+}m^2}{2p_z} \,.
\label{eq:Ep}
\end {equation}
The first three terms in (\ref{eq:Heff}) are just the sum of (\ref{eq:Ep})
over the three particles $e^-\gamma e^+$, using
$p_{z,e^-}{+}k_{z,\gamma}{+}p_{z,e^+} = 0$ from (\ref{eq:pconserved}) and
$p_{z,e^+} \simeq -E$ from fig.\ \ref{fig:lpmflow}.  Alternatively,
one may understand the minus sign in the third term of (\ref{eq:Heff})
as arising from the sign change due to the complex conjugation
$(e^{-i \varepsilon\,\Delta t})^* = e^{+i\varepsilon\,\Delta t}$ of the initial
electron's evolution in the conjugate amplitude (red) in
fig.\ \ref{fig:lpm}a.

The physical interpretation of the ``potential'' term $V$ in
(\ref{eq:Heff}) may be understood by examining the corresponding
Schr\"odinger
equation but ignoring everything \textit{except} the
potential term in (\ref{eq:Heff}):
\begin {equation}
   i \partial_t \Psi(\b,t) = V(\b) \, \Psi(\b,t)
   = -i \int d^2q_\perp \frac{d\Gamma_\el}{d^2q_\perp} \,
     ( 1 - e^{i\q_\perp\cdot\b} ) \, \Psi(\b,t)
\label {eq:FP0}
\end {equation}
where we've used translation invariance to write the equation in
terms of
\begin {equation}
  \b \equiv \b_{e^-} - \b_{e^+} .
\label {eq:bdef}
\end {equation}
Above, ``$\Psi$'' is mathematically analogous to a wave-function,
but remember that the 3-particle description of fig.\ \ref{fig:lpm}
is being used to
describe the calculation of a \textit{probability},
not a probability amplitude.
Fourier transforming (\ref{eq:FP0}) gives
\begin {equation}
   \partial_t \Psi(\p_\perp,t)
   = - \Gamma_\el \Psi(\p_\perp,t)
     + \int d^2q_\perp \frac{d\Gamma_\el}{d^2q_\perp} \, \Psi(\p_\perp{-}\q_\perp,t)
   .
\label {eq:FP}
\end {equation}
This is simply a classical Fokker-Planck description of how a
probability distribution $\Psi(\p_\perp)$ for transverse momentum
evolves with time.  The first term on the right-hand-side of (\ref{eq:FP})
is a a loss term, quantifying decrease of $\Psi(\p_\perp)$
for a given $\p_\perp$ from elastic scattering that changes
$\p_\perp$ to some other transverse momentum $\p_\perp{+}\q_\perp$.
The second term is a gain term, quantifying increase of
$\Psi(\p_\perp)$ for a given $\p_\perp$ from elastic scattering
$\p_\perp{-}\q_\perp \to \p_\perp$.

The Hamiltonian (\ref{eq:Heff}) marries together the quantum mechanical
treatment of the radiation process with a classical Fokker-Planck
treatment of the underlying elastic scatterings, which is the
guiding principle of Migdal's original calculation \cite{Migdal}.
Zakharov's formulation provides a useful and powerful way to organize the
analysis.  We will not need to delve further into the details of its
derivation except for later discussing an alternative interpretation
of the ``potential'' $V(\b)$ in section \ref{sec:Vwilson}.

In the case of QED, it's easiest to choose the $z$ axis to be exactly
in the direction of the photon momentum $\k$, so that
\begin {equation}
  \k_\perp = 0 .
\label {eq:kperp}
\end {equation}
(This is possible because elastic scattering of the photon from the medium
is negligible in the high-energy limit, and so its $\k_\perp$ is not changing
over the formation time.)
With this choice of $z$ axis,
(\ref{eq:pconserved}) gives $\p_\perp \equiv \p_{\perp,e^-} = -\p_{\perp,e^+}$,
and (\ref{eq:Heff}) may be reduced to a one-body problem
\begin {equation}
  {\cal H} =
  \frac{p_\perp^2}{2\Mo}
  + V(\b) + \frac{m^2}{2\Mo} \,,
\label {eq:Heff1}
\end {equation}
where%
\footnote{
  \label{foot:Pperp}
  We have put the bar over $\Mo$ in order to distinguish it from the
  symbol $M_0 \equiv x_\gamma(1{-}x_\gamma)E$ used in
  ref.\ \cite{qedNf} (which we refer to a number of times in
  this paper).  The difference comes from normalization of the momentum
  variable, which is $\p_\perp$ here but was
  $\P_\perp \equiv (1{-}x_\gamma) \k_\perp - x_\gamma \p_{\perp,e^-}$ in
  ref.\ \cite{qedNf}.  For our choice of $\k_\perp = 0$ in (\ref{eq:kperp}),
  the relation is $\p_\perp = -\P_\perp/x_\gamma$, and so our
  $p_\perp^2/2\Mo$ in (\ref{eq:Heff1}) here equals
  $P_\perp^2/2M_0$ there.
}
\begin {equation}
  \Mo \equiv \frac{(1{-}x_\gamma)E}{x_\gamma}
\label {eq:M0}
\end {equation}
plays the role that ``mass'' would have in the analogy with
non-relativistic quantum mechanics.

\begin {figure}[t]
\begin {center}
  \includegraphics[scale=0.6]{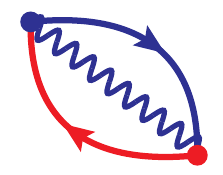}
  \caption{
    \label{fig:lpmpinch}
    The diagram in fig.\ \ref{fig:lpm}b redrawn to indicate that
    elastic scatterings need not be tracked before the first vertex nor after
    the last.
  }
\end {center}
\end {figure}

For differential rates $d\Gamma/dx$ that are
integrated over the transverse momenta of the (on-shell) final-state high-energy
particles, one may use unitarity (before medium averaging) to show that it is
unnecessary to follow elastic scattering of those final-state particles
\textit{after} they have been emitted in both the amplitude and
conjugate amplitude.%
\footnote{
  These points were well understood by those who came before,
  but one may find an explicit discussion of them
  in section IV.A of ref.\ \cite{2brem}
  in language similar to that used
  here, though in the context of
  a more complicated example that will be relevant
  to us later in this paper.
  For discussions of how to handle the case where integration over
  final $\p_\perp$'s is \textit{not} desired, see, for example, refs.\
  \cite{LOptZakharov,LOptWiedemann1,LOptWiedemann2,LOptBlaizot,LOptApolinario}.
}
Similarly, one need not follow elastic scattering
of the (on-shell)
initial state electron before it has radiated in one or the other.
Fig.\ \ref{fig:lpm}a has been redrawn to emphasize this point in
fig.\ \ref{fig:lpmpinch}.  The advantage of this representation is that
it visually suggests that $\b \equiv \b_{e^-} - \b_{e^+}$ must vanish
at the times $t$ and $\bar t$ of the two vertices, which will be
a feature of Zakharov's rate formula below.

Here, for simplicity, we will ignore masses ($m{=}0$)
as elsewhere in this paper, in which case Zakharov's
version of the LPM bremsstrahlung rate formula is
\begin {equation}
  \left[ \frac{d\Gamma}{dx_\gamma} \right]_\LPM
  =
  \frac{ \alpha\,P_{e\to\gamma}(x_\gamma) }{ \Mo^2 }
  \Re \int_0^\infty d(\Delta t) \>
  \grad_{\b'} \cdot \grad_{\b} \,
  G(\b',\Delta t;\b,0) \Bigl|_{\b'=\b=0}
\label {eq:Zrate}
\end {equation}
in our notation.
$\Delta t \equiv \bar t {-} t$ is the time separation
of the vertices in fig.\ \ref{fig:lpm}, and $G(\b',t';\b,t)$ is the
propagator for the Schr\"odinger-like equation with Hamiltonian
(\ref{eq:Heff1}) (here with $m{=}0$).
The gradients $\grad_{\b}$ and $\grad_{\b'}$ are the $b$-space versions
of one factor of transverse momentum associated with each
high-energy, nearly-collinear photon emission
vertex.
As promised,
transverse separations vanish at the start and end of the
evolution (once the vertex $\grad$'s are taken).
The rest of the factors, including the
DGLAP splitting function $P_{e\to\gamma}(x_\gamma)$
given by (\ref{eq:Peg}), also come from the QED matrix elements
associated with the photon emission vertices.
The origin of eq.\ (\ref{eq:Zrate}), including the overall
normalization, is reviewed in a little more detail
in appendix \ref{app:LObrem}.


\subsection{The multiple scattering (\boldmath$\qhat$) approximation}

High energy bremsstrahlung is nearly collinear and so corresponds to
small values of transverse displacements $b_i$ during the bremsstrahlung
process and so to small values of $b$ in the effective Hamiltonian
(\ref{eq:Heff1}).
Taking the small-$b$ limit, one finds formally that
the potential (\ref{eq:V}) simplifies to
\begin {equation}
   V(\b) \simeq -\tfrac{i}{4} b^2
     \int d^2q_\perp \frac{d\Gamma_\el}{d^2q_\perp} \, q_\perp^2 ,
\label {eq:VHO}
\end {equation}
which is like a harmonic oscillator potential but with imaginary-valued
spring constant.
The integral above represents the contribution per unit time from each
individual random elastic collision with the medium to increase the
total $(\Delta p_\perp)^2$ transferred from the medium to a high-energy
particle, as in the definition (\ref{eq:qhatdef}) of $\qhat$.%
\footnote{
  In the limit of a large number of scatterings, $\Delta p_\perp$ becomes
  large compared to the $q_\perp$ transferred by any single scattering.
  Taking the corresponding limit $q_\perp \ll p_\perp$, the Fokker-Plank equation
  (\ref{eq:FP}) reduces to a simple $\p_\perp$-space diffusion equation
  $\partial_t \Psi(\p_\perp)
   \simeq \tfrac14 \qhat \,\nabla_{\p_\perp}^2 \! \Psi(\p_\perp)$ with
  diffusion coefficient proportional to $\qhat$.
  Because of the factor $\tfrac14$ in this expression,
  an initial distribution
  $\Psi(\p_\perp) = \delta^{(2)}(\p_\perp)$ will evolve into a Gaussian with
  $\langle (\Delta p_\perp)^2 \rangle = \qhat\, \Delta t$, as in
  the definition (\ref{eq:qhatdef}) of $\qhat$.
}
So
\begin {equation}
   V(\b) \simeq -\tfrac{i}{4} \qhat b^2
\label {eq:Vqhat}
\end {equation}
with
\begin {equation}
   \qhat = \int d^2q_\perp \frac{d\Gamma_\el}{d^2q_\perp} \, q_\perp^2 .
\label {eq:qhatdef2}
\end {equation}
However,
the integral (\ref{eq:qhatdef2}) is actually UV logarithmically divergent.
Cutting off large $q_\perp$ is parametrically equivalent to cutting off
small impact parameters $b$, which is the origin of the logarithmic
dependence of $\qhat$ on $\bmin$ discussed in section \ref{sec:Elpm}.
As mentioned then,
the relevant physical scales for this cutoff in the application to
LPM bremsstrahlung are reviewed in appendix \ref{app:bmin}.

The upshot is that, in this high-energy, multiple scattering approximation,
the effective
Hamiltonian ${\cal H}$ of (\ref{eq:Heff1}) takes the form of a 2-dimensional
harmonic oscillator
\begin {subequations}
\label {eq:HO}
\begin {equation}
  {\cal H} = \frac{p_\perp^2}{2 \Mo} + \tfrac12 \Mo\Omega_0^2 b^2
\label {eq:HeffHO}
\end {equation}
with $\Mo$ given by (\ref{eq:M0}) and
\begin {equation}
  \Omega_0 \equiv \sqrt{ - \frac{i\qhat}{2\Mo} }
  = \sqrt{ - \frac{i x_\gamma \qhat}{2(1{-}x_\gamma)E} } \,.
\label {eq:Omega0}
\end {equation}
\end {subequations}
The time scale $1/|\Omega_0|$ is parametrically the bremsstrahlung
formation time (\ref{eq:tform})
[except that (\ref{eq:tform}) was specialized to the case where $1{-}x_\gamma$
is not small].

The propagator for a 2-dimensional harmonic oscillator is simple:
\begin {equation}
  G(\b',\Delta t;\b,0) =
  \frac{\Mo\Omega_0 \csc(\Omega_0\,\Delta t)}{2\pi i}
  \exp\Bigl( i\Mo\Omega_0 \bigl[
    \tfrac12(b^2 + b'^2) \cot(\Omega_0\,\Delta t)
    - \b\cdot\b' \csc(\Omega_0\,\Delta t)
  \bigr] \Bigr) .
\label {eq:Gprop}
\end {equation}
Plugging this into the version (\ref{eq:Zrate}) of the
LPM bremsstrahlung rate formula gives
\begin {subequations}
\label {eq:ZrateHO}
\begin {equation}
  \left[ \frac{d\Gamma}{dx_\gamma} \right]_\LPM
  =
  - \frac{\alpha}{\pi} \,P_{e\to\gamma}(x_\gamma)
  \Re \int_0^\infty d(\Delta t) \> \Omega_0^2 \csc^2(\Omega_0\,\Delta t) .
\label {eq:ZrateHO1}
\end {equation}

The above integral is convergent for $\Delta t \to \infty$ because
$\Omega_0$ has an imaginary part,
but the integral has a linear UV divergence associated with $\Delta t \to 0$.
Fortunately, this divergence is easy to take care of.
The simplest method is to sidestep the problem by realizing that
(on-shell) bremsstrahlung $e \to e\gamma$ is impossible in the absence
of something to scatter from.  So, if we subtract the vacuum contribution
from (\ref{eq:ZrateHO1}), we have not subtracted anything at all.
Empty space corresponds to $\qhat = 0$ and so to the limit
$\Omega_0 \to 0$, and so we may replace (\ref{eq:ZrateHO1}) by its
vacuum-subtracted version
\begin {equation}
  \left[ \frac{d\Gamma}{dx_\gamma} \right]_\LPM
  =
  - \frac{\alpha}{\pi} \,P_{e\to\gamma}(x_\gamma)
  \Re \int_0^\infty d(\Delta t) \>
  \left[ \Omega_0^2 \csc^2(\Omega_0\,\Delta t) - \frac{1}{(\Delta t)^2} \right]
  =
  \frac{\alpha}{\pi} \, P_{e\to\gamma}(x_\gamma)
  \Re (i\Omega_0) .
\label {eq:ZrateHO2}
\end {equation}
\end {subequations}
Using the definition (\ref{eq:Omega0}) of $\Omega_0$, this formula
reproduces the deep-LPM rate formula (\ref{eq:LPM}) quoted earlier.

Alternatively, it is also possible to introduce a UV regulator and
work with the original integral (\ref{eq:ZrateHO1}) directly.%
\footnote{
\label{foot:UV}
  One may regulate the original
  integral (\ref{eq:ZrateHO1}) directly
  by replacing $\Delta t$ by
  $\Delta t - i\varepsilon$, in which case the regulated divergence is
  killed when taking the real part of the integral in (\ref{eq:ZrateHO1}).
  That method is hard to consistently generalize to overlap calculations
  (see \cite{dimreg}).
  Or one may use dimensional regularization \cite{dimreg}.
  The case of (\ref{eq:ZrateHO1}) is then simple because dimensional
  regularization discards power-law divergences, and (\ref{eq:ZrateHO1})
  has no sub-leading logarithmic divergence.  So for this particular
  integral dimensional regularization is equivalent to a vacuum subtraction.
}
That can be necessary for calculating overlap effects in general
situations \cite{qedNf,dimreg}, but we will later be able to sidestep
that necessity in the soft photon limit of interest here.


\subsection{\boldmath$V(b)$ and time-like Wilson loops}
\label {sec:Vwilson}

\begin {figure}[t]
\begin {center}
  \includegraphics[scale=0.4]{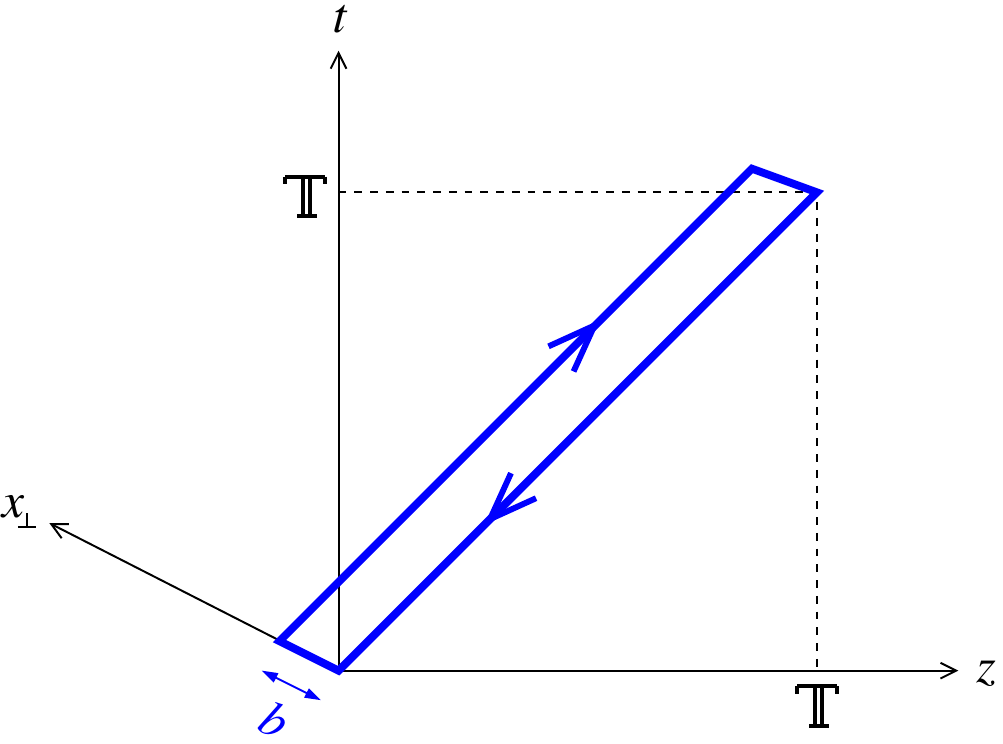}
  \caption{
    \label{fig:Wilson}
    A Wilson loop with long light-like edges that may be used to define $V(\b)$.
  }
\end {center}
\end {figure}

It will be useful for later to review that the ``potential'' $V(\b)$ of
(\ref{eq:V}) also has a simple relation to long, light-like Wilson
loops evaluated in the background of the electromagnetic fields present
in the medium.  Fig.\ \ref{fig:Wilson} depicts a Wilson loop with
parallel light-like sides separated by a transverse displacement $\b$
and extending for a long duration $\Time$ in time.  $V(\b)$ can be defined
in terms of the $\Time \to \infty$ behavior as \cite{LRW1,LRW2}
\begin {equation}
  \Bigdlangle e^{-i e \oint dx^\mu A_\mu^{\rm bkgd} (x)} \Bigdrangle
  \sim
  e^{-i\,V(\b)\,\Time} ,
\label {eq:Wilson}
\end {equation}
where $A_\mu^{\rm bkgd}(x)$
is the background electromagnetic field from the medium and
$\dlangle \cdots \drangle$ denotes medium averaging.

For media where the elastic scattering rate can be computed
perturbatively
(such as in the present case),%
\footnote{
  \label{foot:perturbative}
  We should clarify that in this context a ``perturbative'' treatment
  means
  perturbation theory in the interaction of the high-energy parton
  with the electromagnetic field in the medium.  It does not mean, for
  example, that
  the screening of nuclear electric fields by atomic
  electrons in Gold has to be calculable in perturbation theory.
  Also, ``perturbative'' refers here to the calculation
  of the rate for scattering from the medium; for LPM
  bremsstrahlung and pair production, we still sum over arbitrary
  numbers of elastic scatterings. 
}
the $V(\b)$ extracted
from (\ref{eq:Wilson}) turns out to be the same as the one presented in
(\ref{eq:V}).
In particular, in the perturbative case, (\ref{eq:Wilson}) is dominated by
2-point correlations of the background field, so that
\begin {align}
  \Bigdlangle e^{-i e \oint dx^\mu A_\mu^{\rm bkgd} (x)} \Bigdrangle
  &=
  \exp\left( \frac12 \Bigdlangle 
    {\scriptstyle
      \left[ -i e \oint dx^\mu A_\mu^{\rm bkgd} (x) \right]^2
    }
  \Bigdrangle \right)
\nonumber\\
  &=
  \exp\left(
     -\frac{e^2}{2} \oint dx^\mu \oint dy^\nu
     \bigdlangle A_\mu^{\rm bkgd} (x) \, A_\nu^{\rm bkgd} (y) \bigdrangle
  \right) ,
\label {eq:Wilson2point}
\end {align}
which gives
\begin {equation}
  V(\b) =
  -i e^2 \lim_{\Time\to\infty}
  \frac{1}{2\Time}
  \oint dx^\mu \oint dy^\nu
     \bigdlangle A_\mu^{\rm bkgd} (x) \, A_\nu^{\rm bkgd} (y) \bigdrangle .
\label {eq:V2point}
\end{equation}
Now break up the double integral in (\ref{eq:V2point}) into the various
contributions
coming from $x$ and $y$ each either on the
$e^-$ or $e^+$ light-like lines in fig.\ \ref{fig:Wilson}, writing
\begin {equation}
  V(\b) = -\frac{i}{2} \sum_j \sum_k Q_j Q_k \, \gamma(\b_j{-}\b_k) ,
\label {eq:Vqqsum}
\end {equation}
where the sums are over $j$ and $k \in \{e^-,e^+\}$ and where
\begin {equation}
  \gamma(\b) \equiv
  e^2 \int d(\delta t) \>
     \bigdlangle
        v^\mu A_\mu^{\rm bkgd}(\delta t,\b,\delta t) \,
        v^\nu A_\nu^{\rm bkgd}(0,{\bm 0},0)
     \bigdrangle .
\label {eq:gamma2point}
\end {equation}
Above $Q_j = \pm 1$ is the charge of particle $j$ in units of the
electron charge $e$, and
$v^\mu = (v^0,v^\perp,v^z) \equiv (1,{\bm 0},1)$.
Though we will not review the argument here,%
\footnote{
  See, for example, the discussion in section II.A and appendix A of
  ref.\ \cite{Vqhat}.  The $C_R$ and $g^2$ in that (QCD) discussion should
  be interpreted as $C_R=1$ and $g^2 = e^2 = 4\pi\alpha$ in the
  QED application here.
}
the $\gamma(\b)$ above can be shown to equal the $\gamma(\b)$ of (\ref{eq:V}).

The fact that formation times are long compared
to the electric-field correlation length of the medium allows
the time scale $\delta t$ in (\ref{eq:gamma2point}) to be treated
as effectively ``instantaneous''
\textit{compared to} the time scales for bremsstrahlung,
which is what allows the
Hamiltonian-like formalism (\ref{eq:Heff}) that implicitly treats
$V$ as local in time.

$V(\b)$ can be defined by the more general formula (\ref{eq:Wilson}) even
when the interactions between high-energy particles and the medium
are strong, as in applications to strongly-coupled quark-gluon plasmas.
But the fact that $V$ is determined (up to small corrections)
by just 2-point correlations for the QED application [and
so can be decomposed into a sum (\ref{eq:Vqqsum}) over which lines
those 2-point correlators connect] will be extremely useful
for our analysis of overlap effects later in section \ref{sec:origin}.


\subsection{Pair production}

\begin {figure}[t]
\begin {center}
  \includegraphics[scale=0.6]{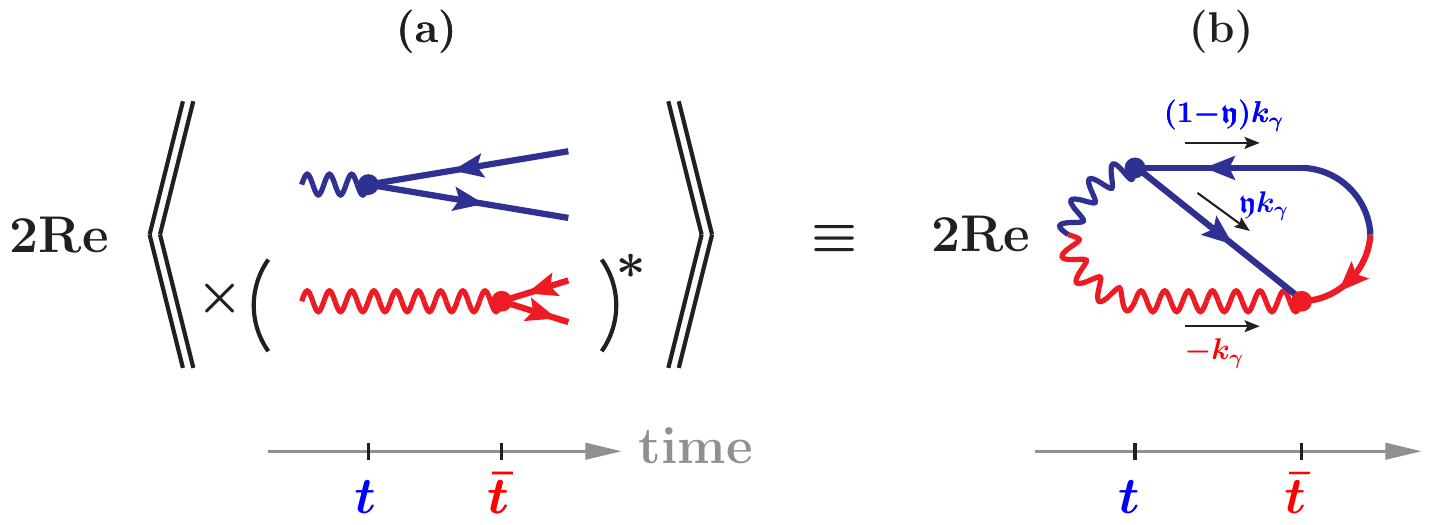}
  \caption{
    \label{fig:pair}
    Like figs.\ \ref{fig:lpm} and \ref{fig:lpmflow} but for pair production
    $\gamma \to e^- e^+$ from a photon with energy $k_\gamma$ to
    daughters with energy fractions $\yfrak$ and $1{-}\yfrak$ of $k_\gamma$.
  }
\end {center}
\end {figure}

The analog of figs.\ \ref{fig:lpm} and \ref{fig:lpmflow} are shown in
fig.\ \ref{fig:pair} for pair production.
Analogous to (\ref{eq:Heff}), the medium-averaged
3-particle evolution
between the two vertices in fig.\ \ref{fig:pair}b is described by
an effective Hamiltonian
\begin {equation}
  {\cal H}_\pair =
  \frac{p_{\perp e^-}^2{+}m^2}{2(1{-}\yfrak) k_\gamma}
     + \frac{p_{\perp e^+}^2{+}m^2}{2\yfrak k_\gamma}
  -\frac{k_{\perp \gamma}^2}{2k_\gamma}
  + V(\b_{e^-} - \b_{e^+}) ,
\label {eq:Hpair}
\end {equation}
but it is now the photon that appears in the conjugate amplitude and so
it is the photon free-particle energy term
$k_{\perp\gamma}^2/2k_{\gamma,z} \simeq k_{\perp\gamma}^2/2k_\gamma$
that is negated above.  $V(\b)$ above is the same as (\ref{eq:V}),
and for pair production the relevant DGLAP splitting function is
\begin {equation}
  P_{\gamma\to e}(\yfrak) = \yfrak^2 + (1{-}\yfrak)^2 .
\end {equation}

If we again choose the $z$-axis to be exactly
in the direction of the photon, and again use medium-averaged
momentum conservation, we get $\p_{e^+} = -\p_{e^-}$ as before and
(\ref{eq:Hpair}) reduces to the 1-body problem
\begin {equation}
  {\cal H}_\pair =
  \frac{p_\perp^2}{2\Mpr}
  + V(\b) + \frac{m^2}{2\Mpr} \,,
\label {eq:Hpair1}
\end {equation}
where
\begin {equation}
  \Mpr \equiv \yfrak(1{-}\yfrak)k_\gamma .
\label {eq:Mpr}
\end {equation}
The subscript ``$\pr$'' in (\ref{eq:Mpr}) is shorthand for pair production.
Neglecting the mass $m$, the
LPM rate formula analogous to (\ref{eq:Zrate}) is
\begin {equation}
  \left[ \frac{d\Gamma}{d\yfrak} \right]^\LPM_\pair
  =
  \frac{ \Nf\alpha\,P_{\gamma\to e}(\yfrak) }{ \Mpr^2 }
  \Re \int_0^\infty d(\Delta t) \>
  \grad_{\b'} \cdot \grad_{\b} \,
  G_\pr(\b',\Delta t;\b,0) \Bigl|_{\b'=\b=0} ,
\label {eq:Zpair}
\end {equation}
(where the factor of
$\Nf$ counts the number of flavors the pair could have whose
mass can be ignored).

Again making the multiple-scattering
approximation (\ref{eq:Vqhat}), the Hamiltonian (\ref{eq:Hpair1}) becomes
\begin {subequations}
\label {eq:HOpair}
\begin {equation}
  {\cal H}_\pair = \frac{p_\perp^2}{2\Mpr} + \tfrac12 \Mpr\Omegapr^2 b^2
\end {equation}
with
\begin {equation}
  \Omegapr \equiv \sqrt{ - \frac{i\qhat}{2\Mpr} }
  = \sqrt{ - \frac{i\qhat}{2\yfrak(1{-}\yfrak)k_\gamma} } \,.
\label {eq:Omegapr}
\end {equation}
\end {subequations}
The analog to (\ref{eq:ZrateHO}) is
\begin {equation}
  \left[ \frac{d\Gamma}{d\yfrak} \right]_\pair^\LPM
  =
  - \frac{\Nf\alpha}{\pi} \,P_{\gamma\to e}(\yfrak)
  \Re \int_0^\infty d(\Delta t) \> \Omegapr^2 \csc^2(\Omegapr\,\Delta t)
  =
  \frac{\Nf\alpha}{\pi} \,P_{\gamma\to e}(\yfrakE)
  \Re (i\Omegapr) ,
\label {eq:ZpairHO}
\end {equation}
again resolving the UV divergence of the integral using vacuum subtraction.
Eq.\ (\ref{eq:ZpairHO}) reproduces the deep-LPM limit
\begin {equation}
  \left[ \frac{d\Gamma}{d\yfrak} \right]_\pair^\LPM \simeq 
  \frac{\Nf\alpha}{2\pi} \, P_{\gamma\to e}(\yfrak)
     \sqrt{ \frac{\qhat}{\yfrak(1{-}\yfrak)\kgamma} }
\label{eq:LPMpair}
\end {equation}
of Migdal's original result \cite{Migdal}.
Integrating (\ref{eq:LPMpair}) over $\yfrak$ gives the total LPM
pair production rate
\begin {equation}
  \Gamma_\pair^\LPM \simeq
  \frac{3\Nf\alpha}{8} \sqrt{ \frac{\qhat}{\kgamma} }
\label{eq:GammaPair}
\end {equation}
(for $k_\gamma \gg \Elpma$).

It will be useful later to have a parametric estimate for the pair
production formation time.
Similar to bremsstrahlung, the time scale in (\ref{eq:ZpairHO}) is
set by $1/|\Omega_\pr|$, and so
\begin {equation}
  \tform^\pair \sim \frac{1}{|\Omega_\pr|}
  \sim 
  \sqrt{ \frac{\yfrak(1{-}\yfrak)k_\gamma}{\qhat} } \,.
\label {eq:tpair}
\end {equation}
In terms of this formation time,
\begin {equation}
  \Gamma_\pair^\LPM \sim \frac{\Nf\alpha}{\tform^\pair}~\Bigg|_{\rm democratic} \,.
\label{eq:GammaPairEst}
\end {equation}


\subsection{Parametric sizes of \boldmath$p_\perp$ and $b$}

It will also be useful for later arguments to have a parametric understanding
of the typical sizes of transverse momentum
$\p_\perp \equiv \p_{\perp e^-} = -\p_{\perp e^+}$
(relative to the photon direction)
and separation $\b \equiv \b_{e^-} {-} \b_{e^+}$ during the formation
time for ordinary LPM bremsstrahlung or pair production.
In both cases, the definition $(p_\perp^2)_{\rm typical} \simeq \qhat t$ of
$\qhat$ gives $p_\perp \sim (\qhat t_\form)^{1/2}$ over time scales of order
the formation time $t_\form \sim 1/|\Omega|$.
We'll later need this information specifically for soft bremsstrahlung
($x_\gamma \ll 1$) and for democratic pair production
($\yfrakE(1{-}\yfrakE) \sim 1$),
in which case (\ref{eq:tform}) and (\ref{eq:tpair}) give
\begin {align}
   p_\perp &\sim
   \begin {cases}
     \bigl( \qhat E/x_\gamma \bigr)^{1/4} &
       \mbox{soft bremsstrahlung} ; \\
     \bigl( \qhat \kgamma \bigr)^{1/4} &
       \mbox{democratic pair production} .
   \end {cases}
\\
\intertext{
  By the uncertainty principle, typical $e^-e^+$ separations in the
  3-particle evolution picture are just $b \sim 1/p_\perp$, and so
}
   |\b_{e^-} {-} \b_{e^+}| &\sim
   \begin {cases}
     \bigl( \qhat E/x_\gamma \bigr)^{-1/4} &
       \mbox{soft bremsstrahlung} ; \\
     \bigl( \qhat \kgamma \bigr)^{-1/4} &
       \mbox{democratic pair production} .
   \end {cases}
\label {eq:bscales}
\end {align}


\section{Warmup: Overlap corrections for \boldmath$\Nf\alpha \ll x_\gamma \ll 1$}
\label {sec:NLO}

In this section, we review the time-ordered
interference diagrams that were needed in refs.\ \cite{qedNf,qedNfenergy}
to calculate the overlap correction of bremsstrahlung followed by pair
production in the region $\Nf\alpha \ll x_\gamma \le 1$ where the overlap
correction is small, and then we will show how to obtain analytic
results in the soft-photon limit $\Nf\alpha \ll x_\gamma \ll 1$.
An understanding of the diagrams needed, and how to efficiently
calculate them in the soft-photon limit, will set us up to
explore the transition to the
very-soft photon emission case $x_\gamma \lesssim \Nf\alpha$ in
section \ref{sec:mainresult}.

\begin {figure}[t]
\begin {center}
  \includegraphics[scale=0.7]{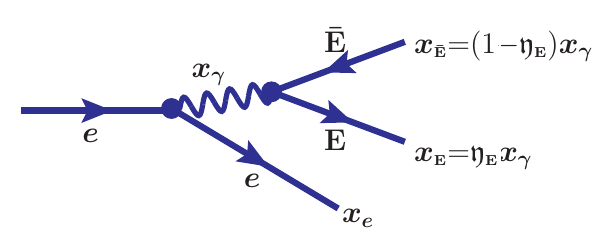}
  \caption{
    \label{fig:eEnotation}
    Our notation ($\E$ vs.\ $e$)
    for distinguishing pair-produced electrons from the
    original electron in the large-$\Nf$ limit.
    The $x$'s are energy fractions compared to the original electron;
    $\yfrakE$ is the energy fraction of the pair electron $\E$ compared
    to the photon that produces it;
    and $\xEbar = 1 - \xe - \xE$.
  }
\end {center}
\end {figure}

Refs.\ \cite{qedNf,qedNfenergy} studied the QED problem mainly as a stepping
stone to understanding the corresponding QCD problem.
In that context, the large-$\Nf$ limit of QED was invoked simply to
reduce the number of QED interference diagrams that needed to be calculated.
The pair-produced leptons can be of any flavor, and so, in the large-$\Nf$
limit, the chance that the pair-produced ``electron'' has the same flavor
as the original electron is suppressed by $1/\Nf$.  So we may
treat the pair-produced lepton as distinguishable from the initial
electron.
Following ref.\ \cite{qedNfenergy},
we will emphasize this distinguishability by referring to the
produced pair as $\E\Ebar$ instead of $e^-e^+$, as in
fig.\ \ref{fig:eEnotation}.

In this section, we will refer to the ``perturbative order''
of a calculation as counting
the number of high-energy splitting vertices but implicitly
summing over arbitrarily
many elastic scatterings from the medium.  In particular,
``leading-order'' (LO) bremsstrahlung or pair production will refer
to the corresponding ordinary LPM rates, while next-to-leading-order (NLO)
will refer to the first correction due to overlapping splittings.


\subsection{Review of relevant diagrams}

When they overlap,
bremsstrahlung $e\to e\gamma$ combined with pair production
$\gamma \to \E\Ebar$ leads to $e \to e\E\Ebar$ overall.
In order to simplify the initial discussion of diagrams, we will focus
first on diagrams where the intermediate photon is nearly on-shell
and transversely polarized.  We discuss afterward similar-size
contributions from longitudinally polarized photons.


\subsubsection{Transversely polarized photons}
\label{sec:NLOtransverse}

\begin {figure}[tp]
\begin {center}
  \includegraphics[scale=0.5]{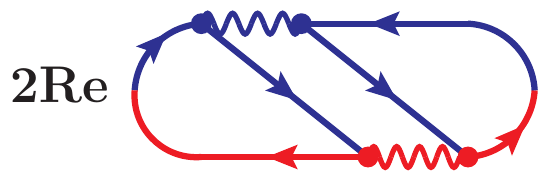}
  \caption{
     \label{fig:real1}
     A time-ordered interference diagram for overlapping
     $e \to e\gamma \to e \E\Ebar$ in large-$\Nf$ QED \cite{qedNf}
     via a transversely-polarized photon.
  }
\end {center}
\end {figure}

Fig.\ \ref{fig:real1} shows a time-ordered interference diagram
that contributes to overlapping $e \to e \gamma \to e \E \Ebar$.
In the large-$\Nf$ limit,
for diagrams involving only transversely-polarized photons,
this is the only contribution
(at this order in $\Nf\alpha$)
for which the diagram includes a time interval
(between the middle two vertices) where
the bremsstrahlung and pair production visually overlap.

\begin {figure}[tp]
\begin {center}
  \includegraphics[scale=0.5]{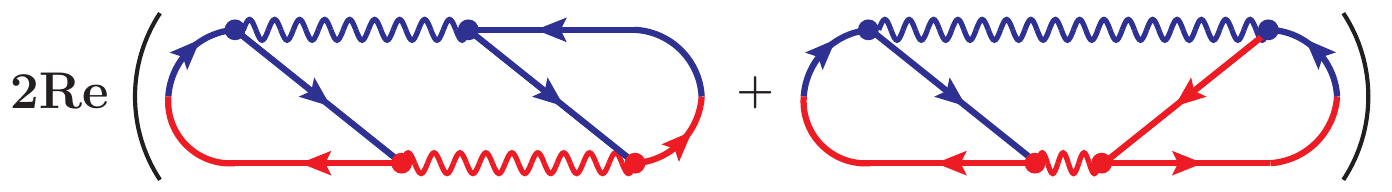}
  \caption{
     \label{fig:real1b}
     Additional time-ordered interference diagrams contributing to
     $e \to e \E\Ebar$ in large-$\Nf$ QED \cite{qedNf}.
  }
\end {center}
\end {figure}

There is, however, a less obvious contribution that we should mention,
even though it will ultimately not impact the calculation in this
paper.  The diagrams of fig.\ \ref{fig:real1b} do not contain any
time interval where the bremsstrahlung and pair production parts of the
diagram overlap.  And yet, because the time-ordering in these diagrams
imposes ordering restrictions on the range of times that the vertices can be
integrated over,
(i) the diagrams of fig.\ \ref{fig:real1b}
are not equivalent to
(ii) simply combining the
LPM bremsstrahlung and pair production probabilities computed from
the diagrams of figs.\ \ref{fig:lpm} and \ref{fig:pair}.
Because of this mismatch, the difference of (i) and (ii) also generates
an overlapping formation time correction to
modeling shower development with just ordinary LPM bremsstrahlung and
pair production rates.  This contribution is discussed and calculated in
refs.\ \cite{seq,qedNf},%
\footnote{
  If interested, see in particular the qualitative discussion in section 1.1
  of ref.\ \cite{seq} and corresponding calculation (for QCD)
  in section 2.1 of that paper.
}
but we will not need the details here.

\begin {figure}[tp]
\begin {center}
  \includegraphics[scale=0.5]{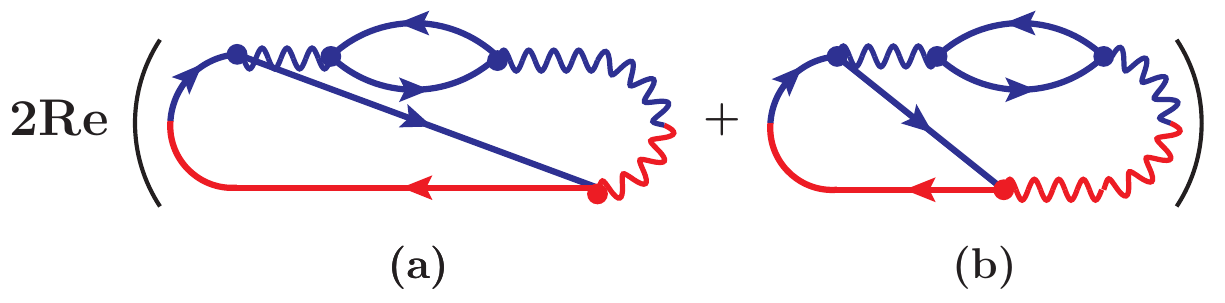}
  \caption{
     \label{fig:virt1}
     Time-ordered interference diagrams for
     NLO bremsstrahlung $e \to e\gamma$ in large-$\Nf$ QED \cite{qedNf}.
  }
\end {center}
\end {figure}

\begin {figure}[tp]
\begin {center}
  \includegraphics[scale=0.5]{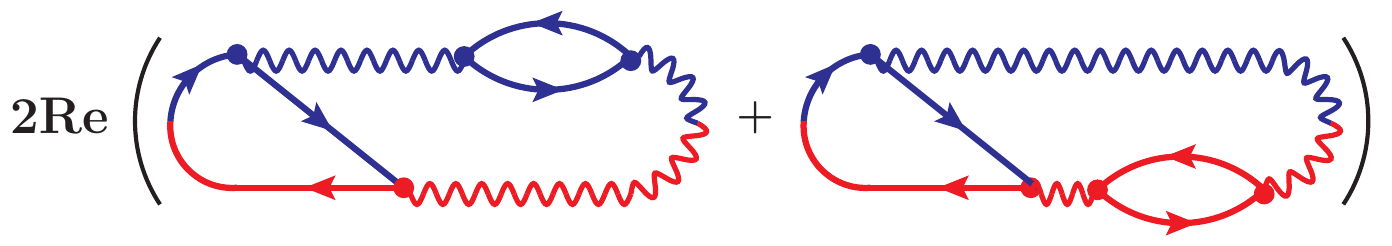}
  \caption{
     \label{fig:virt1b}
     Additional diagrams for NLO bremsstrahlung,
     analogous to additional $e{\to}e\E\Ebar$ diagrams of
     fig.\ \ref{fig:real1b}.
  }
\end {center}
\end {figure}

At the same order in $\Nf\alpha$ as the overlapping
$e \to e\gamma \to e\E\Ebar$ of fig.\ \ref{fig:real1},
there is also a next-to-leading-order
(NLO) correction to bremsstrahlung alone
from (i)
bremsstrahlung with \textit{virtual} pair production
$e \to e\gamma \to e\E\Ebar \to e\gamma$
interfering with
(ii) leading-order (LO)
$e \to e\gamma$.  Two such time-ordered interference diagrams are
shown in fig.\ \ref{fig:virt1}.  In these diagrams, bremsstrahlung
visually overlaps the virtual pair production, but there is no real
pair production.  (To have the bremsstrahlung overlap with both
virtual
pair production \text{and} a subsequent real pair production would
be yet-higher order in $\Nf\alpha$.)
Similar to fig.\ \ref{fig:real1b}, fig.\ \ref{fig:virt1b} shows
other diagrams involving virtual pair production that should also
be accounted for, but again we will not need details.

\begin {figure}[t]
\begin {center}
  \includegraphics[scale=0.6]{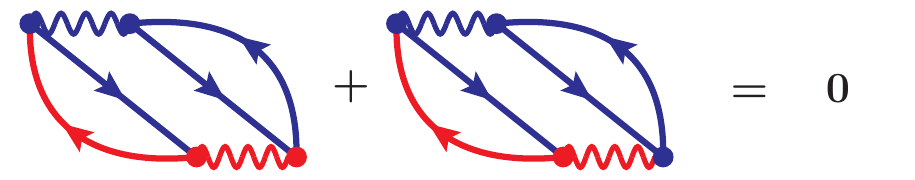}
  \caption{
    \label{fig:cancel}
    The sum of the diagrams of fig.\ \ref{fig:real1} and
    fig.\ \ref{fig:virt1}b, redrawn in the style of fig.\ \ref{fig:lpmpinch}.
    Only the color of the last vertex is different.
    This cancellation is the analog, for our application, of the
    Largest Time Equation in t'Hooft and Veltman's discussion of
    cutting equations for Feynman diagrams \cite{diagrammar}.%
    \protect\footnotemark
  }
\end {center}
\end {figure}

In this paper, we have focused on the rate $d\Gamma/dx_\gamma$ for the
initial electron of energy $E$ to lose energy $x_\gamma E$.  At leading order,
that's just the LPM bremsstrahlung rate.  At NLO, it gets contributions
from the overlap of bremsstrahlung with either real pair production
($e{\to}e\E\Ebar$ as in figs.\ \ref{fig:real1}--\ref{fig:real1b})
or virtual pair production (NLO $e{\to}e\gamma$ as in
figs.\ \ref{fig:virt1}--\ref{fig:virt1b}):%
\footnote{
  Eq.\ (\ref{eq:netrate}) above is referred to in ref.\ \cite{qedNfenergy}
  as the ``net rate $[d\Gamma/dx]_\uee$'' for $e{\to}e$ with
  $x = \xe = 1{-}x_\gamma$.
  See eqs.\ (3.1a) and (3.3a) of ref.\ \cite{qedNfenergy}.
  See also our later discussion in section \ref{sec:smallNf} of
  the current paper.
}
\begin {equation}
  \frac{d\Gamma}{dx_\gamma}
  = \left[ \frac{d\Gamma}{dx_\gamma} \right]_{e\to e\gamma}^\LPM
    + \left(
        \int_0^1 d\yfrakE \>
        \left[ \Delta \frac{d\Gamma}{dx_\gamma \> d\yfrakE} \right]_{e\to e\E\Ebar}
      \right)
    + \left[ \Delta \frac{d\Gamma}{dx_\gamma} \right]_{e\to e\gamma}^\NLO .
\label {eq:netrate}
\end {equation}
(The $\Delta$'s in front of $d\Gamma/dx$'s on the right-hand side
is a notation
used by refs.\ \cite{seq,qedNf} merely as a reminder
that these terms implement corrections to
a purely LPM description of shower evolution.%
\footnote{
  The motivation for that reminder comes from the detailed treatment
  in ref.\ \cite{seq,qedNf} of
  the diagrams of figs.\ \ref{fig:real1b} and \ref{fig:virt1b} above.
}%
)
There is a tremendous simplification in the sum (\ref{eq:netrate}) of
the time-ordered diagrams: the contributions from
figs.\ \ref{fig:real1}--\ref{fig:real1b} cancel, diagram by diagram,
all but the first diagram of figs.\ \ref{fig:virt1}--\ref{fig:virt1b}.
For example, fig.\ \ref{fig:cancel} shows the sum of the diagrams of
\addtocounter{footnote}{-1}%
\footnotetext{
  Specifically, see section 6.2 of ref.\ \cite{diagrammar}.
}
fig.\ \ref{fig:real1} and fig.\ \ref{fig:virt1}b
redrawn in the compressed style of fig.\ \ref{fig:lpmpinch}, which makes
explicit the vanishing of separations at the first and last vertices
for rates that are integrated over final-state $\p_\perp$'s.
The only difference between the two diagrams in fig.\ \ref{fig:cancel}
is whether the very last vertex is colored red (conjugate amplitude vertex)
or blue (amplitude vertex).  The cancellation occurs because the
last vertex in the first diagram corresponds to (i) a factor of
$+i\delta H$ from treating that splitting vertex perturbatively
in the evolution $(e^{-iHt})^* = e^{+iHt}$ of the conjugate amplitude,
versus (ii) a factor of $-i\delta H$ for the last vertex of the second diagram
from perturbative treatment of the evolution $e^{-iHt}$ of the
amplitude.%
\footnote{
  The terminology of ref.\ \cite{qedNfenergy} is that the diagrams
  are related by a ``back-end transformation,'' which graphically
  consists of sliding the latest-time vertex of a diagram (such as
  drawn in figs.\ \ref{fig:real1}--\ref{fig:real1b}) around the back end of
  the diagram from amplitude to conjugate-amplitude or vice versa,
  without changing the time of the vertex.
}
Other than that, the diagrams are the same provided, as in
eq.\ (\ref{eq:netrate}), one
integrates over the same real-particle energy fractions $(\yfrakE)$ for the
final-state pair $\E\Ebar$ in
the first diagram as
are integrated in the virtual-pair loop in the second diagram
(also $\yfrakE$).

This raises a technical point.  We've loosely referred to $x_\gamma$ and
$\yfrakE$ as
energy fractions, but in time-ordered perturbation theory it's better
to think of them as, say, $p_z$ momentum fractions.
For real pair production, $\yfrakE$ is
integrated over $0 \le \yfrakE \le 1$ because of conservation of momentum.
For \textit{virtual} pair production,
loop momenta would be integrated over all values
in standard Hamiltonian perturbation theory,
and so in particular over $-\infty < \yfrakE < \infty$.
However, in Light-Cone Perturbation Theory (LCPT)
\cite{LB,BL,BPP}%
\footnote{
  For readers not familiar with time-ordered
  LCPT who would like
  the simplest possible example of how it reassuringly
  reproduces the results of
  ordinary Feynman diagram calculations,
  we recommend section
  1.4.1 of Kovchegov and Levin's monograph \cite{KL}.
}
the lightcone components $p^+$ of the momenta of
fermions and transversely-polarized
photons must be non-negative. Energy-momentum
conservation in LCPT
then forces $0 \le \yfrakE \le 1$ in our virtual pair production
loops if $\yfrakE$ is defined as
the $p^+$ momentum fraction (which is also approximately equal to the
energy and $p_z$ fractions in the high-energy limit).  Following
refs.\ \cite{qedNf,qedNfenergy}, our calculations will be carried out
in the framework of LCPT.  This has no significant impact
on the calculation of fig.\ \ref{fig:virt1}a except to restrict
the integration over the virtual momentum fraction
$\yfrakE$ to $0 \le \yfrakE \le 1$, which justifies the cancellations
we have discussed.  Also, the left-to-right time ordering in
our figures should now be interpreted as time ordering of
\textit{lightcone} time
$x^+ \propto t{+}z$ instead of normal time $t$, but, in the
high-energy limit relevant to our problem, this is not an important
qualification.
The choice of using LCPT will have implications for the
evaluation of contributions involving longitudinally polarized photons,
which we take up shortly.

Because of the cancellations, fig.\ \ref{fig:virt1}a is the only diagram from
figs.\ \ref{fig:real1}--\ref{fig:virt1b} that contributes to the
differential rate (\ref{eq:netrate}) for the initial electron to
lose energy $x_\gamma E$.
This can seem confusing because
energy loss can occur
through both (i) overlapping processes $e\to e\E\Ebar$ involving real pair
production and
(ii) bremsstrahlung processes $e\to e\gamma$ involving overlap with
virtual pair-production,
but fig.\ \ref{fig:virt1}a does not at first appear to have anything
to do with real pair production.  Nonetheless, fig.\ \ref{fig:virt1}a
is equal to the desired sum of \textit{all} the real and virtual
pair production processes in figs.\ \ref{fig:real1}--\ref{fig:virt1b}.


\subsubsection{Longitudinally polarized photons}

\begin {figure}[t]
\begin {center}
  \includegraphics[scale=0.6]{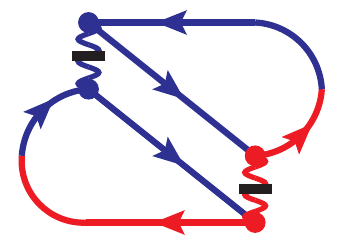}
  \caption{
    \label{fig:II}
    The contribution to $e\to e\E\Ebar$ involving longitudinally-polarized
    photons in both the amplitude and conjugate amplitude.
    The diagram is lightcone-time ($x^+$) ordered from left to right.
  }
\end {center}
\end {figure}

At the same order of $\Nf\alpha$ as figs.\ \ref{fig:real1}--\ref{fig:virt1b},
there are also processes that involve longitudinally polarized photons.
LCPT is formulated in lightcone gauge $A^+{=}0$.
In lightcone gauge, the longitudinal polarization of the photon propagates
instantaneously in lightcone time $x^+$ (reminiscent of Coulomb gauge,
where the electric potential $A^0$ propagates instantaneously in
ordinary time $x^0$).  LCPT implements time-ordered perturbation
theory in lightcone time, and so such interactions are instantaneous
in that formalism.  Fig.\ \ref{fig:II} shows a diagram that contributes
to energy loss through $e{\to}e\E\Ebar$ via longitudinally-polarized photons.
The LCPT convention is to denote longitudinal polarization by drawing
the photon line vertically (since
the interaction is instantaneous in lightcone time) and crossing it
with a bar through its middle.

\begin {figure}[t]
\begin {center}
  \includegraphics[scale=0.6]{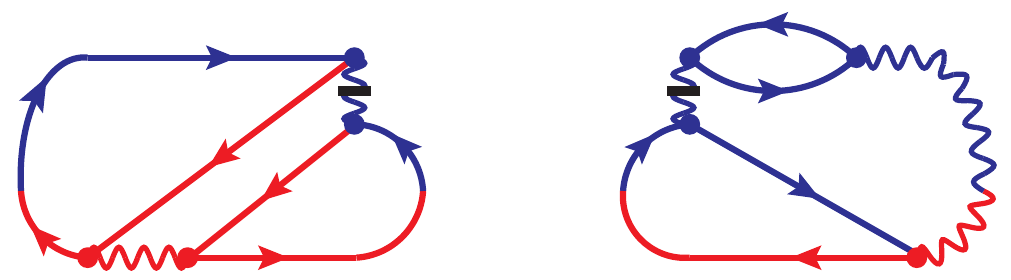}
  \caption{
    \label{fig:Iexamples}
    Two examples of lightcone-time ordered interference diagrams with one
    longitudinally-polarized photon and
    one transversely-polarized photon.
    The first diagram contributes to $e\to e\E\Ebar$ and the second to
    NLO $e\to e\gamma$.
    [In fact, other than adding the above diagrams to
    their conjugates by taking $2\Re(\cdots)$,
    all other diagrams of this type
    cancel each other in the net rate (\ref{eq:netrate}) for reasons
    similar to fig.\ \ref{fig:cancel}.]
  }
\end {center}
\end {figure}

For general $x_\gamma$, there are also diagrams such as
fig.\ \ref{fig:Iexamples} that involve combination of longitudinally
and transversely polarized photons.
For reasons explained in appendix \ref{app:I},
diagrams of this type are unimportant in the
soft photon limit of interest in this paper, and so we will ignore
them here.

\begin {figure}[t]
\begin {center}
  \includegraphics[scale=0.45]{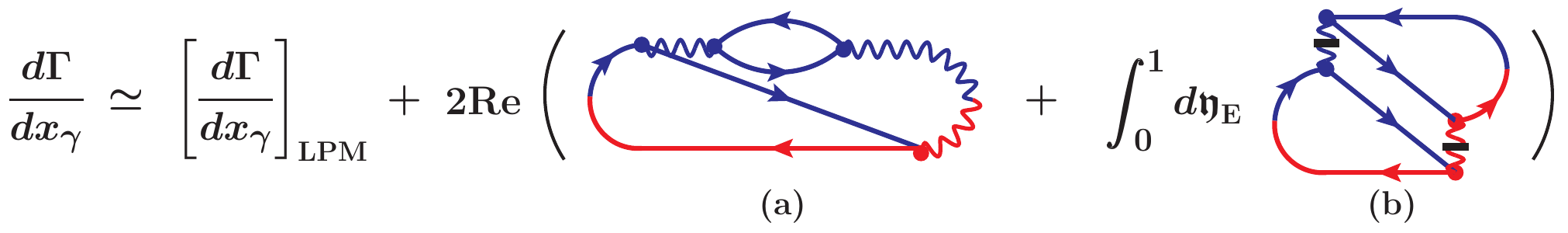}
  \caption{
    \label{fig:netrate}
    The differential rate for the initial electron to lose energy
    by $x_\gamma E$, through NLO for $\Nf\alpha \ll x_\gamma \ll 1$.
    The right-hand side accounts for all overlap effects at this order
    (ignoring diagrams like fig.\ \ref{fig:Iexamples} that are suppressed for
    $x_\gamma \ll 1$).
  }
\end {center}
\end {figure}

The upshot is that, in the soft photon limit $x_\gamma \ll 1$, the only
contributions to the rate (\ref{eq:netrate}) through NLO are those of
fig.\ \ref{fig:netrate}.

\subsection {Calculation in soft-photon limit}

We now turn to making a fully analytic calculation of
fig.\ \ref{fig:netrate}a in the soft-photon limit.
(We will relegate a similar calculation of fig.\ \ref{fig:netrate}b to
appendix \ref{app:II}.)

We will start by previewing the basic approximation and then
discuss the origin of that approximation.  In the soft photon approximation
$x_\gamma \ll 1$, we find that
\begin {equation}
  \left[ \frac{d\Gamma}{dx_\gamma} \right]_{\rm (a)}
  \equiv 2\Re\bigl( \mbox{fig.\ \protect\ref{fig:netrate}a} \bigr)
  = \int_0^1 d\yfrakE \>
    \left[ \frac{d\Gamma}{dx_\gamma \, d\yfrakE} \right]_{\rm (a)}
\label {eq:Atotal}
\end {equation}
where
\begin {subequations}
\label{eq:calGa}
\begin {equation}
  \left[ \frac{d\Gamma}{dx_\gamma \, d\yfrakE} \right]_{\rm (a)}
  \simeq
  - 2\Re
  \int d({\rm times}) \>
  \left[ \frac{d{\cal G}}{dx_\gamma\,d(\Delta t_\br)} \right]_\brem
  \left[ \frac{d{\cal G}}{d\yfrakE\,d(\Delta t_\pr)} \right]_\pair
\label {eq:Afactorize}
\end {equation}
and where the $d{\cal G}/{dx\,d(\Delta t)}$ represent the ordinary
LPM bremsstrahlung or pair rate formula (\ref{eq:ZrateHO}) or (\ref{eq:ZpairHO})
\textit{before} integrating over the time separation $\Delta t$ or
taking $2\Re(\cdots)$.  That is,
\begin {align}
   \left[ \frac{d{\cal G}}{dx_\gamma\,d(\Delta t)} \right]_\brem
   &=
   - \frac{\alpha}{2\pi} \,P_{e\to\gamma}(x_\gamma)
   \, \Omega_0^2 \csc^2(\Omega_0\,\Delta t) ,
\label {eq:dGbrem}
\\
   \left[ \frac{d{\cal G}}{d\yfrakE\,d(\Delta t)} \right]_\pair ~
   &=
   - \frac{\Nf\alpha}{2\pi} \,P_{\gamma\to e}(\yfrakE)
   \, \Omegapr^2 \csc^2(\Omegapr\,\Delta t) .
\label {eq:dGpair}
\end {align}
\end {subequations}
The overall minus sign in (\ref{eq:Afactorize}) will arise from
the pair production being virtual rather than real.
The integral in (\ref{eq:Afactorize}) is over
the relative times of the vertices in fig.\ \ref{fig:netrate}a:
\begin {equation}
   \int d({\rm times}) \> \cdots
   ~\equiv~
   \int_0^\infty d(t_4{-}t_3) \> d(t_3{-}t_2) \> d(t_2{-}t_1) \> \cdots,
\label {eq:inttimes1}
\end {equation}
where $t_1 < t_2 < t_3 < t_4$ are the time-ordered
vertex times from left
to right in the figure, $\Delta t_\br \equiv t_4{-}t_{1}$ is
the duration of the bremsstrahlung,
and $\Delta t_\pr \equiv t_3{-}t_2$ is the duration of the virtual pair
$\E\Ebar$.
Because the integrand in (\ref{eq:Afactorize}) depends only on
$\Delta t_\br$
and $\Delta t_\pr$, we may reduce the
three time integrals in (\ref{eq:inttimes1}) to two:
\begin {equation}
  \int d({\rm times}) \> \cdots
  ~=~
  \int_0^\infty d(\Delta t_\pr) \, \int_{\Delta t_\pr}^\infty d(\Delta t_\br) \>
  (\Delta t_\br {-} \Delta t_\pr) \> \cdots .
\label {eq:inttimes2}
\end {equation}

The factorization of the rate in (\ref{eq:Afactorize})
[before integrating over times and taking $2\Re(\cdots)$]
into a factor for
bremsstrahlung times a factor for pair production
will be a special property of the soft photon limit.
It does not apply to the more general (and much more complicated) expressions
developed in ref.\ \cite{qedNf} to handle the case $x_\gamma \sim 1$.


\subsection {Origin of the approximation}
\label {sec:origin}

\begin {figure}[t]
\begin {center}
  \includegraphics[scale=0.6]{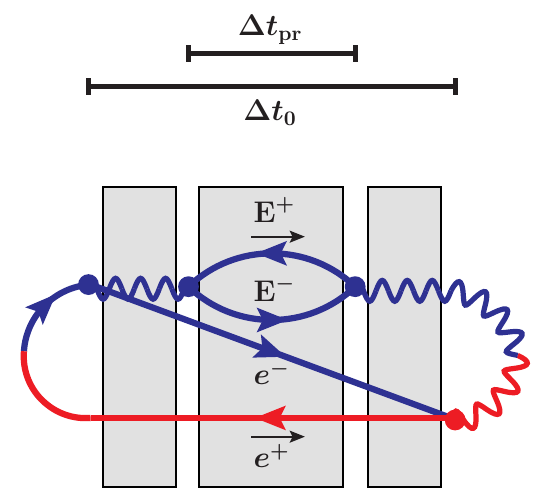}
  \caption{
    \label{fig:fund}
    The diagram of fig.\ \ref{fig:netrate}a separated into regions
    of 3-particle and 4-particle evolution.
  }
\end {center}
\end {figure}

Fig.\ \ref{fig:fund} shows that
the diagram divides into two regions (left and right)
of 3-particle evolution and one region (middle) of 4-particle
evolution.  In the 3-particle regions, the ``particles'' are
$e^-e^+\gamma$, and the effective medium-averaged
Hamiltonian is exactly the same
as in the LPM bremsstrahlung case of (\ref{eq:Heff}).
In the 4-particle region, we have $e^- e^+ \E^- \E^+$, and
the effective Hamiltonian is of the form
\begin {equation}
  {\cal H}_4 =
  \frac{p_{\perp e^-}^2{+}m^2}{2(1{-}x_\gamma)E}
     + \frac{p_{\perp \ssE^-}^2{+}m_\ssE^2}{2\yfrakE x_\gamma E}
     + \frac{p_{\perp \ssE^+}^2{+}m_\ssE^2}{2(1{-}\yfrakE)x_\gamma E}
     - \frac{p_{\perp e^+}^2{+}m^2}{2E}
  + V_4(\b_{e^-},\b_{e^+},\b_{\ssE^-}, \b_{\ssE^+}) ,
\label {eq:H4}
\end {equation}
where the free-particle terms follow from the same arguments as
in section \ref{sec:basic}, the denominators are twice
the energy $\simeq$ $p_z$ of the corresponding particles,
and the $k_\perp^2/2x_\gamma E$ photon term of (\ref{eq:Heff}) has
been replaced by two terms for $\E^-$ and $\E^+$.

The 4-body potential $V_4$ above is given by the generalization of
(\ref{eq:Vqqsum}) from two particles $e^-e^+$ to four particles
$e^-e^+\E^-\E^+$:%
\footnote{
  For a short review, see, for example, section II.B of ref.\ \cite{Vqhat},
  where the color matrices $\underline{\Time}_i$ for QCD there
  correspond to the charges $Q_i$ in the QED case here.
}
\begin {equation}
  V_4(\b_{e^-},\b_{e^+},\b_{\ssE^-}, \b_{\ssE^+}) =
    -\frac{i}{2}
    \hspace{-1.6em}
      \sum_{\substack{ j,k \in \\ \{e^-\!,e^+\!,\ssE^-\!,\ssE^+\} }}
    \hspace{-1.6em}
       Q_j Q_k \, \gamma(\b_j{-}\b_k) ,
\label {eq:V4sum}
\end {equation}
where $\gamma(\b)$ is the same as in (\ref{eq:V}).
Using $\sum_j Q_j = 0$, the sum (\ref{eq:V4sum}) can be written as
simply a sum over 2-body ``potentials'' (\ref{eq:V}):
\begin {multline}
  V_4(\b_{e^-},\b_{e^+},\b_{\ssE^-}, \b_{\ssE^+}) =
     V(\b_{e^-}{-}\b_{e^+}) + V(\b_{e^-}{-}\b_{\ssE^+})
     + V(\b_{\ssE^-}{-}\b_{e^+}) + V(\b_{\ssE^-}{-}\b_{\ssE^+})
\\
     - V(\b_{e^-}{-}\b_{\ssE^-}) - V(\b_{e^+}{-}\b_{\ssE^+}) ,
\label {eq:V4sum2}
\end {multline}
where the signs of each term are given by $-Q_j Q_k$.
This decomposition of 4-body $V_4$ into a sum of 2-body $V$'s relies
on being able to perturbatively
treat individual elastic scatterings of high-energy particles
with the medium.%
\footnote{
  In QCD, there is also a (slightly different) decomposition of 4-body $V_4$'s
  into 2-body $V$'s in the large-$\Nc$ limit, even when the interactions with
  the medium are strong.  See, for example, section 4.3 of ref.\ \cite{2brem}
  or section 4 of ref.\ \cite{LOptBlaizot}.
  But, \textit{in the case} of specializing to the $\qhat$ approximation,
  section 3 of ref.\ \cite{Vqhat} argues generally that
  $N$-body $V$'s should decompose into a sum of 2-body $V$'s.
}

Making the multiple-scattering ($\qhat$) approximation (\ref{eq:Vqhat})
in the deep LPM regime, (\ref{eq:V4sum2}) can be simplified to
\begin {equation}
  V_4(\b_{e^-},\b_{e^+},\b_{\ssE^-}, \b_{\ssE^+})
  = -\tfrac{i}{4} \qhat \, (\b_{e^-}{+}\b_{\ssE^-}{-}\b_{\ssE^+}{-}\b_{e^+})^2 .
\label {eq:V4qhat}
\end {equation}

We are now ready to make the soft-photon approximation.  The virtual pair
production loop in fig.\ \ref{fig:fund} will be integrated over all
$p^+$ momentum fractions $\yfrakE$ of the pair electron $\E^-$
relative to the photon, and that integral (as one may verify later)
will be dominated by democratic splittings.  From (\ref{eq:bscales}),
the typical separation between the pair electron and positron
during democratic pair
production is
\begin{equation}
  |\b_{\ssE^-} {-} \b_{\ssE^+}| \sim
  (\qhat \kgamma)^{-1/4} \sim (x_\gamma \qhat E)^{-1/4} ,
\end {equation}
where we've switched to our labeling $\E^-\E^+$ of the pair in the
present discussion, in the context of which also $k_\gamma = x_\gamma E$.
For the underlying bremsstrahlung, (\ref{eq:bscales}) gives
\begin {equation}
  |\b_{e^-} {-} \b_{e^+}| \sim
  \left( \frac{\qhat E}{x_\gamma} \right)^{-1/4} ,
\end {equation}
and so, in the soft photon limit,
\begin {equation}
  |\b_{e^-} {-} \b_{e^+}| \ll |\b_{\ssE^-} {-} \b_{\ssE^+}| .
\label {eq:bhierarchy}
\end {equation} 
Because of this, a good approximation to the 4-body potential
(\ref{eq:V4qhat}) is
\begin {equation}
  V_4(\b_{e^-},\b_{e^+},\b_{\ssE^-}, \b_{\ssE^+})
  = -\tfrac{i}{4} \qhat\, (\b_{\ssE^-}{-}\b_{\ssE^+})^2
  \qquad
  \mbox{(for $x_\gamma \ll 1$)}.
\label {eq:V4soft}
\end {equation}
Notice that the evolution of the soft virtual pair $\E^-\E^+$ is now
decoupled from the hard $e^- e^+$ in the soft photon limit.
This is what justifies factorizing out the pair-production part
$[d{\cal G}/d\yfrakE\,d(\Delta t_\pr)]_\pair$ of
the calculation in (\ref{eq:Afactorize}).

We will find it convenient to use (\ref{eq:bhierarchy}) to
modify (\ref{eq:V4soft}) to
the equally good (but no better) soft-photon
approximation
\begin {equation}
  V_4(\b_{e^-},\b_{e^+},\b_{\ssE^-}, \b_{\ssE^+})
  = -\tfrac{i}{4} \qhat\, (\b_{\ssE^-}{-}\b_{\ssE^+})^2
    -\tfrac{i}{4} \qhat\, (\b_{e^-}{-}\b_{e^+})^2
  \qquad
  \mbox{(for $x_\gamma \ll 1$)}.
\label {eq:V4soft2}
\end {equation}
The advantage is that the $e^- e^+$ associated with bremsstrahlung then
evolves with the exact same Hamiltonian (\ref{eq:Heff})
in all three regions of fig.\ \ref{fig:fund}, covering the
entire bremsstrahlung process.%
\footnote{
\label{foot:kperpzero}
  This argument implicitly assumes that choosing
  the $z$ axis to be in the direction of the photon ($\k_\perp{=}0$)
  in the first region of fig.\ \ref{fig:fund} will \textit{also}
  make the $z$ axis align with the direction of the photon
  (and so $\k_\perp{=}0$) in
  the last region of fig.\ \ref{fig:fund}.  That's only true if
  the total transverse momentum $\p_{\perp\ssE^-}{+}\p_{\perp\ssE^+}$ of
  the $\E^-\E^+$ pair is conserved during the
  medium-averaged 4-particle evolution
  in the middle region of fig.\ \ref{fig:fund}.
  That wouldn't be true in general, where only
  $\p_{\perp e^-}{+}\p_{\perp e^+}{+}\p_{\perp\ssE^-}{+}\p_{\perp\ssE^+} = 0$
  would be conserved,
  but it \textit{is} true in the
  soft-photon limit,
  where the medium-averaged evolution of the $\E^-\E^+$ pair
  decouples from everything else.
}
In consequence, what remains after
factorizing out the pair production in (\ref{eq:Afactorize}) is simply
the corresponding formula $[d{\cal G}/dx_\gamma\,d(\Delta t_\br)]_\brem$
for ordinary (i.e.\ non-overlapping) LPM bremsstrahlung.

One might worry about whether one can ignore the differences in
the hard $e^- e^+$ evolution described by
(\ref{eq:V4qhat}), (\ref{eq:V4soft}), and (\ref{eq:V4soft2}) during
the time $\Delta t_\pr$ when the virtual pair is present.
First, from the soft bremsstrahlung formation time (\ref{eq:tform})
and the democratic case of the pair production formation time
(\ref{eq:tpair}), note that
\begin{equation}
   t_\form^\brem \sim \sqrt{ \frac{E}{x_\gamma\qhat} }
   \gg
   \sqrt{ \frac{x_\gamma E}{\qhat} } \sim t_\form^\pair
\label {eq:thierarchy}
\end {equation}
in the soft-photon limit.
So the exact treatment of 4-particle evolution on the $e^-e^+$ only
represents a negligible fraction of $e^- e^+$ evolution over an
ordinary LPM soft bremsstrahlung formation time $\tform^\brem$.  However, the
integral over the duration $\Delta t_0$ of bremsstrahlung in
(\ref{eq:Afactorize}) extends all the way down to the lower limit
$\Delta t_0{=}\Delta t_\pr$, and we will find in the overlap calculation
that $\Delta t_0 \sim \Delta t_\pr$ contributes as much as
$\Delta t_0 \sim t_\form^\brem$, with a logarithm arising from all scales
between, and so we need to correctly handle the case
$\Delta t_0 \ll t_\form^\brem$ including $\Delta t_0 \sim \Delta t_\pr$.
Fortunately, in the limit $\Delta t_0 \ll t_\form^\brem$, the
ordinary LPM bremsstrahlung contribution (\ref{eq:dGbrem}) to our factorized
formula (\ref{eq:Afactorize}) becomes insensitive to the medium:
\begin {equation}
   \left[ \frac{d{\cal G}}{dx_\gamma\,d(\Delta t)} \right]_\brem
   \simeq
   - \frac{\alpha\,P_{e\to\gamma}(x_\gamma)}{2\pi (\Delta t)^2}
   \qquad \mbox{(for $|\Omega_0|\,\Delta t \ll 1$)} .
\end {equation}
So in that case the kicks to the medium do not matter, and so the
differences between describing correlations of those kicks using
(\ref{eq:V4qhat}), (\ref{eq:V4soft}), or (\ref{eq:V4soft2})
still do not matter.

\begin {figure}[t]
\begin {center}
  \includegraphics[scale=0.6]{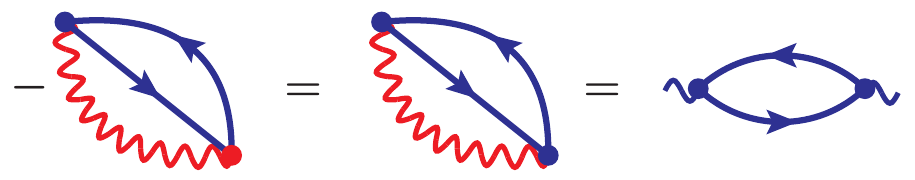}
  \caption{
    \label{fig:pairbackend}
    Like the relation in fig.\ \ref{fig:cancel}, the first equality here
    relates
    (left) LPM pair production $\gamma\to e^- e^+$ to
    (middle) the interference of virtual pair production
    $\gamma \to e^- e^+ \to \gamma$ with no pair production $\gamma \to \gamma$.
    The final equality reflects the fact that the conjugate amplitude photon
    line plays no role in the calculation.
  }
\end {center}
\end {figure}

Finally, we should say a little more about the overall minus sign
in our soft-photon factorization (\ref{eq:Afactorize}).
By the same arguments that gave the cancellation of diagrams in
fig.\ \ref{fig:cancel}, the processes of (non-overlapping)
real LPM pair production
and (non-overlapping) virtual LPM pair production are related by
a minus sign as in
the first equality of fig.\ \ref{fig:pairbackend}.
Having always chosen the $z$ axis in the direction of the photon
and so $k_\perp = 0$,
the photon played no role in evolution
based on the effective Hamiltonian (\ref{eq:Hpair}) for pair production,
and so it makes no difference if we remove the conjugate amplitude photon line,
as indicated by the last equality of fig.\ \ref{fig:pairbackend}.
This is why, once factorized, the virtual pair production loop corresponds
to a factor of $-[d{\cal G}/d\yfrakE\,d(\Delta t_\pr)]_\pair$ in
(\ref{eq:Afactorize}) instead of
$+[d{\cal G}/d\yfrakE\,d(\Delta t_\pr)]_\pair$.

\subsection {Evaluating the integrals}

Return now to the factorized expression (\ref{eq:calGa}), and use
(\ref{eq:inttimes2}) for the time integrations:
\begin {multline}
  \left[ \frac{d\Gamma}{dx_\gamma \, d\yfrakE} \right]_{\rm (a)}
  \simeq
  - \frac{\Nf\alpha^2}{2\pi^2} \, P_{e\to\gamma}(x_\gamma) \,P_{\gamma\to e}(\yfrakE)
  \,
  \Re
  \int_0^\infty d(\Delta t_\pr) \> \Omegapr^2 \csc^2(\Omegapr\,\Delta t_\pr)
\\ \times
  \int_{\Delta t_\pr}^\infty d(\Delta t_\br) \>
  (\Delta t_\br {-} \Delta t_\pr) \,
   \Omega_0^2 \csc^2(\Omega_0\,\Delta t_\br) .
\label {eq:Afactorize2}
\end {multline}


\subsubsection{The \boldmath$\Delta t_\br$ integral}

The $\Delta t_\br$ integral gives
\begin {equation}
  \int_{\Delta t_\pr}^\infty d(\Delta t_\br) \>
    (\Delta t_\br {-} \Delta t_\pr) \,
    \Omega_0^2 \csc^2(\Omega_0\,\Delta t_\br)
  = -\ln\bigl( 1 - e^{-2i\Omega_0\,\Delta t_\pr} \bigr) .
\label {eq:intpr}
\end {equation}
Because $\Omega_\pr$ is complex, the $\csc^2(\Omegapr\,\Delta t_\pr)$
in (\ref{eq:Afactorize2}) falls exponentially for
$\Delta t_\pr \gg 1/|\Omegapr|$, and so the
$\Delta t_\pr$ integral is dominated by $\Delta t_\pr \lesssim 1/|\Omegapr|$.
Then
\begin {equation}
   |\Omega_0\,\Delta t_\pr| \lesssim \frac{|\Omega_0|}{|\Omegapr|}
   \sim \frac{\tform^\pair}{\tform^\brem} \,.
\end {equation}
Eq.\ (\ref{eq:thierarchy}) then gives
\begin {equation}
   |\Omega_0\,\Delta t_\pr| \lesssim x_\gamma \ll 1
\end {equation}
in the soft-photon limit.  So we may approximate the result of
(\ref{eq:intpr}) by $-\ln(2i\Omega_0\,\Delta t_\pr)$, and
(\ref{eq:Afactorize2}) becomes
\begin {equation}
  \left[ \frac{d\Gamma}{dx_\gamma \, d\yfrakE} \right]_{\rm (a)}
  \simeq
  \frac{\Nf\alpha^2}{2\pi^2} \, P_{e\to\gamma}(x_\gamma) \,P_{\gamma\to e}(\yfrakE)
  \,
  \Re
  \int_0^\infty d(\Delta t_\pr) \> \Omegapr^2 \csc^2(\Omegapr\,\Delta t_\pr)
  \ln (2i\Omega_0\,\Delta t_\pr) .
\label {eq:Afactorize3}
\end {equation}
Unfortunately, the remaining integral over $\Delta t_\pr$ has a UV divergence
associated with $\Delta t_\pr \to 0$.


\subsubsection{Sidestepping the UV divergence}

Unlike the LPM rate (\ref{eq:ZrateHO1}), attempting to sidestep
the UV divergence of
(\ref{eq:Afactorize3}) by simply subtracting
the full vacuum limit $\qhat\to 0$
(i.e.\ $\Omega_\br\to 0$ and $\Omega_\pr \to 0$) does not yield
a finite expression.
Instead, consider subtracting the vacuum limit of \textit{just}
the virtual pair loop ($\Omegapr \to 0$ only) in fig.\ \ref{fig:fund},
and then add it back in:
\begin {multline}
  \left[ \frac{d\Gamma}{dx_\gamma \, d\yfrakE} \right]_{\rm (a)}
  \simeq
\\
  \frac{\Nf\alpha^2}{2\pi^2} \, P_{e\to\gamma}(x_\gamma) \,P_{\gamma\to e}(\yfrakE) \,
  \Re
    \int_0^\infty d(\Delta t_\pr) \>
      \left[ \Omegapr^2 \csc^2(\Omegapr\,\Delta t_\pr)
             - \frac{1}{(\Delta t_\pr)^2} \right]
      \ln (2i\Omega_0\,\Delta t_\pr)
\\
  +
  \left[ \frac{d\Gamma}{dx_\gamma \, d\yfrakE} \right]_{\rm (a)}^\vacloop .
\label {eq:Afactorize4}
\end {multline}

It is possible, but unpleasant, to evaluate the vacuum-loop contribution
$d\Gamma_{\rm (a)}^\vacloop$ to fig.\ \ref{fig:fund} using
a consistent UV regularization scheme.%
\footnote{
  See, for example, the discussion of dimensional regularization in
  footnote \ref{foot:UV}.
}
We will instead give qualitative arguments for why that is
unnecessary.  First, remember that our interest in this paper is focused
on how the overlap effects of subsequent pair production on bremsstrahlung
become important for sufficiently soft photons, and that the qualitative
reason for that is that the soft pair can be much more easily deflected
by the medium than the original high-energy electron.
But, by definition, the soft pair in $d\Gamma_{\rm (a)}^\vacloop$
does \textit{not}
interact with the medium, and so should not have a large effect on the
underlying bremsstrahlung process.

A second crude argument is to think about the total evolution of the photon
in $d\Gamma_{\rm (a)}^\vacloop$, which (virtual pair loop included) involves no
interactions with the medium.  So consider the photon evolution in
terms of ordinary Feynman rules instead of in the time-ordered formalism
we have been using.  The cost of adding
a \textit{vacuum} self-energy loop to a vacuum photon propagator should just
be of order $\Nf\alpha$, up to renormalization logarithms; there is no
ratio of scales involved that could give an order $\Nf\alpha/x_\gamma$
effect, as in (\ref{eq:LPMqualplus}),
that would become important when $x_\gamma \sim \Nf\alpha$.%
\footnote{
  In a little more detail, the vacuum photon self-energy will have the form
  \[
     \Pi_{\alpha\beta} =
     \Nf\alpha (g_{\alpha\beta} Q^2 - Q_\alpha Q_\beta) \times
     \bigl[ \mbox{renormalization log} + O(1) \bigr]
  \]
  and inserting a vacuum self-energy changes the photon propagator from
  \[
     G_{(0)} \longrightarrow G_{(0)} \Pi G_{(0)}
     = \Nf\alpha\, G_{(0)} \times
       \bigl[ \mbox{renormalization log} + O(1) \bigr]  .
  \]
}
This is related to the fact that
the UV divergence of the pair loop contributes to the renormalization
of the coupling $\alpha$, which means that the divergent contribution
of $d\Gamma_{\rm (a)}^\vacloop$ must be of order the ordinary LPM
rate times $\Nf\alpha$ times renormalization logarithms.
The much more complicated but more general calculation of ref.\ \cite{qedNf}
(without assuming $x_\gamma \ll 1$)
verified that this renormalization works out correctly.


\subsubsection{The \boldmath$\Delta t_\pr$ and $\yfrakE$ integrals}

Changing integration variable to $\tau \equiv i\Omegapr \,\Delta t_\pr$,
the integral in (\ref{eq:Afactorize4}) can be rewritten as
\begin {multline}
   \int_0^\infty d(\Delta t_\pr) \>
     \left[ \Omegapr^2 \csc^2(\Omegapr\,\Delta t_\pr)
            - \frac{1}{(\Delta t_\pr)^2} \right]
      \ln (2i\Omega_0\,\Delta t_\pr)
\\
    =
    i\Omegapr
    \int_0^\infty d\tau \>
    \left( \frac{1}{\sh^2\tau} - \frac{1}{\tau^2} \right)
    \ln\Bigl( \frac{2\Omega_0\tau}{\Omegapr} \Bigr) .
\end {multline}
Using%
\footnote{
  For a derivation of the integral (\ref{eq:integral1}), see, for example,
  the derivation of eq.\ (E.18) of ref.\ \cite{qcd} in the corresponding
  paragraph of appendix B of ref.\ \cite{qcd}.
}
\begin {equation}
    \int_0^\infty d\tau \>
      \left( \frac{1}{\sh^2\tau} - \frac{1}{\tau^2} \right) \ln(c\tau)
    =
    -\ln(c\pi) + \gammaE ,
\label {eq:integral1}
\end {equation}
(\ref{eq:Afactorize4}) becomes
[dropping the $d\Gamma^\vacloop$ in the soft-photon limit]
\begin {equation}
  \left[ \frac{d\Gamma}{dx_\gamma \, d\yfrakE} \right]_{\rm (a)}
  \simeq
  \frac{\Nf\alpha^2}{2\pi^2} \, P_{e\to\gamma}(x_\gamma) \,P_{\gamma\to e}(\yfrakE) \,
  \Re(i\Omega_\pr)
    \left[
      \ln\Bigl(\frac{|\Omegapr|}{2\pi|\Omega_0|}\Bigr) + \gammaE
  \right] ,
\label {eq:Afactorize5}
\end {equation}
where we've used the fact that $\Omegapr$ and $\Omega_0$ have the same
complex phase to rewrite $\Omegapr/\Omega_0$ as $|\Omegapr|/|\Omega_0|$.

Now we need to integrate over the loop variable $\yfrakE$ as
in (\ref{eq:Atotal}).  Using (i) the explicit formula
(\ref{eq:Omegapr}) for $\Omegapr$, (ii) the soft-photon approximation
$\Omega_0 \simeq \sqrt{-i x_\gamma \qhat/2E}$ to (\ref{eq:Omega0}),
and (iii) the integrals
\begin {equation}
  \int_0^1 d\yfrakE \>
     \frac{ P_{\gamma\to e}(\yfrakE) }{ \sqrt{\yfrakE(1{-}\yfrakE)} }
  = \frac{3\pi}{4}
\end {equation}
and
\begin {equation}
  \int_0^1 d\yfrakE \>
     \frac{ P_{\gamma\to e}(\yfrakE) }{ \sqrt{\yfrakE(1{-}\yfrakE)} }
     \, \ln\bigl(\yfrakE(1{-}\yfrakE)\bigr)
  = -\frac{\pi}{4} - 3\pi\ln 2 ,
\end {equation}
gives
\begin {equation}
  \left[ \frac{d\Gamma}{dx_\gamma} \right]_{\rm (a)}
  =
  \left[ \frac{d\Gamma}{dx_\gamma} \right]_\LPM
  \frac{\Nf\alpha}{2 x_\gamma} \, f_\Plus^{\rm(a)}(x_\gamma)
\label {eq:fplusAdef}
\end {equation}
with
\begin {equation}
  f_\Plus^{\rm(a)}(x_\gamma) \simeq
  \tfrac34 \ln\bigl( \tfrac{2}{\pi x_\gamma} \bigr)
  + \tfrac34 \gammaE + \tfrac18
  \,.
\label {eq:fplusA}
\end {equation}


\subsection{Result for net rate \boldmath$d\Gamma/dx$}

The longitudinal-photon interference diagram of fig.\ \ref{fig:netrate}b
for $e\to e\E\Ebar$
is computed in the soft-photon limit in appendix \ref{app:II} and gives
\begin {equation}
  \left[ \frac{d\Gamma}{dx_\gamma \, d\yfrakE} \right]_{\rm (b)}
  \simeq
  \frac{\Nf\alpha^2}{\pi^2} \, \frac{\yfrakE(1{-}\yfrakE)}{x_\gamma}
  \Re(i\Omega_\pr) \, 4\ln2 \,.
\label {eq:Bfactorize1}
\end {equation}
Integrating over the final state momentum fraction $\yfrakE$
as in the figure gives the analog of (\ref{eq:fplusAdef}) with
\begin {equation}
  f_\Plus^{\rm(b)}(x_\gamma) \simeq
  \tfrac12 \ln 2 .
\label {eq:fplusB}
\end {equation}
Finally, adding together (\ref{eq:fplusA}) and (\ref{eq:fplusB}) gives
the NLO version of the $\LPMplus$ rate,
\begin {subequations}
\label {eq:NLOrate}
\begin {equation}
  \left[ \frac{d\Gamma}{dx_\gamma} \right]_{\LPMplus}^\NLO
  \simeq
  \left[ \frac{d\Gamma}{dx_\gamma} \right]_\LPM
  \left\{
    1 + \frac{\Nf\alpha}{2 x_\gamma}  \, f_\Plus^\NLO(x_\gamma)
  \right\}
  \qquad \mbox{(for $\Nf\alpha \ll x_\gamma$)}
\end {equation}
with
\begin {equation}
  f_\Plus^\NLO(x_\gamma) \simeq
  \tfrac34 \ln\bigl( \tfrac{2}{\pi x_\gamma} \bigr)
  + \tfrac34 \gammaE + \tfrac18 + \tfrac12 \ln 2
  \qquad \mbox{(for $x_\gamma \ll 1$)}
\label {eq:fNLO}
\end {equation}
\end {subequations}
which was the result previewed in (\ref{eq:fplusAB}).

For readers with reservations about any of the
various soft-photon approximations made so far, the agreement
of (\ref{eq:fNLO}) with the numerical result (\ref{eq:fplusold}),
extracted from numerical results \cite{qedNfenergy}
that made \textit{no} soft-photon approximations, should
be reassuring.


\section{Main result: Soft photon emission \boldmath$x_\gamma \ll 1$
  including $x_\gamma \lesssim \Nf\alpha$}
\label{sec:mainresult}

In section \ref{sec:disruption} we argued qualitatively
(like Galitsky and Gurevich) that pair production will disrupt the
bremsstrahlung formation time when the ordinary LPM formation
time $\tform^\LPM$ becomes longer than the mean time $1/\Gamma_\pair$
for the photon to convert to an $\E^-\E^+$ pair, so that
the actual bremsstrahlung formation time
(\ref{eq:tformplus1}) in that case would be $\tform^\LPMplus \sim 1/\Gamma_\pair$.


\subsection{Diagrams}

How does this happen diagrammatically?  As a related example, consider the
ordinary Feynman propagator for a particle propagating
through the vacuum.  For the free propagator
$G_{(0)}(p) = i/(p^2-m^2+i\varepsilon)$,
Fourier transforming frequency to time
gives oscillating behavior $G_{(0)}(\p,t) \propto e^{-i E_\p t}$ at
large times.  But, if the particle is unstable, then the self-energy
$\Pi(p)$ has an imaginary part related by the optical theorem
to the decay rate of the particle.
The \textit{full} propagator $G(p) = i/(p^2-m^2-\Pi(p)+i\varepsilon)$ then
has time dependence of the form
$G(\p,t) \propto e^{-i E'_\p t} e^{-\gamma_\p t}$ at
large times.  In diagrammatic language, one must resum
the self-energy $\Pi$ into the propagator to obtain the
factor $e^{-\gamma t}$ accounting for the probability amplitude that
the particle has \textit{not} already decayed before time $t$ elapses.
Note as well that the real part of the self-energy can have the effect
of modifying the oscillation $e^{-i E_\p t}$ of the amplitude.

\begin {figure}[t]
\begin {center}
  \includegraphics[scale=0.6]{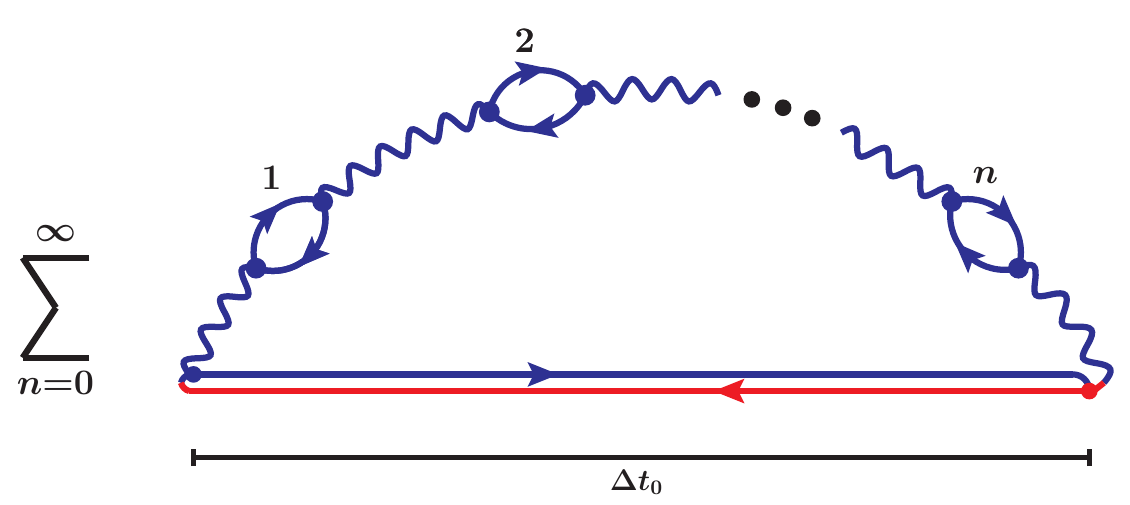}
  \caption{
    \label{fig:bubbles}
    Resummation of virtual pair production bubbles for a soft
    photon line in figs.\ \ref{fig:lpm}b and \ref{fig:netrate}a.
    Unlike previous figures, we have drawn the diagram here
    to visually emphasize that the separation of the hardest
    electrons $e^-$ (amplitude) and $e^+$ (conjugate amplitude)
    is relatively small, as in (\ref{eq:bhierarchy}).
  }
\end {center}
\end {figure}

So now consider the contribution of fig.\ \ref{fig:netrate}a to
the NLO $e \to e\gamma$ rate.  In order to end up with the final-state
photon, the photon cannot have actually converted to a real (as opposed
to virtual and so temporary) $\E^-\E^+$ pair during the time $\Delta t_0$
of the bremsstrahlung process.
In order to account for the probability amplitude that such a conversion
has not taken place, we need to resum self-energy bubbles in the
photon line of the diagram, and so we need to compute
fig.\ \ref{fig:bubbles}.  As we shall eventually see explicitly,
it is this resummation that will limit
the $\LPMplus$ bremsstrahlung formation time to be no larger than
order $1/\Gamma_\pair$.  The resummation of photon self-energy bubbles
will not only introduce a decay factor of the form $e^{-\gamma\,\Delta t_0}$
into the calculation, but in this case
the self-energy will also have a comparable
real part affecting the oscillation of the amplitude as well.

We claim that this is the only modification that needs to be made to
generalize the net rate of fig.\ \ref{fig:netrate} from
$\Nf\alpha \ll x_\gamma \ll 1$ to also include $x_\gamma \lesssim \Nf\alpha$.
The $n{=}0$ (no bubble) term in
fig.\ \ref{fig:bubbles} corresponds, after taking $2\Re(\cdots)$, to
the ordinary LPM bremsstrahlung rate $[d\Gamma/dx_\gamma]_\LPM$ on the
right-hand side of fig.\ \ref{fig:netrate}.  The $n{=}1$ (one bubble)
term in fig.\ \ref{fig:bubbles} corresponds to fig.\ \ref{fig:netrate}a.
So all we need do for the upgrade
is to add $2\Re(\cdots)$ of the $n\ge 2$ bubble terms
of fig.\ \ref{fig:bubbles} to the right-hand side of fig.\ \ref{fig:netrate}
and so to the right-hand side of (\ref{eq:NLOrate}).

Here, the large-$\Nf$ limit helps simplify the discussion by reducing the
number of diagrams to consider.  Adding (i) a fermion loop
to an existing photon line is formally suppressed by a factor of $\Nf\alpha$,
while adding (ii) a new photon line to a diagram is formally suppressed by a
factor of $\alpha$ and so is suppressed compared to (i) by a factor of
$1/\Nf$.  One might worry about whether an additional photon line could
be so infrared enhanced as to compensate the $1/\Nf$ suppression,
but we do not believe this is an issue.%
\footnote{
  As discussed in section \ref{sec:disruption}, during bremsstrahlung,
  very soft
  ($x_\gamma \ll \Nf\alpha$) photons convert to very soft pairs which can
  then interact directly with a QED medium like gluons do with a QCD medium.
  So it is reasonable to turn to analogous work on in-medium
  QCD showers for qualitative guidance.
  In the QCD case, adding a super-soft gluon emission to, say, an
  underlying $g\to gg$ bremsstrahlung process leads to
  power-law infrared divergences in the $\qhat$ approximation, but those
  divergences \textit{cancel} between real emission and virtual emission when
  combined into equations for shower evolution.
  See section 3 and appendix E of ref.\ \cite{qcd} for the gory details.
  The net effect of super-soft emissions after this cancellation can be
  packaged as a double-logarithmic contribution to
  the effective value of $\qhat$
  due to the fact that the super-soft emissions carry away transverse
  momentum and so affect the rate of change of the $(\Delta\p_\perp)^2$
  for the harder partons.  See ref.\ \cite{LMW} and its consequences for
  the QCD LPM effect in \cite{Blaizot,Iancu,Wu}.
  In QED, we expect that these
  effects are all further suppressed by an extra factor of $\Nf\alpha$ to
  account for the cost of converting the soft photon to the soft pair.
}

\begin {figure}[t]
\begin {center}
  \includegraphics[scale=0.6]{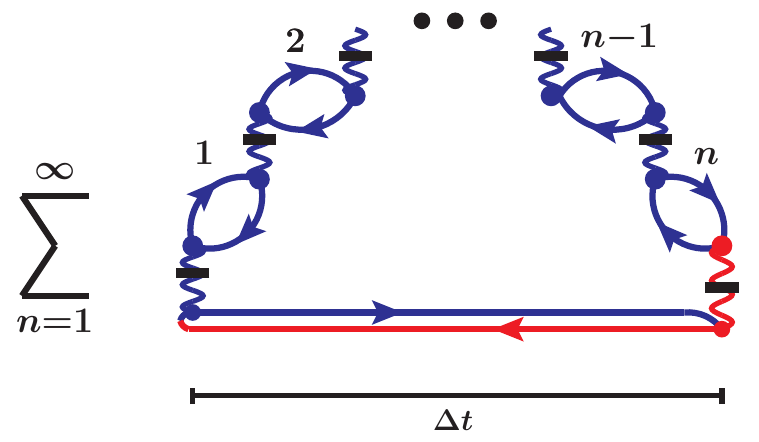}
  \caption{
    \label{fig:IIbubbles}
    Analogous to fig.\ \ref{fig:bubbles} except that all of the photons are
    longitudinally-polarized photons.
  }
\end {center}
\end {figure}

Adding new fermion-loop bubbles to the photon lines of the various diagrams
of figs.\ \ref{fig:real1}--\ref{fig:virt1b} will not affect the cancellation
of diagrams
discussed in section \ref{sec:NLOtransverse}, and so, as far as those diagrams
are concerned, we are left with
fig.\ \ref{fig:bubbles} in place of fig.\ \ref{fig:netrate}a.
There are many similar cancellations among diagrams involving
one or more longitudinally-polarized photon exchanges,%
\footnote{
  Also, any fermion loop where one end
  is connected to a transverse photon
  and the other to a longitudinal photon may be ignored for reasons
  similar to the arguments of appendix \ref{app:I} to ignore
  diagrams with a single longitudinal photon.
}
but there is also a class of diagrams for which a different argument
is needed.  Consider the generalization of fig.\ \ref{fig:netrate}b
to fig.\ \ref{fig:IIbubbles}.  There is a fundamental
qualitative difference
between adding virtual pair-production bubbles to the
photon line of the original LPM bremsstrahlung diagram of
fig.\ \ref{fig:lpm}b and adding them to the $e\to e\E\Ebar$ diagram
of fig.\ \ref{fig:netrate}b.  Ordinary LPM bremsstrahlung
for $x_\gamma \ll \Nf\alpha$ has a
formation time (\ref{eq:tform}) that is \textit{long} compared to
the mean free time $1/\Gamma_\pair$ for pair production, and so it is
important to account for the probability $\exp(-\Gamma_\pair t)$ for
the photon to survive that long without producing a real
$\E\Ebar$ pair.  The time scale in the longitudinal diagram
fig.\ \ref{fig:netrate}b is completely different.
In appendix \ref{app:II}, we show by explicit calculation that this
process only lasts for a much shorter time of order the \textit{pair}
production time $t_\form^\pair \sim \Nf\alpha/\Gamma_\pair$
(\ref{eq:GammaPairEst}).
Over that period of time,
$\exp(-\Gamma_\pair t) = \exp\bigl(-O(\Nf\alpha)\bigr) \simeq 1$ up to
corrections that are suppressed by $\Nf\alpha \ll 1$ instead of by
$\Nf\alpha/x_\gamma$.  So in this case
we may ignore the bubble sum that gives rise
to the $\exp(-\Gamma_\pair t)$ factor, and we need only retain the
leading term corresponding to fig.\ \ref{fig:netrate}b.

\subsection{Calculating \boldmath$n{\ge}2$ fermion bubble diagrams}

As discussed above, we need to add the $n{\ge}2$ part of the bubble
sum of fig.\ \ref{fig:bubbles} to the earlier NLO net rate
(\ref{eq:NLOrate}).  The fact that the only new thing
needed is $n{\ge}2$ will be important to simplifying the calculation,
but it will be useful to first formally consider the
sum over all $n{\ge}0$.  The soft-photon factorization of
(\ref{eq:Afactorize}) will generalize to
\begin {equation}
  \left[ \frac{d\Gamma}{dx_\gamma} \right]_{n\ge 0}
  \equiv 2\Re\bigl( \mbox{fig.\ \protect\ref{fig:bubbles}} \bigr)
  =
  \sum_{n=0}^\infty
  2\Re
  \int d({\rm times}) \> 
  \left[ \frac{d{\cal G}}{dx_\gamma\,d(\Delta t_\br)} \right]_\brem
  \prod_{j=1}^n \biggl(
    -\biggl[ \frac{d{\cal G}}{d(\Delta t_\pr^{(j)})} \biggr]_\pair
  \biggr) ,
\label {eq:Abubble}
\end {equation}
where
\begin {equation}
  \left[ \frac{d{\cal G}}{d(\Delta t_\pr)} \right]_\pair
  \equiv \int_0^1 d\yfrakE \> 
    \left[ \frac{d{\cal G}}{d\yfrakE\,d(\Delta t_\pr)} \right]_\pair
\end {equation}
and where $[d{\cal G}/dx_\gamma\,d(\Delta t_\br)]_\brem$
and $[d{\cal G}/d\yfrakE\,d(\Delta t_\pr)]_\pair$
are the same as in (\ref{eq:calGa}).
Letting $t_j$ and $\Delta t_j \equiv \Delta t_\pr^{(j)}$ be the
start time and duration of bubble $j = 1\cdots n$, the constraints
on the time integrations are
\begin {equation}
  0 < t_1 < t_1{+}\Delta t_1 < t_2 < t_2{+}\Delta t_2 < \cdots
    < t_n < t_n{+}\Delta t_n < \Delta t_0
\end {equation}
(where recall that $\Delta t_0$ is our notation for the duration of
the bremsstrahlung).  The integral over this time range can be written
as
\begin {equation}
   \int d({\rm times})
   ~=~
   \int_{0<t_1<t_2<\cdots<t_n<\Delta t_0}
   \int_0^{t_2-t_1} d(\Delta t_1)
   \int_0^{t_3-t_2} d(\Delta t_2)
   \cdots
   \int_0^{\Delta t_0-t_n} d(\Delta t_n)
   .
\label {eq:bubbletimes1}
\end {equation}
We claim (and will verify \textit{a posteriori})
that the sum over the $n \ge 2$ terms will be dominated by
\begin {equation}
  \Delta t_0 \sim \tform^\LPMplus \gtrsim \frac{1}{\Gamma_\pair}
\label {eq:dt0assumption}
\end {equation}
and that spacing between consecutive bubbles should be set by the
mean free time for pair production so that
\begin {equation}
  t_{j+1}{-}t_j \sim \frac{1}{\Gamma_\pair}
\label {eq:separated}
\end {equation}
(with the definition that $t_{n+1} \equiv \Delta t_0$).
Because
\begin {equation}
  \frac{1}{\Gamma_\pair} \sim \frac{\tform^\pair}{\Nf\alpha}
  \gg \tform^\pair ,
\end {equation}
and because
(\ref{eq:dGpair}) for
$[d{\cal G}/d\yfrakE\,d(\Delta t_\pr)]_\pair$ falls off exponentially
for $\Delta t_\pr \gg \tform^\pair$, it is a good approximation to replace
the upper limits of the $d(\Delta t_j)$ integrals in
(\ref{eq:bubbletimes1}) by $\infty$:
\begin {equation}
   \int d({\rm times})
   ~\simeq~
   \int_{0<t_1<t_2<\cdots<t_n<\Delta t_0}
   \int_0^\infty d(\Delta t_1)
   \int_0^\infty d(\Delta t_2)
   \cdots
   \int_0^\infty d(\Delta t_n)
\label {eq:bubbletimes2}
\end {equation}
for $n\ge 2$.  However, for now, it will be educational to use
(\ref{eq:bubbletimes2}) for $n{=}1$ as well to see how it is different,
though we will drop that and the $n{=}0$ term later.

So, proceeding with using (\ref{eq:bubbletimes2}) in (\ref{eq:Abubble}),
each $\Delta t$ integral factorizes from the other integrals, giving
\begin {equation}
  \left[ \frac{d\Gamma}{dx_\gamma} \right]_{n\ge 0}
  \simeq  \sum_{n=0}^\infty
  2\Re \left\{
    [-{\cal G}_\pair]^n
    \int_{0<t_1<t_2<\cdots<t_n<\Delta t_0} \> 
    \left[ \frac{d{\cal G}}{dx_\gamma\,d(\Delta t_\br)} \right]_\brem
  \right\}
  ,
\label {eq:Abubble2}
\end {equation}
where
\begin {equation}
   {\cal G}_\pair =
   - \frac{\Nf\alpha}{2\pi} \int_0^1 d\yfrakE \> P_{\gamma\to e}(\yfrakE)
     \int_0^\infty d(\Delta t) \> \Omegapr^2 \csc^2(\Omegapr\,\Delta t)
\label {eq:calG1}
\end {equation}
is a complex number related to the total pair production rate
by%
\footnote{
   Eq.\ (\ref{eq:calGrate}) relates the pair production rate to the
   real part of the one-loop bubble ($-{\cal G}$),
   whereas the optical theorem relates
   the pair production rate to the imaginary part of the self energy.
   These statements are consistent because diagrams and self-energies
   are related by a factor of $i$, and $-{\cal G} \propto -i\Pi$ so
   that $\Re(-{\cal G}) \propto \Im\Pi$.  More specifically,
   \[
    -{\cal G}_{\gamma\to\E\Ebar} =
    \lim_{K_\gamma^2 \to 0}
    \frac{ -i \vareps_\mu \Pi^{\mu\nu}(K_\gamma) \vareps_\nu }{ 2\kgamma } \,,
   \]
   where $\kgamma$ is the photon energy, the factor of $1/(2\kgamma)$ is
   because we use non-relativistic
   instead of relativistic normalization of states in this paper,
   $K_\gamma$ is the photon 4-momentum that appears in a Feynman diagram
   treatment of the photon self-energy, $\vareps$ is any transverse
   photon polarization, and the
   limit $K_\gamma^2 \to 0$ is because the photon is approximately on-shell
   in the limits studied here.
}
[see (\ref{eq:ZpairHO})] 
\begin {equation}
   \Gamma_\pair = 2\Re({\cal G}_\pair) .
\label{eq:calGrate}
\end {equation}
When evaluating the real pair production rate, we subtracted out the
(vanishing) vacuum contribution.  In our previous analysis of virtual
pair production loops, we also argued that we could subtract out
the vacuum contribution to the loop in the soft-photon approximation.
So, as before, we may resolve the $\Delta t\to 0$ divergence
by replacing (\ref{eq:calG1}) by its vacuum-subtracted version
\begin {multline}
   {\cal G}_\pair \simeq
   - \frac{\Nf\alpha}{2\pi} \int_0^1 d\yfrakE \> P_{\gamma\to e}(\yfrakE)
     \int_0^\infty d(\Delta t) \>
     \left[
       \Omegapr^2 \csc^2(\Omegapr\,\Delta t) - \frac{1}{(\Delta t)^2}
     \right]
\\
  = \frac{\Nf\alpha}{2\pi} \int_0^1 d\yfrakE \>
      P_{\gamma \to e}(\yfrakE) \, i\Omegapr
  = \frac{3\Nf\alpha}{8} \sqrt{ \frac{i\qhat}{2\kgamma} } \,.
\label {eq:calG2}
\end {multline}

The integrals over the times $t_1,\cdots,t_n$ in
(\ref{eq:Abubble2}) just give a factor of $(\Delta t_0)^n/n!$ so that
\begin {equation}
    \int_{0<t_1<t_2<\cdots<t_n<\Delta t_0} \cdots
    = \int_0^\infty d(\Delta t_0) \> \frac{(\Delta t_0)^n}{n!} \, \cdots .
\end {equation}
The sum over $n$ in (\ref{eq:Abubble2}) is then simply the Taylor series
of an exponential, and so
\begin {align}
  \left[ \frac{d\Gamma}{dx_\gamma} \right]_{n\ge 0}
  &\simeq
  2\Re \int_0^\infty d(\Delta t_0) \>
    \left[ \frac{d{\cal G}}{dx_\gamma\,d(\Delta t_\br)} \right]_\brem
    e^{-{\cal G}_\pair \Delta t_0}
\nonumber\\
  &= 
  - \frac{\alpha}{\pi} \,P_{e\to\gamma}(x_\gamma)
    \Re \int_0^\infty d(\Delta t_0) \> \Omega_0^2 \csc^2(\Omega_0\,\Delta t_0)\,
    e^{-{\cal G}_\pair \Delta t_0}
  .
\label {eq:Abubble3}
\end {align}
This is equivalent to the ordinary LPM formula (\ref{eq:ZrateHO1}) but with
an additional factor $e^{-{\cal G}_\pair \Delta t_0}$ that cuts off
bremsstrahlung durations $\Delta t_0$ larger than (roughly) the
mean free time for pair production.

The remaining time integral in (\ref{eq:Abubble3}) is UV divergent from
$\Delta t_0 \to 0$.  In this limit,
\begin {equation}
   e^{-{\cal G}_\pair \Delta t_0}
    = 1 - {\cal G}_\pair \Delta t_0 + O\bigl((\Delta t_0)^2\bigr) .
\label {eq:expExpand}
\end {equation}
The first term corresponds to the $n{=}0$ term in our original sum
(\ref{eq:Abubble}) and is responsible for linear UV divergence of our integral
(\ref{eq:Abubble3}).  That term corresponds to the ordinary LPM rate,
which has been computed previously, and so we can drop it here.
The second term in (\ref{eq:expExpand}) corresponds to the $n{=}1$
term in (\ref{eq:Abubble}) and generates a logarithmic UV divergence.
The $n{=}1$ term corresponds to the one-bubble diagram in
fig.\ \ref{fig:netrate}a, which we have also already calculated.
In that earlier calculation, the logarithmic divergence was cut off
by the lower limit $\Delta t_\pr$ on the $d(\Delta t_0)$ integration
in (\ref{eq:intpr}).  At that lower limit,
$\Delta t_0 = \Delta t_\pr \ll 1/\Gamma_\pair$ and the
assumption (\ref{eq:dt0assumption}) that we made in approximating
the time integrals (\ref{eq:bubbletimes1}) by (\ref{eq:bubbletimes2})
is invalid.  But since we've already correctly computed the
$n{=}1$ diagram, we can drop the $n{=}1$ term in (\ref{eq:expExpand}) as
well, corresponding in total to replacing
\begin {equation}
   e^{-{\cal G}_\pair \Delta t_0} \longrightarrow
   e^{-{\cal G}_\pair \Delta t_0} - 1 + {\cal G}_\pair \Delta t_0
\end {equation}
in (\ref{eq:Abubble3}), which will give an expression
\begin {equation}
  \left[ \frac{d\Gamma}{dx_\gamma} \right]_{n\ge 2}
  \simeq
  - \frac{\alpha}{\pi} \,P_{e\to\gamma}(x_\gamma)
    \Re \int_0^\infty d(\Delta t_0) \> \Omega_0^2 \csc^2(\Omega_0\,\Delta t_0)\,
    \left[ e^{-{\cal G}_\pair \Delta t_0} - 1 + {\cal G}_\pair \Delta t_0 \right]
\label {eq:Abubble3b}
\end {equation}
for just the
$n{\ge}2$ terms of the original bubble sum (\ref{eq:Abubble}).
\textit{This} integral is completely convergent and is dominated by
$\Delta t_0 \sim \tform^{\LPMplus}$ (the smaller of $1/|\Omega_0|$ and
$1/\Gamma_\pair$), as was assumed in (\ref{eq:dt0assumption}).%
\footnote{
  Now that we've come to (\ref{eq:Abubble3b}), we can also give a little more
  argument, \textit{a posteriori}, for our assumption (\ref{eq:separated})
  that the bubbles are well separated.
  Define $\xi \equiv {\cal G}_\pair t$ and imagine rewriting the
  exponential factor in (\ref{eq:Abubble3b}) once again as a Taylor series
  $e^{-\xi} = \sum_n (-\xi)^n/n!$.  That sum is dominated by
  $n \lesssim |\xi|$ and so $n \lesssim \Gamma_\pair t$.  That means that
  the typical separations $t/n$ of the bubbles in a time interval $t$
  are no less than order $1/\Gamma_\pair$,
  just as assumed in (\ref{eq:separated}).
}

In appendix \ref{app:integration}, we show that
\begin {equation}
  \int_0^\infty dt \> \Omega^2 \csc^2(\Omega t) \,
       \left[ e^{-{\cal G} t} - 1 + {\cal G} t \right]
  = {\cal G} \left[
        \psi\bigl(1{+}\tfrac{\cal G}{2i\Omega}\bigr) + \gammaE
    \right] .
\label {eq:integral}
\end {equation}
Let's take a moment to recast results in terms of the pair production
rate $\Gamma_\pair$ instead of the complex number ${\cal G}_\pair$.
Since ${\cal G}_\pair$ has complex phase $e^{i\pi/4} = (1{+}i)/\sqrt2$
in (\ref{eq:calG2}),
we can use (\ref{eq:calGrate}) to write
\begin {equation}
  {\cal G}_\pair = \frac{e^{i\pi/4}}{\sqrt2} \, \Gamma_\pair \,.
\end {equation}
Since $\Omega_0$ has complex phase $e^{-i\pi/4}$ in (\ref{eq:Omega0}),
we can then rewrite
\begin {equation}
  \frac{{\cal G}_\pair}{2i\Omega_0}
  =
  \frac{\Gamma_\pair}{2^{3/2}|\Omega_0|} \,,
\end {equation}
which is a real number.  Then, again using (\ref{eq:calGrate}),
we can use (\ref{eq:integral}) in (\ref{eq:Abubble3b}) to get
\begin {equation}
  \left[ \frac{d\Gamma}{dx_\gamma} \right]_{n\ge 2}
  \simeq
  - \frac{\alpha}{2\pi} \,P_{e\to\gamma}(x_\gamma) \,
  \Gamma_\pair
  \left[
     \psi\bigl(1{+}\tfrac{\Gamma_\pair}{2^{3/2}|\Omega_0|}\bigr) + \gammaE
  \right] .
\label {eq:Abubble4}
\end {equation}
Recall that $1/|\Omega_0|$ is of order the ordinary LPM formation time.
Eq.\ (\ref{eq:Abubble4})
will exhibit a transition in behavior between the case where
pair production is unlikely to occur during bremsstrahlung
($1/\Gamma_\pair \gg \tform^\brem$ and so $\Gamma_\pair/|\Omega_0| \ll 1$)
and the case where
it is likely (where $\Gamma_\pair/|\Omega_0| \gg 1$).

So far, we have kept the result in terms of $P_{e\to\gamma}(x_\gamma)$, $\Omega_0$,
and $\Gamma_\pair$ to emphasize the physical origin of the
various scales and factors.  But now we use the explicit formulas
(\ref{eq:Peg}), (\ref{eq:Omega0}), and (\ref{eq:GammaPair})
in the soft-photon limit to rewrite (\ref{eq:Abubble4}) as
\begin {subequations}
\label {eq:nge2rate}
\begin {equation}
  \left[ \frac{d\Gamma}{dx_\gamma} \right]_{n\ge 2}
  \equiv
  \left[ \frac{d\Gamma}{dx_\gamma} \right]_\LPM \times
    \frac{\Nf\alpha}{2 x_\gamma}  \, f_{n\ge2}(x_\gamma)
\end {equation}
with
\begin {equation}
  f_{n\ge2}(x_\gamma) \simeq
  -\tfrac34 \left[
    \psi\bigl(
      1{+}\tfrac{3\Nf\alpha}{16x_\gamma}
    \bigr)
    + \gammaE
  \right] .
\label {eq:fnge2}
\end {equation}
\end {subequations}


\subsection {Final LPM\boldmath$+$ result}

For the final result, we just need to add (\ref{eq:nge2rate}) to the
NLO result (\ref{eq:NLOrate}) previously calculated in
section \ref{sec:NLO}.  This yields the $\LPMplus$ result (\ref{eq:LPMplus})
previewed in the introduction.


\section{Arguments for why our results do not depend on large \boldmath$\Nf$}
\label {sec:smallNf}

Though the assumption of the large-$\Nf$ limit was convenient for
our earlier discussion, we believe that it is unnecessary.
We outline our arguments in this section but do not claim
a fully systematic and rigorous diagrammatic analysis.

\begin {figure}[t]
\begin {center}
  \includegraphics[scale=0.5]{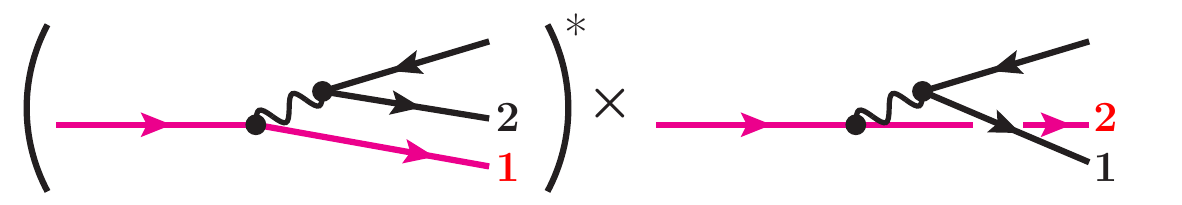}
  \caption{
    \label{fig:ambiguity}
    An interference contribution showing that, when bremsstrahlung
    and pair production overlap for
    $e^- \to e^-\gamma \to e^-e^-e^+$, one cannot unambiguously identify
    which of the two final-state electrons is the continuation
    (magenta line) of the original electron.
    (Adapted from fig.\ 11 of ref.\ \cite{qedNfstop}.)
  }
\end {center}
\end {figure}

One of the advantages of the large-$\Nf$ limit was that the pair-produced
lepton $\E$ had flavor distinguishable (up to corrections suppressed by
$1/\Nf$) from the initial electron $e$.  That allowed us to follow the
initial electron through the $e{\to}e\gamma{\to}e\E\Ebar$ process of
fig.\ \ref{fig:eEnotation} and unambiguously identify which final-state
daughter is the direct descendant of the initial electron.  That in turn
allowed for an unambiguous definition of the rate
$d\Gamma/dx_\gamma$ characterized in this paper as the differential rate for
``the original electron to lose energy $x_\gamma E$,''
which means that the
unambiguously identified final-state daughter $e$ of $e{\to}e\E\Ebar$
has energy $(1{-}x_\gamma)E$.  Outside of the large-$\Nf$ approximation,
however, identification of this final-state descendant of the initial
electron is ambiguous, when bremsstrahlung overlaps with pair
production, because of interference contributions such as
fig.\ \ref{fig:ambiguity}.  To handle similar ambiguities for
gluon showers with overlapping $g {\to} gg {\to} ggg$, ref.\ \cite{qcd}
introduced the concept (stated more generally here)
of a net rate $[d\Gamma/dx]_{i\to j}^\net$ for a particle of energy $E$
and species $i$ to produce a daughter of energy $xE$ and species $j$
(along with any number of other daughters of any species).
The evolution equations
for shower development can be formulated in terms of such net rates.%
\footnote{
  Specifically, see the discussion in sections 3.1 and 8
  of ref.\ \cite{finale2} for the net $g{\to}g$
  rate in the purely gluonic case,
  where all daughters are of the same species.
  For the other extreme, see section 3.1 of ref.\ \cite{qedNfenergy}
  for definitions of net rates
  in large-$\Nf$ QED, where all daughters are of different species.
}
In this language, one may replace the rate
$d\Gamma/dx_\gamma$ discussed in this paper by
\begin {equation}
  \frac{d\Gamma}{dx_\gamma} ~~\longrightarrow~~
  \left[ \frac{d\Gamma}{d\xe} \right]^\net_{e^- \to e^-}
  \mbox{evaluated at $\xe=1{-}x$}
  ,
\label {eq:replacenet}
\end {equation}
where the ``$x$'' on the right-hand side plays the role of $x_\gamma$ on the
left-hand side.  In the large-$\Nf$ limit, the two sides are equal and
$x = x_\gamma$.
Outside of the large-$\Nf$ limit, the right-hand side of (\ref{eq:replacenet})
is well defined but the left-hand side is not (not least because there is
no single value of $x_\gamma$ in fig.\ \ref{fig:ambiguity}).

Having made this distinction, we now argue that it is unimportant for
our purposes here.  Consider separately the cases of hard and
soft bremsstrahlung.

For hard bremsstrahlung ($x_\gamma{\sim}1$), there is no soft-photon
enhancement, and overlap
effects are suppressed by order $\Nf\alpha$ (and so by order $\alpha$ for
$\Nf{=}1$) compared to the ordinary LPM bremsstrahlung rate.
For ordinary LPM bremsstrahlung $e^- \to e^-\gamma$, there is
only one electron daughter and no ambiguity in identifying
the descendant of the original electron.  The conceptual difference
between the two sides of (\ref{eq:replacenet}) in this case is a higher-order
effect that can be ignored to good approximation.

\begin {figure}[t]
\begin {center}
  \includegraphics[scale=0.5]{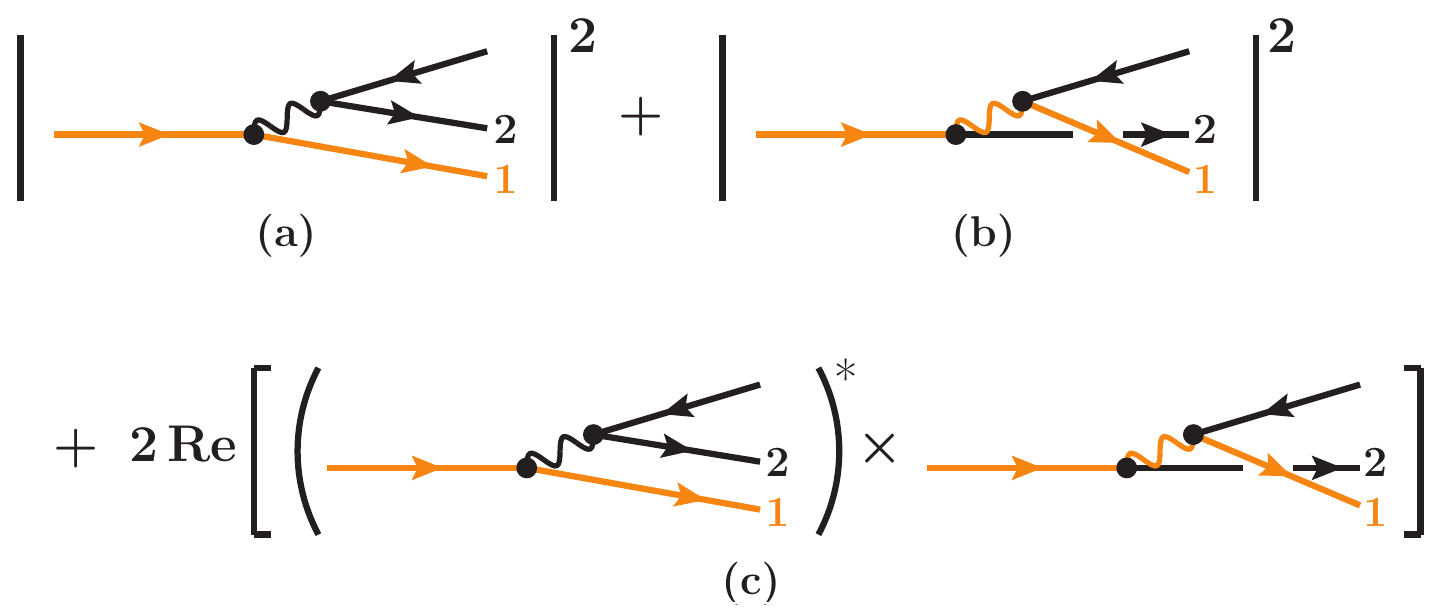}
  \caption{
    \label{fig:softambiguity}
    Contributions to $e^- \to e^-\gamma \to e^-e^-e^+$ when one of the
    final-state electrons has energy very close to the initial energy $E$.
    The orange lines denote hard particles carrying energy $\simeq E$.
    (For the purpose of this figure, we have not specialized to any
    particular time ordering of conjugate amplitude vertices
    relative to the amplitude vertices.)
  }
\end {center}
\end {figure}

For soft bremsstrahlung ($x_\gamma{\ll}1$), we \textit{can} distinguish
the two final-state electrons by the fact that one is soft
($\xe{\ll}1$) and the other is hard ($\xe{\simeq}1$).
In particular, consider the contributions of
fig.\ \ref{fig:softambiguity} (for $e^- \to e^-\gamma \to e^-e^-e^+$)
to the right-hand side of (\ref{eq:replacenet}) in the limit
$x \equiv 1{-}\xe \ll 1$, e.g. $\xe=0.99$.  Let's label the electron
with $\xe=0.99$ as ``electron 1'' and the other
as ``electron 2.''  The orange lines in
fig.\ \ref{fig:softambiguity} must then all be hard, with energies
$\ge 0.99\, E$, whereas the black lines must be soft, with energies
$\le 0.01\, E$.  The first term (a) in the figure has a soft photon line
and will be associated with a DGLAP splitting factor
$P_{e\to\gamma}(x_\gamma) \simeq 2/x_\gamma$ that is enhanced by the small value of
$x_\gamma$.  The second term (b) has a hard photon and so no such
soft-photon limit; it will be suppressed compared to (a).
The last term (c) is in between but still suppressed compared to
(a).  So, in this limit (which is the only limit where overlap corrections
can be important), the pair-produced electron is soft and the other
electron hard, and so the pair-produced electron can be unambiguously
distinguished (up to small corrections) just like in the large-$\Nf$ case.

This means that the important time-ordered interference graphs will be
the same as we have already calculated in this paper, and so our result
(\ref{eq:LPMplus}) should apply even to $\Nf{=}1$.  One may either
(i) consider our original definition of $[d\Gamma/dx_\gamma]_\LPMplus$ as
``{}the rate for the original electron to lose
energy by $x_\gamma E$\,''
to
be a description that is accurate enough at the order of our calculation
or (ii) reinterpret our result as a formula for the
right-hand side of (\ref{eq:replacenet}).


\section{Physical interpretation of logarithm in LPM+ rate}
\label {sec:log}

For very soft photons, the $\LPMplus$ rate
(\ref{eq:previewLPMplus}) has a logarithm:
\begin {equation}
  \left[ \frac{d\Gamma}{dx_\gamma} \right]_{\LPMplus}
  \approx
  \frac{3\Nf\alpha^2}{8\pi x_\gamma^{3/2}}
  \sqrt{ \frac{\qhat}{E} } \,
  \ln \Bigl( \frac{1}{\Nf\alpha} \Bigr)
  \qquad \mbox{(for $x_\gamma \ll \Nf\alpha$)}
\label {eq:LLOlpm+}
\end {equation}
at leading-log order in $\Nf\alpha$.
This logarithm has a simple physical interpretation, independent of
the formalism we have developed for calculating overlap rates.

It will be helpful to first review a similar origin
of the logarithm in the NLO rate (\ref{eq:LPMplusold}),
\begin {equation}
  \left[ \frac{d\Gamma}{dx_\gamma} \right]_{\LPMplus}
  -
  \left[ \frac{d\Gamma}{dx_\gamma} \right]_{\LPM}
  \approx
  \frac{3\Nf\alpha^2}{8\pi x_\gamma^{3/2}}
  \sqrt{ \frac{\qhat}{E} } \,
  \ln \Bigl( \frac{1}{x_\gamma} \Bigr)
  \qquad \mbox{(for $\Nf\alpha \ll x_\gamma \ll 1$)} .
\label {eq:LLOnlo}
\end {equation}
This result was interpreted in ref.\ \cite{qedNfenergy}%
\footnote{
  Specifically, see appendix B.1.1 of ref.\ \cite{qedNfenergy},
  which draws in turn from appendix B.1 of ref.\ \cite{seq}.
}
as
ordinary LPM pair production $\gamma \to \E\Ebar$ from a photon
originating from vacuum-like DGLAP splitting $e \to e\gamma$ of the
initial electron.  To say the same thing in different words: the
photon comes from the Weizs\"acker-Williams distribution for the
probability of finding a photon of longitudinal fraction
$x_\gamma$ in the highly-boosted Coulomb field of the initial electron,
except that we need to account
for a medium modification to the collinear logarithm.
This distribution takes the form
\begin {equation}
  F_\gamma(x_\gamma) \approx
  \frac{\alphas}{2\pi} \, P_{e\to\gamma}(x_\gamma) \,
  \ln\Bigl( \frac{(p_\perp^{\rm max})^2}{(p_\perp^{\rm min})^2} \Bigr) ,
\label {eq:Fgamma}
\end {equation}
where $\p_\perp$ is the transverse momentum variable we have used
previously for $e\to e\gamma$ splittings;%
\footnote{
  So our discussion following (\ref{eq:Fgamma}) takes
  $\p_\perp$ to be the electron transverse momentum in the case that
  the $z$ axis is chosen in the direction of the photon ($\k_\perp = 0$).
  Textbook discussions of DGLAP splitting
  (such as ref.\ \cite{Peskin}) usually instead choose the $z$
  axis to be the direction of the initial electron and then work with
  the transverse momentum of the photon or final electron.
  This makes no difference to the ratio of max over min transverse momentum
  scales appearing in (\ref{eq:Fgamma}).  In particular, in that case
  the transverse momentum is what we call $\pm\P_\perp$ in footnote
  \ref{foot:Pperp} and appendix
  \ref{app:LObrem}, related to our variable $\p_\perp$ here
  by $\p_\perp = -\P_\perp/x_\gamma$, giving
  $p_\perp^{\rm max}/p_\perp^{\rm min} = P_\perp^{\rm max}/P_\perp^{\rm min}$.
}
$p_\perp^{\rm max}$ and $p_\perp^{\rm min}$ are the limits of the $p_\perp$
range that gives rise to a collinear logarithm; and we write
$F(x)$ for the photon distribution instead of
traditionally $f(x)$ simply to avoid confusion with unrelated
uses of the notation $f(x)$ elsewhere in this paper.  In vacuum,
$p_\perp^{\rm min} \sim m_e$.  In our case, it is useful to 
translate from
$p_\perp$ to the duration $\Delta t$ of the virtual photon, which in
vacuum is given by%
\footnote{
\label{foot:pperp}
  In (\ref{eq:pperpTOdt}), $E$ is the initial electron energy and
  $\Delta E = (E_{e^-}{+}\kgamma) - E$ is the off-shellness in energy
  of the intermediate $e^-\gamma$ state, given by the first term in
  (\ref{eq:Heff1}), where we've assumed that $x_\gamma$ is small
  and so approximated $\Mo \simeq E/x_\gamma$ in (\ref{eq:M0}).
}
\begin {equation}
  p_\perp^2 \simeq \frac{2E\,\Delta E}{x_\gamma}
    \sim \frac{2E}{x_\gamma\,\Delta t} \,,
\label{eq:pperpTOdt}
\end {equation}
the details of which are unimportant other than
$p_\perp^2 \propto \Delta E \sim 1/\Delta t$.
We may then rewrite the collinear logarithm in (\ref{eq:Fgamma}) as
\begin {equation}
  \ln\biggl( \frac{(p_\perp^{\rm max})^2}{(p_\perp^{\rm min})^2} \biggr)
  \approx
  \ln\biggl( \frac{(\Delta t)_{\rm max}}{(\Delta t)_{\rm min}} \biggr) .
\label {eq:colog}
\end {equation}

In medium, however, the electron continually receives kicks to
$p_\perp$ from the medium and so cannot sustain a very small $p_\perp$ for a
very long time.  The minimum $p_\perp$ corresponds to virtual photon
duration $\Delta t$ of order the LPM bremsstrahlung formation time, and so%
\footnote{
  Following up on footnote \ref{foot:pperp}, one may also see this from
  (\ref{eq:Heff1}) by estimating $b \sim 1/p_\perp$ and determining the
  condition on $p_\perp$
  for the (medium-dependent) potential term in
  (\ref{eq:HeffHO})
  to be ignorable comparable to the (vacuum) kinetic term.
  The corresponding condition on time scales is then
  given by $\Delta t \sim 1/\Delta E$.
}
\begin {equation}
  (\Delta t)_{\rm max} \sim \tform^\brem .
\end {equation}
At the other extreme, the off-shellness $\Delta E$ of the system
just before LPM pair production $\gamma\to\E\Ebar$ begins must be less than the
typical off-shellness $\Delta E$ during LPM pair production in order
to avoid suppressing the pair production rate compared to the LPM
result (\ref{eq:GammaPair}). 
So we need
$(\Delta E)_\brem \lesssim (\Delta E)_\pair^\LPM$, and so
the minimum duration $\Delta t \sim 1/\Delta E$ for the virtual photon is
\begin {equation}
  (\Delta t)_{\rm min} \sim \tform^\pair .
\end {equation}
The corresponding pair production rate is then
\begin {equation}
  \frac{d\Gamma}{dx_\gamma} \approx
    F_\gamma(x_\gamma) \, \Gamma_\pair
  \simeq
  \frac{\alphas}{2\pi} \, P_{e\to\gamma}(x_\gamma) \,
  \ln\biggl( \frac{\tform^\brem}{\tform^\pair} \biggr)
  \, \Gamma_\pair .
\label {eq:LLOrate}
\end {equation}
This reproduces the leading-log result (\ref{eq:LLOnlo}) using
(\ref{eq:tform}) and (\ref{eq:tformpair}) for the formation
times, (\ref{eq:GammaPair}) for the LPM pair production rate, and
the soft-photon approximation $P_{e\to\gamma}(x_\gamma) \simeq 2/x_\gamma$
for the DGLAP splitting function.

We can now make the same analysis for the $\LPMplus$ rate
when $x_\gamma \ll \Nf\alpha$.
The only difference is that we should then use
the $\LPMplus$ bremsstrahlung formation time
$\tform^\brem \sim 1/\Gamma_\pair$ in (\ref{eq:LLOrate}).
This reproduces (\ref{eq:LLOlpm+}) at leading-log
order in $\Nf\alpha$, which is a reassuring cross-check of our analysis.

Before leaving this topic, we should clarify a possible confusion.
The rates $d\Gamma/dx_\gamma$ quoted at leading-log order in
(\ref{eq:LLOlpm+}) and (\ref{eq:LLOnlo}) included virtual pair
production as well as real pair production.  We have discussed before
how, in that sum, most real pair production diagrams such as
figs.\ \ref{fig:real1}--\ref{fig:real1b} (which include all
at that order that generate the collinear logarithm) are canceled by most of
the virtual pair production diagrams such as
figs.\ \ref{fig:virt1}--\ref{fig:virt1b}.  So one might imagine that the
leading-log rate (\ref{eq:LLOrate})
associated with real pair production will
correspondingly cancel in our final real$+$virtual formula
for $d\Gamma/dx_\gamma$.
The loophole is that there is \textit{no} such vacuum-like
collinear log enhancement of
the \textit{total} virtual contribution to $e \to e\gamma$;
if the pair production never really happened, it can't be enhanced by
a vacuum-like collinear logarithm associated with radiation
from the initial electron.  This harks back to the
clarification at the end of section \ref{sec:NLOtransverse}
that even though the diagram of fig.\ \ref{fig:virt1}a (and its later
generalization to fig.\ \ref{fig:bubbles}) does not appear to have
anything to do with real pair production, it nonetheless captures
all of the physics of real plus virtual pair production.  One may
see all of this play out explicitly in results of
ref.\ \cite{qedNfenergy}
for the NLO case of $\Nf\alpha \ll x_\gamma \ll 1$, where (like here) the
diagram of our fig.\ \ref{fig:virt1}a generates exactly the same
small-$x_\gamma$ logarithmic behavior (\ref{eq:LLOrate}) that we
have discussed above for real pair production.%
\footnote{
  Specifically, the $-\tfrac34 \ln(1{-}\xe) = -\tfrac34 \ln x_\gamma$
  term in eq. (3.15a) of ref.\ \cite{qedNfenergy} for
  $f_{e\to e}^{\rm real}$ represents the appearance of the logarithm
  in real pair production.  But there is no such log behavior in
  eq.\ (3.15b) of ref.\ \cite{qedNfenergy} for $f_{e\to e}^{\rm virt}$,
  and so no such logarithm for total virtual
  pair production.  When the two are added to get $f_{e\to e}$ in
  eq.\ (3.10a) of ref.\ \cite{qedNfenergy}, the logarithm associated
  with real pair production remains, regardless of the fact that, at the
  level of individual diagrams, one can point to a \textit{subset} of virtual
  diagrams that, in isolation, would cancel all the real diagrams
  that generated the logarithm.
}


\section{Conclusion}
\label {sec:conclusion}

We have verified through explicit calculation our qualitative claim that
pair production disrupts the LPM effect for
very soft ($k_\gamma \lesssim \Nf\alpha\kern1pt E$ but still extremely
high energy)
brem\-sstrahlung, leading to significantly
\textit{larger} bremsstrahlung rates than predicted by standard LPM formulas.
However, for the sake of simplicity, our explicit calculations in this paper
were restricted to $k_\gamma \gg \Elpma$.  That means, for example,
electron energies $E \gg 2.5/\alpha$ TeV in Gold, or $\gg 234/\alpha$ PeV for
air, for our quantitative results to be applicable in a region of
$(E,\kgamma)$ where the effect is large.  The obvious next step is
to extend our calculations to lower energies by including the mass of the
electron in our calculations.  At even low energies, the dielectric
effect (the medium-induced mass of the photon) should also be included.
Our work on these extensions is in progress.

If desired,
incorporating this effect into a complete description of shower development
(rather than following only the descendant of the original electron)
will require not only the $e^-{\to}e^-$ net rate described in
(\ref{eq:replacenet}) but also $e^-{\to}e^+$ and $e^-{\to}\gamma$ net rates,
which we leave for future work.

Finally, it will be interesting to investigate the feasibility
of verifying this effect experimentally, as
the ordinary LPM effect has been.


\begin{acknowledgments}

The work of Arnold and Bautista was supported,
in part, by the National Science Foundation under Grant No.~2412362.
Elgedawy was supported by
Guangdong Major Project of Basic and Applied Basic Research No.\ 2020B0301030008
and by the
National Natural Science Foundation of China under Grant Nos.\ 12035007 and
12447145.
We also thank Ulrik Uggerh\o j and Larry McLerran for helpful communications
regarding footnotes \ref{foot:uggerhoj} and \ref{foot:mclerran}.

\end{acknowledgments}

\appendix

\section{The scale \boldmath$\bmin$ of $\qhat(\bmin)$}
\label{app:bmin}

\subsection{The scale and its relation to Migdal's logarithms}

In the context of bremsstrahlung and pair production, $\qhat$ formally
arises from the small-$\b$ harmonic oscillator approximation (\ref{eq:VHO})
to the potential $V(\b)$ of (\ref{eq:V}).  As mentioned in the main text,
(\ref{eq:VHO}) gives rise to a UV-divergent $\q_\perp$-integral
(\ref{eq:qhatdef2})
for $\qhat$,
\begin {equation}
   \qhat = \int d^2q_\perp \frac{d\Gamma_\el}{d^2q_\perp} \, q_\perp^2 ,
\label {eq:qhatdef2b}
\end {equation}
because $d\Gamma_\el/d^2\q_\perp$ falls like $1/q_\perp^4$
at large $q_\perp$.  However, the original definition (\ref{eq:V}) of $V(\b)$
involves $q_\perp$ integrals with no such large-$q_\perp$ divergence.
At leading-log order, the actual small-$b$ limit
of $V(\b)$ is given by
\begin {equation}
   V(\b) \simeq -\tfrac{i}{4} \qhat(b) \, b^2
\label {eq:Vqhat3}
\end {equation}
with
\begin {equation}
   \qhat(b) \approx
   \int_{|\q_\perp| \lesssim 1/b} d^2q_\perp \frac{d\Gamma_\el}{d^2q_\perp} \, q_\perp^2 .
\label {eq:qhatdef3}
\end {equation}

We should emphasize that the $\b$ here,
appearing in $V(\b)$,
is the transverse
separation between two high-energy charged particles in Zakharov's formalism,
such as $\b=\b_{e^-}{-}\b_{e^+}$ in (\ref{eq:Heff}) for bremsstrahlung or
(\ref{eq:Hpair}) for pair production.  This $\b$ is \textit{not} the impact
parameter for elastic collisions, but it does determine the upper bound on
the $q_\perp$ integration in (\ref{eq:qhatdef3}).  In that integration,
$q_\perp$ is the transverse momentum transfer from a single collision
and corresponds to impact parameter $\sim 1/q_\perp$ for that collision.
The upper bound $1/b$ on $q_\perp$ in (\ref{eq:qhatdef3}) means
that the separation $b$ is (parametrically) also the
\textit{minimum} impact parameter for the range of
elastic collisions relevant to computing $V(\b)$.

At leading-log order for $\qhat$, we may also (i) further restrict the
integration region in (\ref{eq:qhatdef3}) to
$1/a_Z \lesssim |\q_\perp| \lesssim 1/b$, where $a_Z$ is the atomic
screening length for the Coulomb field of the nucleus, and then
(ii) ignore atomic screening in the integrand by taking
$d\Gamma_\el/d^2q_\perp \simeq n|Z e^2/q_\perp^2|^2/(2\pi)^2$.  The integral
(\ref{eq:qhatdef3}) then gives
\begin {equation}
   \qhat(b) \approx 8\pi n Z^2 \alpha^2 \ln\Bigl( \frac{a_Z}{b} \Bigr) ,
\label {eq:qhat3}
\end {equation}
and so the small-$b$ behavior of $V(b)$ is proportional to
$b^2 \ln(a_Z/b)$ instead of $b^2$.
To make the harmonic oscillator
approximation, one treats the logarithm as large and replaces the $b$
in the logarithm by a fixed parametric scale, which is the
minimum relevant impact parameter scale
$\bmin$ of (\ref{eq:qhat}).  That scale should be of order
$1/p_\perp$ where, by (\ref{eq:qhatdef}),
$p_\perp \sim (\qhat \tform)^{1/2}$ is the typical total transverse momentum
picked up during the relevant time scale $\tform \sim 1/|\Omega|$
for the process.  So
\begin {equation}
   \bmin \sim (\qhat \tform)^{-1/2} .
\label {eq:bmintform}
\end {equation}
With (\ref{eq:Omega0}) and (\ref{eq:Omegapr}), that gives
\begin {equation}
  \bmin \sim \bmin^{(0)} \equiv
  \begin{cases}
    \bigl[ (1{-}x_\gamma)\qhat E/x_\gamma \bigr]^{-1/4} ,
       & \mbox{ordinary LPM bremsstrahlung} ; \\
    \bigl[ \yfrakE(1{-}\yfrakE)\qhat\kgamma \bigr]^{-1/4} ,
       & \mbox{LPM pair production} .
  \end {cases}
\label {eq:bminval}
\end {equation}
Two caveats to (\ref{eq:bminval}) are that (i) we need
$\bmin \ll 1/m_e$ in our paper so that we are deep in the LPM regime and
therefore may ignore (as we have throughout this paper)
the effects of $m_e$ in our estimates and (ii) we need
$\bmin \gg R_A$ so that
we remain in the $Z$-enhanced electric field outside of the nucleus
(assuming medium-sized or large nuclei).  To connect to Migdal \cite{Migdal},
it is convenient to rewrite (\ref{eq:qhat}) as
\begin {equation}
   \qhat \approx 8\pi n Z^2 \alpha^2 \zeta \ln(m_e a_Z) ,
\label {eq:qhat4}
\end {equation}
where, at leading log order,
\begin {equation}
  \zeta \equiv \frac{ \ln(a_Z/\bmin) }{ \ln(m_e a_Z) }
  \approx
  \begin{cases}
    1 , & \mbox{for}~ \zeta_0 \le 1 \phantom{{}\le 2}
        ~ \mbox{(corresponding to $\bmin \sim 1/m_e$)}; \\
    \zeta_0 , & \mbox{for}~ 1 \le \zeta_0 \le 2
        ~ \mbox{(corresponding to $R_A \lesssim \bmin \lesssim 1/m_e$)}; \\
    2 , & \mbox{for}~ \zeta_0 \ge 2 \phantom{{} \le 1}
        ~ \mbox{(corresponding to $\bmin \sim R_A$)} \\
  \end {cases}
\label {eq:zetadef}
\end {equation}
with
\begin {equation}
  \zeta_0 \equiv \frac{ \ln(a_Z/\bmin^{(0)}) }{ \ln(m_e a_Z) }
  .
\end {equation}
The ``2'' in (\ref{eq:zetadef}), which is the ``2'' in (\ref{eq:accident}),
is really a numerical approximation
\begin {equation}
  \frac{ \ln(a_Z/R_A) }{ \ln(m_e a_Z) } \simeq 2
\end {equation}
used by Migdal.
Our (\ref{eq:zetadef}) above is equivalent at leading-log order
(where only a rough estimate of the argument of the logarithm is
important) to Migdal's definition of $\xi(s)$
for bremsstrahlung and $\xi(\bar s)$ for pair production.%
\footnote{
  Migdal sometimes works in units where $m_e = 1$, and so it is easier
  to compare our $\zeta$ to the definition of $\xi$ shown in
  eqs.\ (75--76) and (86) of the review \cite{SpencerReview},
  where all $m_e$ dependence is explicit.
  Also, since our discussion here only covers a parametric analysis of the
  argument of the logarithm in the formula for $\qhat$,
  we do not worry about multiplicative $O(1)$ factors
  inside the logarithm when comparing to Migdal.
}

Since we have ignored the electron mass in our LPM analysis in this paper,
our analysis breaks down when $m^2$ cannot be ignored compared to $p_\perp^2$
in (\ref{eq:Heff}) and (\ref{eq:Hpair}).  The boundary $p_\perp \sim m$
corresponds to $\bmin \sim 1/m$ and so $\zeta_0 \simeq 1$ above, which is
where the deep LPM rate formulas transition toward ordinary
Bethe-Heitler formulas.  Migdal arranged for his Bethe-Heitler limit
[the first case in (\ref{eq:zetadef})] to
be better than merely a leading log approximation, and his version of our
$\ln(m_e a_Z)$ in (\ref{eq:qhat4}) is%
\footnote{
   The much later review \cite{SpencerReview} uses $\ln(184\,Z^{-1/3})$.
}
\begin {equation}
   \ln(m_e a_Z) \longrightarrow \ln(190\,Z^{-1/3}) .
\end {equation}


\subsection{
  Appropriate scale for \boldmath$\qhat$ in LPM/BH ratios in
  fig.\ \ref{fig:overBH}
}

In the Bethe-Heitler rate formula (\ref{eq:BH1}), the appropriate scale
for $\qhat(\bmin)$ is $\bmin \sim m_e^{-1}$, corresponding to
$\zeta = 1$ in (\ref{eq:zetadef}).  For the ordinary
LPM bremsstrahlung result, the value
of $\zeta$ depends on $E$ and $\kgamma$ according to (\ref{eq:zetadef})
and (\ref{eq:bminval}).  Those choices of $\bmin$ will be the same
close to the left ($k_\gamma{=}E$)
edge of the LPM/BH graph in fig.\ \ref{fig:overBH},
but will differ as one moves to the
right side of that graph.  For the sake of simplicity, we simply ignored
the scale dependence of $\qhat(\bmin)$ when making this graph
and took the $\qhat$'s in
the numerator and denominator to be the same when evaluating the
LPM/BH ratio in fig.\ \ref{fig:overBH}.
That means there can be up to a factor of $\sqrt2$ error in the
values of the ratio as one moves to the right side of the graph
since the deep LPM rate (\ref{eq:LPM}) is proportional to
$[\qhat(\bmin)]^{1/2}$ and the BH rate (\ref{eq:BH1}) is proportional
to $\qhat(m_e^{-1})$.
Similarly, we also ignored this mild dependence on scale when plotting
the $\LPMplus$/BH ratio on the right-hand side of fig.\ \ref{fig:overBH}.


\subsection{
  Appropriate scale for \boldmath$\qhat$
   in LPM+/LPM ratio in fig.\ \ref{fig:LPM+overLPM}
}

The main issue above was that the BH rate had a fixed scale $\bmin \sim 1/m_e$
while the LPM rate did not.  One might hope that the scale choice cancels
in fig.\ \ref{fig:LPM+overLPM}.  It does for $\alpha E \ll \kgamma \ll E$,
where the LPM and $\LPMplus$ rates agree.  But once one gets into the
range $k_\gamma \ll \alpha E$ where the $\LPMplus$ rate differs significantly,
then the $\LPMplus$
bremsstrahlung formation time (\ref{eq:tformplus1}) is different
from the ordinary LPM formation time, and so
the scale (\ref{eq:bmintform}) for $\bmin$ is different.
To complicate matters more, the argument of the
logarithm in the $\LPMplus$ rate
(\ref{eq:previewLPMplus}) for $x_\gamma \ll \Nf\alpha$ arises parametrically
as the ratio of the pair formation time
$\tform^\pair \sim 1/|\Omega_\pr|$ to the
$\LPM+$ bremsstrahlung formation time, which is $1/\Gamma_\pair$ in
this limit.  This involves more than one time scale and so also somewhat
different values for $\qhat$.  We have again ignored the mild
scale dependence of $\qhat$ when making fig.\ \ref{fig:LPM+overLPM}.
Variations by factors of $\sqrt2$ will not affect our demonstration that
the $\LPMplus$ rate is significantly less suppressed than the ordinary
LPM rate when $x_\gamma \ll \Nf\alpha$.


\section{An additional but minor difference with Galitsky and Gurevich}
\label{app:minor}

Though we disagree on the direction of the
effect, one might expect us to agree with Galitsky and Gurevich
on the region of $(\kgamma,E)$ where overlapping
pair production is important since
we all agree that it becomes important when $1/\Gamma_\pair$ becomes
$\lesssim$ the ordinary LPM bremsstrahlung formation time, as in
(\ref{eq:tformplus1}).  In this paper, we have focused on $E \gg \Elpma$
and found that overlapping pair production is important for
$\kgamma \lesssim \alpha E$ but that the ordinary LPM effect is unmodified
for
\begin {equation}
   \alpha E \lesssim \kgamma \lesssim E
   \qquad \mbox{(ordinary LPM when $E \gtrsim \Elpm$)} .
\label {eq:ordinaryLPM}
\end {equation}
Galitsky and Gurevich's paper is focused on
$E \ll \Elpma$, and because of this $\kgamma \ll \Elpma$, and so
the pair production rate $\Gamma_\pair$ that they use in their analysis
is approximately constant ($\sim 1/X_0 \sim \qhat/m^2$ in our notation)
instead of the deep-LPM formula (\ref{eq:pairqual}).
It is because of their $\kgamma \ll \Elpma$ approximation to the
pair production rate that
their concluding figure depicts the ordinary LPM effect
region completely disappearing at $E \sim \Elpm$ instead of
remaining for (\ref{eq:ordinaryLPM}).


\section{Sketch of the derivation of the LPM bremsstrahlung rate
  (\ref{eq:Zrate})}
\label {app:LObrem}

In this appendix, we sketch a derivation of the LPM bremsstrahlung rate
in the form of (\ref{eq:Zrate}).  We will first outline the derivation in
terms of a different transverse-momentum variable%
\footnote{
   The variable $\P_\perp$ of (\ref{eq:P}), or something proportional to it,
   was used by BDMPS-Z in, for example, refs.\ \cite{BDMS,Zakharov1}.
   For comparisons of notation, see the appendix of ref.\ \cite{simple}.
}
\begin {equation}
   \P_\perp \equiv x_\gamma \p_{\perp e^+} + \k_\perp
\label {eq:P}
\end {equation}
(without choosing to set $\k_\perp=0$)
and a corresponding conjugate transverse separation $\B$ typically used in
discussions of the QCD LPM effect, leading to a notationally
different version
\begin {subequations}
\label {eq:ZrateB}
\begin {equation}
  \left[ \frac{d\Gamma}{dx_\gamma} \right]_\LPM
  =
  \frac{ \alpha\,P_{e\to\gamma}(x_\gamma) }{ M_0^2 }
  \Re \int_0^\infty d(\Delta t) \>
  \grad_{\B'} \cdot \grad_{\B} \,
  G(\B',\Delta t;\B,0) \Bigl|_{\B'=\B=0}
\end {equation}
of the rate formula (\ref{eq:Zrate}), where
\begin {equation}
  M_0 \equiv x_\gamma(1-x_\gamma)E
\label {eq:M0B}
\end {equation}
\end {subequations}
differs by a factor of $x_\gamma^2$ from the harmonic oscillator
``mass'' $\Mo$ (\ref{eq:M0}) used
in the main text.
Then, by relating $\B$ to our $\b \equiv \b_{e^-}-\b_{e^+}$, we will verify
that (\ref{eq:ZrateB}) is equivalent to the formula (\ref{eq:Zrate})
used in the main text.
One of our reasons for this review is that in this paper we find it very
useful (in the soft photon limit)
to work in a coordinate system where the $z$ axis
is chosen to be in the direction of the photon so that $\k_\perp = 0$.
But
we need to explain a subtlety that arises were one to
\textit{directly} derive (\ref{eq:Zrate}) by first fixing $\k_\perp=0$
[as in (\ref{eq:kperp}) and (\ref{eq:Heff1})] and working from the start
in terms of
the transverse momentum variable $\p_\perp \equiv \p_{\perp,e^-} = -\p_{\perp,e^+}$
used in the main text.
The same subtlety will later arise in our derivation in appendix \ref{app:II}
of the diagram of fig.\ \ref{fig:II}, and the review of the
LPM bremsstrahlung
rate (\ref{eq:Zrate}) here
provides a simpler example in which to first address it.


\subsection{Sketch of derivation of (\ref{eq:ZrateB}) and equivalence to
   (\ref{eq:Zrate})}
\label {app:LOderive}

Consider the effective 3-particle Hamiltonian (\ref{eq:Heff}) for any
choice of $z$ axis that is nearly collinear with the high-energy
bremsstrahlung process (i.e.\ \textit{without} specializing to the
choice $\k_\perp = 0$ made in the main text).  Using
medium-averaged momentum conservation (\ref{eq:pconserved}),
one may algebraically rewrite the kinetic ($p_\perp^2$)
terms of ${\cal H}$ as $P_\perp^2/2M_0$, with $\P_\perp$ and $M_0$
given by (\ref{eq:P}) and (\ref{eq:M0B}).
It will be useful to note that, also using momentum conservation
(\ref{eq:pconserved}), one may
rewrite (\ref{eq:P}) in several equivalent forms
\begin {equation}
   \P_\perp  = x_\gamma \p_{\perp e^+} + \k_\perp
   = (1{-}x_\gamma) \k_\perp - x_\gamma \p_{\perp e^-}
   = - \p_{\perp e^-} - (1{-}x_\gamma) \p_{\perp e^+} .
\label {eq:Palt1}
\end {equation}
This may be written more symmetrically as
\begin {equation}
   \P_\perp = x_2\p_{\perp 1} - x_1\p_{\perp 2}
   = x_3\p_{\perp 2} - x_2\p_{\perp 3}
   = x_1\p_{\perp 3} - x_3\p_{\perp 1} ,
\label {eq:Palt}
\end {equation}
where particles $(1,2,3)$ refer to $(e^+,\gamma,e^-)$,
\begin {equation}
   (x_1,x_2,x_3) = (-1,x_\gamma,1{-}x_\gamma) ,
\end {equation}
and the kinetic terms of ${\cal H}$ (\ref{eq:Heff}) in this notation are
\begin {equation}
  \frac{p_{\perp 1}^2}{2x_1 E}
  + \frac{p_{\perp 2}^2}{2x_2 E}
  + \frac{p_{\perp 3}^2}{2x_3 E}
  = \frac{P_\perp^2}{2M_0^2}
\label {eq:HeffP}
\end {equation}
with $M_0 = -x_1 x_2 x_3 E$.  The last version
\begin {equation}
  \P_\perp = x_1\p_{\perp 3} - x_3\p_{\perp 1}
\label {eq:P31}
\end {equation}
of (\ref{eq:Palt}) will be the
most convenient for our purposes.  One can check by computing
$[\P_\perp,\B]$ that
\begin {equation}
   \B \equiv \frac{\b_3 - \b_1}{x_3 + x_1}
\label {eq:B31}
\end {equation}
is conjugate to (\ref{eq:P31}).  As an aside, consistency with the
permutation symmetry of (\ref{eq:Palt}) implies that one should
correspondingly have
\begin {equation}
   \B = \frac{\b_3 - \b_1}{x_3 + x_1}
   = \frac{\b_1 - \b_2}{x_1 + x_2}
   = \frac{\b_2 - \b_3}{x_2 + x_3} \,,
\label {eq:Balt}
\end {equation}
and a discussion of how these constraints on $(\b_1,\b_2,\b_3)$ are consistent
with the quantum evolution of the 3-particle system may be found, for example,
in section 3 of ref.\ \cite{2brem}.
Here we focus on (\ref{eq:B31}), which is related to the variable
$\b \equiv \b_{e^-} {-} \b_{e^+}$ of the main text by
\begin {equation}
   \B = - \frac{\b}{x_\gamma} \,.
\label {eq:Bvsb}
\end {equation}

The variables $\P_\perp$ and $\B$ are useful because they are independent
of the exact choice of $z$ axis, as long as the $z$ axis remains nearly
collinear with the bremsstrahlung process.
Specifically, consider small rotations
about the $x$ or $y$ axis, which would induce small changes in the
direction of the $z$ axis.  Under such small rotations, transverse momenta
and transverse positions of high-energy particles transform as
\begin {equation}
  \p_{\perp j} \to \p_{\perp j} + p_{z,j} \bxi \simeq \p_{\perp j} + x_j E \bxi
\end {equation}
and
\begin {equation}
  \b_{\perp j} \to \b_{\perp j} + z \bxi \simeq \b_{\perp j} + t \bxi ,
\end {equation}
where
\begin {equation}
  \bxi = (\xi_x,\xi_y) = (\theta_y,-\theta_x)
\end {equation}
for (active) infinitesimal rotations about the $x$ and $y$ axes by
$(\theta_x,\theta_y)$.  The definitions (\ref{eq:Palt}) and (\ref{eq:Balt})
are invariant under these transformations.
Above, $t \simeq z$ is time.

From (\ref{eq:Heff}), (\ref{eq:Vqhat}), (\ref{eq:Omega0}),
(\ref{eq:HeffP}), and (\ref{eq:Bvsb}),
we have
\begin {equation}
  {\cal H} =
  \frac{P_\perp^2}{2M_0} + V(x_\gamma \B)
  = \frac{P_\perp^2}{2M_0} + \tfrac12 M_0\Omega_0^2 B^2 .
\label {eq:HeffB}
\end {equation}

\begin {figure}[t]
\begin {center}
  \includegraphics[scale=0.6]{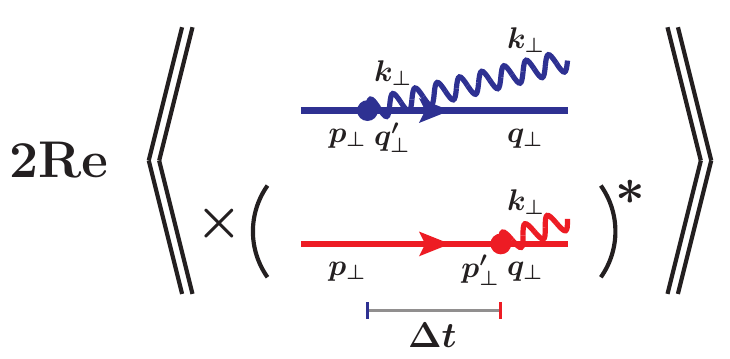}
  \caption{
    \label{fig:lpmp}
    Like fig.\ \ref{fig:lpm}a but depicting the notation used
    in (\ref{eq:lpmp1}).
    Note that the transverse momentum $\k_\perp$ of the photon does not
    change with time because the photon does not interact significantly
    with the medium over the formation time.  Above, $\p_\perp$ denotes
    the transverse momentum of the initial electron an instant before
    the splitting in the amplitude.  $\q_\perp$ denotes the electron's
    transverse momentum an instant after the later splitting in the
    conjugate amplitude.  It must be the same in the amplitude at that time
    if the final state is on-shell.  (One could just as well
    define $\q_\perp$ as the transverse momentum at an even later time,
    but that makes no difference if the rate is
    integrated over all final transverse momenta.)
  }
\end {center}
\end {figure}

Now turn attention to calculating the LPM bremsstrahlung rate corresponding
to fig.\ \ref{fig:lpm}a, which we've labeled
with individual particle transverse momentum variables in fig.\ \ref{fig:lpmp}.
In terms of transverse momenta,
\begin {multline}
  \frac{d\Gamma}{dx_\gamma} =
    \frac{E}{2\pi}
    \int_{\q_{\perp},\k_{\perp}}
    \int_0^\infty d(\Delta t) \>
    2\Re
    \Bigdlangle
       \Bigl(
          \int_{\q_\perp'}
           \langle \q_\perp \k_\perp;\Delta t | \q_\perp' \k_\perp;0 \rangle
           {\cal M}_{\p_\perp \to \q_\perp' \k_\perp}
       \Bigr)
\\
       \times
       \Bigl(
          \int_{\p_\perp'}
          {\cal M}_{\p_\perp' \to \q_\perp \k_\perp}
          \langle \p_\perp';\Delta t | \p_\perp;0 \rangle
       \Bigr)^*
    \Bigdrangle ,
\label {eq:lpmp1}
\end {multline}
where ${\cal M}_{\p_\perp \to \q_\perp' \k_\perp}$ is the QFT matrix element
corresponding to the vertex in the amplitude (blue) of fig.\ \ref{fig:lpmp}.
The $\int_{\q_{\perp},\k_{\perp}} \equiv \int d^2q_\perp \, d^2k_\perp/(2\pi)^4$
above is the transverse-momentum part of
the integral over the final-state phase space.  We do not integrate
over $dk_z/2\pi$ because the formula is for the differential
cross-section $d\Gamma/dx_\gamma$; however,
$dk_z/2\pi = E\,dx_\gamma/2\pi$ is the source of the overall factor of
$E/2\pi$ on the right-hand side of (\ref{eq:lpmp1}).
Final state polarizations are implicitly summed (and initial state
polarization may optionally be averaged).

The medium average
\begin {equation}
 \Bigdlangle
   \langle \q_\perp \k_\perp;\Delta t | \q_\perp' \k_\perp;0 \rangle
   \langle \p_\perp';\Delta t | \p_\perp;0 \rangle^*
 \Bigdrangle
\label {eq:timeevolve}
\end {equation}
of the time evolution
in (\ref{eq:lpmp1}) is handled by Zakharov's formalism using the
effective Hamiltonian ${\cal H}$.
Textbook formulas for nearly-collinear
$1{\to}2$ splitting matrix elements ${\cal M}$
are usually given for the case where the parent has zero transverse momentum,
in which case the daughters have equal and opposite transverse momentum
$\pm{\bm\pi}_\perp$.
In that case, generically%
\footnote{
  See, for a textbook example, eqs.\ (17.92) and (17.117) of
  ref.\ \cite{Peskin}.  Remember that we are implicitly summing over
  final-state polarizations.  (Whether or not one also averages over
  initial-state polarization makes no difference if, statistically,
  the medium is locally rotational and parity invariant.)
}
\begin {equation}
  \Bigl|
  \langle -\bpi_\perp,(1{-}x)E;\bpi_\perp,xE
     | \delta H | {\bm 0},E \rangle_{\rm rel}
  \Bigr|^2
  = \frac{2e^2 \pi_\perp^2}{x(1 - x)} \, P(x) ,
\label {eq:calMrel}
\end {equation}
where $P(x)$ is the relevant (unregulated) DGLAP splitting function, and
the subscript ``rel'' denotes relativistic normalization of the states.
When the parent of the $1{\to}2$ splitting has zero transverse momentum,
meaning $\p_{\perp e^+} \equiv \p_{\perp 1} = 0$ in the language
that we have used earlier, then
our definitions (\ref{eq:Palt1}) identify the transverse momenta
$\k_\perp$ and $\p_{\perp,e^-}$ as $\pm\P_\perp$. So $\pi_\perp^2$
in (\ref{eq:calMrel}) can be replaced by $P_\perp^2$, giving a
rotational-invariant formula for $|{\cal M}|^2$.
 
Since (following Zakharov) we have packaged the time-evolution problem in
the language of a Schr\"odinger-like equation, we implicitly use
non-relativistic normalization
$\langle \p | \p' \rangle = (2\pi)^d \delta^{(d)}(\p-\p')$ for our states,
and so (i) no explicit relativistic phase space normalization
factors were needed in (\ref{eq:lpmp1}) but
(ii) we must convert (\ref{eq:calMrel})
to non-relativistic normalization by dividing by $2E_j$ for
each particle $j$ in the splitting.  All told,
\begin {equation}
  |{\cal M}_{1\to 2}|^2
  = \frac{e^2 P_\perp^2}{4x^2(1 - x)^2E^3} \, P(x) .
\label {eq:calMnr}
\end {equation}

In the case of (\ref{eq:lpmp1}), the ${\cal M}_{\p_\perp \to \q_\perp' \k_\perp}$
and the ${\cal M}_{\p_\perp' \to \q_\perp \k_\perp}^*$ are not evaluated at the
same time, and so the transverse momenta involved will not be the same
($\p_\perp$ vs.\ $\p_\perp'$ and $\q_\perp'$ vs. $\q_\perp$).  The effect of
this is that the $P_\perp^2$ in (\ref{eq:calMnr}) is really the dot
product $\P_\perp \cdot (\P_\perp')^*$ of the values of $\P_\perp$ at the
two different times, where the star on
$(\P_\perp')^*$ is a reminder that
one of the $\P_\perp$'s is coming from the splitting matrix element in
the conjugate amplitude.  Also, to get to the $\B$-space formula
(\ref{eq:ZrateB}), we need to convert $\P_\perp$'s to $-i\grad_\B$'s,
so that%
\footnote{
  We've performed some sleight-of-hand here writing the complex conjugation
  in $P_\perp^2 \to P_\perp \cdot (P_\perp')^*$
  to arrive at the overall sign of (\ref{eq:calMnrB}).
  In the spirit of merely ``sketching'' the derivation, we will not drill
  down further.
}
\begin {equation}
  {\cal M}_{\p_\perp \to \q_\perp' \k_\perp} {\cal M}_{\p_\perp' \to \q_\perp \k_\perp}^*
  \longrightarrow
  \frac{e^2}{4x^2(1 - x)^2E^3} \, P(x) \, \grad_{\B} \cdot \grad_{\B'} .
\label {eq:calMnrB}
\end {equation}
Putting this together with the description of the evolution
(\ref{eq:timeevolve}) in terms of the effective Hamiltonian
turns (\ref{eq:lpmp1}) into
\begin {multline}
  \left[ \frac{d\Gamma}{dx_\gamma} \right]_\LPM
  =
  \frac{E}{2\pi} \times
  \frac{e^2}{4x_\gamma^2(1{-}x_\gamma)^2E^3} \, P_{e\to\gamma}(x_\gamma)
\\
  \times 2\Re \int_0^\infty d(\Delta t) \>
  \grad_{\B'} \cdot \grad_{\B} \,
  G(\B',\Delta t;\B,0) \Bigl|_{\B'=\B=0} ,
\label {eq:lpmB}
\end {multline}
where
\begin {equation}
  G(\B',\Delta t;\B,0) \equiv \langle \B',\Delta t|\B,0 \rangle
\label {eq:GdefB}
\end {equation}
is the propagator of the effective Hamiltonian (\ref{eq:HeffB}).
Eq.\ (\ref{eq:lpmB}) may then be rewritten in the compact form (\ref{eq:ZrateB})
presented earlier.

We can now use the relation (\ref{eq:Bvsb}) of $\B$ to
$\b \equiv \b_{e^-}-\b_{e^+}$ to obtain the rate formula
(\ref{eq:Zrate}) presented in the main text.
For all of the various forms of the effective Hamiltonian we have presented,
we have implicitly used conventional quantum mechanical normalization for
states, which means $\langle \B'|\B \rangle = \delta^{(2)}(\B'{-}\B)$
when discussing the Hamiltonian (\ref{eq:HeffB}) written in terms
of $\B$ and $\P$, but $\langle \b'|\b \rangle = \delta^{(2)}(\b'{-}\b)$ when
discussing the Hamiltonian (\ref{eq:Heff}) written in terms of $\b$ and $\p$.
So (\ref{eq:Bvsb}) gives
\begin {equation}
  |\B\rangle = x_\gamma |\b\rangle .
\end {equation}
The same relation also gives $\grad_\B = -x_\gamma \grad_\b$, and so
(\ref{eq:ZrateB}) with propagator (\ref{eq:GdefB}) is equivalent to
(\ref{eq:Zrate}) with propagator
\begin {equation}
  G(\b',\Delta t;\b,0) \equiv \langle \b',\Delta t|\b,0 \rangle .
\label {eq:Gdefb}
\end {equation}


\subsection{Subtlety in direct derivation of (\ref{eq:Zrate})}

We now discuss how to make a similar derivation if one wants to work
in a coordinate system where $\k_\perp = 0$ throughout.
Previously, in (\ref{eq:lpmp1}), we took the initial electron's direction
to be fixed ($\p_\perp$ fixed), but integrated over final phase space,
which includes integrating over the $\k_\perp$ (and so direction) of the
photon.  If we want to fix the direction of the photon to be along the
$z$ axis, we will need to integrate over the relative direction of
the photon and the initial electron by integrating over the direction
of the initial electron.  One might at first think this is equivalent
to integrating over the initial electron's $\p_\perp$ (in our
high-energy, collinear approximation), and so imagine that
we just replace the $\int_{\k_\perp}$ in (\ref{eq:lpmp1}) with an
$\int_{\p_\perp}$, i.e.
\begin {multline}
  \frac{d\Gamma}{dx_\gamma} =
    \frac{E}{2\pi}
    {\cal N}\int_{\q_{\perp},\p_{\perp}}
    \int_0^\infty d(\Delta t) \>
    2\Re
    \Bigdlangle
       \Bigl(
          \int_{\q_\perp'}
           \langle \q_\perp \k_\perp;\Delta t | \q_\perp' \k_\perp;0 \rangle
           {\cal M}_{\p_\perp \to \q_\perp' \k_\perp}
       \Bigr)
\\
       \times
       \Bigl(
          \int_{\p_\perp'}
          {\cal M}_{\p_\perp' \to \q_\perp \k_\perp}
          \langle \p_\perp';\Delta t | \p_\perp;0 \rangle
       \Bigr)^*
    \Bigdrangle ,
\label {eq:lpmk1}
\end {multline}
with the seemingly extraneous new normalization factor
${\cal N}$ set to 1 and with the direction of the photon fixed
(e.g. to the $z$ axis in particular).

That's not quite right because the angles
$\bxi = (\theta_y,-\theta_x)$ of the directions of
the initial electron and the photon relative
to the $z$ axis are $\bxi_\k = \k_\perp/k_z = \k_\perp/x_\gamma E$ and
$\bxi_\p = \p_\perp/p_z = \p_\perp/E$, and so the relation between angles and
$\k_\perp$ is different (by a factor of $x_\gamma$)
than the relation between angles and $\p_\perp$.
Rotational invariance means that nothing changes if
we average the rate
$d\Gamma/dx_\gamma$ over rotations, and we may use this to relate
the $\k_\perp$ integral in (\ref{eq:lpmp1}) to the $\p_\perp$ integral
in (\ref{eq:lpmk1}).  If the initial electron direction
was fixed as in (\ref{eq:lpmp1}), then a (superfluous)
averaging over rotations of that
direction can be implemented, in the nearly-collinear approximation
that we use for high-energy bremsstrahlung, as proportional to the operation
\begin {equation}
   \mbox{averaging over rotations}
   \propto \int d^2\xi_\p = \int \frac{d^2p_\perp}{E^2} .
\label {eq:protate}
\end {equation}
If instead the direction of the photon is fixed as in (\ref{eq:lpmk1}),
the same (superfluous)
averaging of the rate over rotations can be implemented as
proportional to the operation
\begin {equation}
   \mbox{averaging over rotations}
   \propto \int d^2\xi_\k = \int \frac{d^2k_\perp}{(x_\gamma E)^2}
   = \frac{1}{x_\gamma^2} \int \frac{d^2k_\perp}{E^2} .
\label {eq:krotate}
\end {equation}
Comparing (\ref{eq:lpmp1}) to (\ref{eq:lpmk1}),
we see that in order for the rotational average of (\ref{eq:lpmp1})
to match the rotational average of (\ref{eq:lpmk1}), we
need to set the normalization factor in (\ref{eq:lpmk1}) to%
\footnote{
  Readers may be disturbed that the proportionality constants
  in (\ref{eq:protate}) and (\ref{eq:krotate}) are  $1/\infty$ since
  the set of real-world rotations are finite but $\int d^2p_\perp$ and
  $\int d^2 k_\perp$ give $\infty$ in this context.
  The argument can be fixed up by
  putting an upper bound $\xi_{\rm max}$ on the range of rotations
  $\bxi$ that are averaged over, choosing $\xi_{\rm max}$ to be parametrically
  small compared to 1 (so that we can make collinear approximations)
  but parametrically large compared to the small relative angles in
  the high-energy bremsstrahlung process.  Alternatively, one may
  generalize the argument to calculations of rates that are \textit{not}
  collinear, and then, after ${\cal N}$ is found, specialize 
  back to the nearly-collinear limit used in (\ref{eq:lpmp1}) and
  (\ref{eq:lpmk1}).  We will not pursue the details.
}
\begin {equation}
   {\cal N} = x_\gamma^2 .
\label {eq:calN}
\end {equation} 

We can now follow the same steps as in section \ref{app:LOderive}.
But note that fixing $\k_\perp{=}0$ and using (\ref{eq:Palt1})
means that the $P_\perp^2$
in the matrix element formula (\ref{eq:calMnr}) is
equal to $|x_\gamma\p_\perp|^2$, where
$\p_\perp \equiv \p_{\perp e^-} = -\p_{\perp e^+}$ is the transverse
momentum variable used in the main text.  Since $\p_\perp$ was
conjugate to $\b$, that means that the analog of (\ref{eq:calMnrB}) is
\begin {equation}
  {\cal M}_{\p_\perp \to \q_\perp' \k_\perp} {\cal M}_{\p_\perp' \to \q_\perp \k_\perp}^*
  \longrightarrow
  \frac{e^2}{4(1 - x_\gamma)^2E} \, P(x) \, \grad_{\b} \cdot \grad_{\b'} .
\label {eq:calMnrb}
\end {equation}
The analog of (\ref{eq:lpmB}) coming from (\ref{eq:lpmk1}) is then
\begin {multline}
  \left[ \frac{d\Gamma}{dx_\gamma} \right]_\LPM
  =
  {\cal N} \frac{E}{2\pi} \times
  \frac{e^2}{4(1{-}x_\gamma)^2E^3} \, P_{e\to\gamma}(x_\gamma)
\\
  \times 2\Re \int_0^\infty d(\Delta t) \>
  \grad_{\b'} \cdot \grad_{\b} \,
  G(\b',\Delta t;\b,0) \Bigl|_{\b'=\b=0} ,
\label {eq:lpmb}
\end {multline}
Using the normalization factor (\ref{eq:calN}) appropriate to having
fixed $\k_\perp{=}0$ from the beginning,
this derivation then also reproduces the rate formula
(\ref{eq:Zrate}) given in the main text.


\section{Diagrams with longitudinally-polarized photons}

\subsection{Diagram with two longitudinally-polarized photons}
\label{app:II}

Here we discuss how to compute the interference
diagram of fig.\ \ref{fig:netrate}b,
which describes $e^-\to e^-\E^-\E^+$
via an instantaneous, longitudinally
polarized photon in both the amplitude and conjugate amplitude.
The general result (not restricted to the soft-photon approximation)
has already been calculated analytically in ref.\ \cite{qedNf}.%
\footnote{
  See eq.\ (A.35) of ref.\ \cite{qedNf} for the final result, and see
  fig.\ 39 and appendix E.3.1 of ref.\ \cite{qedNf} for its derivation.
}
So we could use that result here, and simplify slightly by taking
its soft-photon limit.
[Like fig.\ \ref{fig:netrate}a, this process represents a
small $O(\Nf\alpha)$ correction to the ordinary LPM picture unless
the photon is soft ($x_\gamma \ll 1$).]
However, ref.\ \cite{qedNf} was motivated by the QCD problem and
chose normalizations for position and momentum variables (and their
corresponding eigenstates) that are more convenient
for QCD than for QED.
The derivation in ref.\ \cite{qedNf} also makes use of comparisons
to previous QCD results for $g \to ggg$ (in medium)
via fundamental four-gluon vertices.  These normalizations and comparisons
make the details of the derivation harder to follow than necessary for
our purposes.  For that reason, we will sketch here the steps of
a QED derivation, making use of the fact that (like the other derivations
in this paper) we can simplify the discussion by choosing the $z$ axis
in the soft-photon limit so that $\k_\perp$ of the photon is zero.%
\footnote{
  See footnote \ref{foot:kperpzero} for a discussion of
  why the soft-photon approximation
  is important to setting $\k_\perp = 0$.
}


\subsubsection{Setup}

Consider the differential rate
$[d\Gamma/dx_\gamma\,d\yfrakE]_{\rm(b)}$ corresponding to
$2\Re(\cdots)$ of fig.\ \ref{fig:netrate}b without yet doing the
integral over $\yfrakE$.  We start with
\begin {multline}
  d\Gamma_{\rm(b)} =
  {\cal N}
  \sum_{\rm hel} 
  \frac{dp_{z,e^-}}{2\pi} \> \frac{dp_{z,\ssE^-}}{2\pi} \>
\\ \times
  2\Re
  \int_0^\infty d(\Delta t) \> (-i\delta H_{e^-\to e^-\ssE^-\ssE^+})
     \,G_4(\Delta t)\,
     (-i\delta H_{e^-\to e^-\ssE^-\ssE^+})^* ,
\label {eq:II1}
\end {multline}
where we will mix using momentum space $p_z$ for the $z$ direction
and position space $\b$ for the transverse directions.  (If preferred,
one may write
everything in momentum space and then Fourier transform.)
The $dp_z/2\pi$ factors are from the final-state phase-space measure
(with non-relativistic normalization convention for states),
and we have already used momentum conservation to dispense with a similar
factor for the final-state $\E^+$.
The $\sum_{\rm hel}$ is a sum over final-state helicities.
The normalization factor ${\cal N}{=}x_\gamma^2$ is the factor
(\ref{eq:calN})
associated with having fixed the intermediate momentum
$\k_\perp{=}0$ in our derivation.
$\delta H_{e^-\to e^-\ssE^-\ssE^+}$ represents the transition matrix element
corresponding to longitudinally polarized photon exchange
in the amplitude that instantaneously (in lightcone time)
converts the original electron into $e^-\E^-\E^+$.
The factor $G_4(\Delta t)$
represents the same 4-particle evolution of $e^-e^+\E^-\E^+$
that was
discussed for the 4-particle (middle) region of fig.\ \ref{fig:fund}.
As discussed in section \ref{sec:origin} the potential $V_4$ for that
evolution can be approximated in the soft photon limit by (\ref{eq:V4soft}):
\begin {equation}
  V_4(\b_{e^-},\b_{e^+},\b_{\ssE^-}, \b_{\ssE^+})
  \simeq -\tfrac{i}{4} \qhat\, (\b_{\ssE^-}{-}\b_{\ssE^+})^2
  \qquad
  \mbox{(for $x_\gamma \ll 1$)}.
\end {equation}
Choosing the $z$ axis so that $\k_\perp = 0$, and so
$\p_{\perp e^-} = -\p_{\perp e^+}$ and
$\p_{\perp\ssE^-} = -\p_{\perp\ssE^+}$, the effective Hamiltonian (\ref{eq:H4})
then decomposes into the sum of independent Hamiltonians
(\ref{eq:HOpair}) for the $\E^-\E^+$ pair and (\ref{eq:HeffHO}) for
$e^-e^+$, with the latter in the free-particle ($\Omega_0{\to}0$) limit.
The propagator then decomposes into a product
$G_4 \simeq G_{e^-e^+} G_{\ssE^-\ssE^+}$
of harmonic oscillator
Hamiltonians of the form (\ref{eq:Gprop}).
As we will review in section \ref{app:inst} below, the coupling of
longitudinal photons to electrons does not involve any factors of
transverse momenta $\p_\perp$ and so in position space does not
involve any derivatives with respect to transverse position.
Since the $e^+$ and $e^-$ lines come together at both vertices in
fig.\ \ref{fig:netrate}b, their separation vanishes at the vertices,
and so we should evaluate the formula (\ref{eq:Gprop}) for
$G_{\ssE^-\ssE^+}$ with initial and final separations $\b=\b'=0$.
Similarly for the corresponding propagator $G_{\ssE^-\ssE^+}$,
giving
\begin {align}
  G_4(\Delta t) &\simeq
  \lim_{\Omega_0\to 0}
  \frac{\Mo\Omega_0 \csc(\Omega_0\,\Delta t)}{2\pi i}
  \times
  \frac{M_\pr\Omega_\pr \csc(\Omega_\pr\,\Delta t)}{2\pi i}
\nonumber\\
  &=
  \frac{\Mo}{2\pi i \Delta t} \times
  \frac{M_\pr\Omega_\pr \csc(\Omega_\pr\,\Delta t)}{2\pi i} \,.
\label {eq:G4approx}
\end {align}

Now use (\ref{eq:G4approx}) in (\ref{eq:II1}).  Converting
the $p_z$'s to the variables $(x_\gamma,\yfrakE)$ of fig.\ \ref{fig:eEnotation},
\begin {equation}
  \frac{dp_{z,e^-}}{2\pi} \> \frac{dp_{z,\ssE^-}}{2\pi}
  =
  \frac{x_\gamma E^2}{(2\pi)^2} \, dx_\gamma \, d\yfrakE ,
\end {equation}
then gives
\begin {equation}
  \left[ \frac{d\Gamma}{dx_\gamma\,d\yfrakE} \right]_{\rm(b)} =
  -{\cal N} \frac{\Mo M_\pr x_\gamma E^2}{8\pi^4}
  \sum_{\rm hel} |\delta H_{e^-\to e^-\ssE^-\ssE^+}|^2 \,
  \Re \int_0^\infty \frac{d(\Delta t)}{\Delta t} \>
    \Omega_\pr \csc(\Omega_\pr\,\Delta t) .
\label {eq:II2}
\end {equation}


\subsubsection{\boldmath$\delta H$ matrix element}
\label {app:inst}

One may extract the matrix element $\delta H$ in (\ref{eq:II2}) from
LCPT rules in the literature for longitudinal photon exchange,%
\footnote{
  See Table 10 of ref.\ \cite{BPP}, though we find the normalization conventions
  there to be difficult and obscure.  Eqs.\ (1.60--1.61)
  of ref.\ \cite{KL} give helpful examples using standard relativistic
  normalization conventions
  but does not explicitly show cases related by crossing
  symmetry.
}
adjusting for different conventions for normalization states.
Instead, we will review the origin of that rule here.

LCPT uses lightcone gauge $A^+{=}0$,
for which the ordinary photon Feynman propagator is
\begin {equation}
     G^{\mu\nu}(q) =
     -\frac{i}{q^2} \Bigl[
       g^{\mu\nu} - \frac{q^\mu n^\nu + q^\nu n^\mu}{q\cdot n}
     \Bigr]
\label {eq:Glightcone}
\end{equation}
with
\begin {equation}
  (n^+,n^-,\bm{n}^\perp) = (0,1,\bm{0})
\label {eq:ndef}
\end {equation}
in lightcone coordinates.%
\footnote{
  We will avoid having to make a specific choice of normalization convention for
  the lightcone components $v^\pm$ for a 4-vector $v$.
}
The propagator (\ref{eq:Glightcone}) can be decomposed into a sum
$G^{\mu\nu} = G^{\mu\nu}_{\rm T} + G^{\mu\nu}_{\rm L}$ of separate propagators
\begin {equation}
    G^{\mu\nu}_{\rm T}(q) =
    \frac{i}{q^2}
    \sum_{\lambda=1}^2 \eps_{(\lambda)}^\mu(q) \, \eps_{(\lambda)}^{\nu*}(q) ,
    \qquad
    G^{\mu\nu}_{\rm L}(q) = \frac{i}{(q\cdot n)^2} \,  n^\mu n^\nu
\label {eq:GLT}
\end {equation}
for transverse polarizations (in $A^+{=}0$ gauge)
\begin {equation}
   (\eps^+,\eps^-,\beps^\perp)_{(\lambda)}
   =
   \Bigl( 0, \frac{\beps^\perp_{(\lambda)}\cdot\q}{n\cdot q},
          \beps^\perp_{(\lambda)} \Bigr)
\label {eq:epsT}
\end {equation}
and longitudinal polarization
\begin {equation}
   \eps_{\rm L}^\mu \propto n^\mu .
\end {equation}
$\beps^\perp_{(\lambda)}$ is any orthonormal basis for transverse
vectors.
The longitudinal propagator $G_{\rm L}(q)$ above depends only on
$n\cdot q \propto q^+$ and so is independent of $q^-$ and so
instantaneous in
lightcone time $x^+$ (the Fourier conjugate of $q^-$).
Similarly, $G_{\rm L}(x)$ is local in transverse position $\x^\perp$,
but it is non-local in $x^-$.

\begin {figure}[t]
\begin {center}
  \includegraphics[scale=0.6]{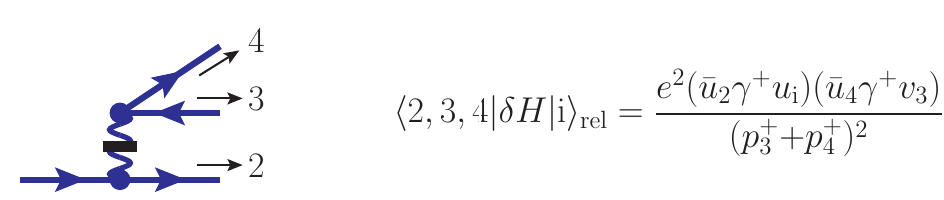}
  \caption{
    \label{fig:dHL}
    Relativistic matrix element corresponding to the amputated Feynman
    diagram $e^- \to e^-\E^-\E^+$ via a longitudinally polarized photon.
    In the formula, we have implicitly assumed that the amputated fermion
    lines should be contracted with on-shell spinors $u$ and $v$.
    One may either crudely view this as an approximation due to the fact that
    the high-energy fermion lines in our application are close to
    on-shell, or one may justify it more generally and more rigorously by
    delving into the derivation of Light Cone Perturbation Theory
    \cite{BL,BPP}.
  }
\end {center}
\end {figure}

Applying standard Feynman rules, but taking just the
longitudinal piece $G_{\rm L}$ of the photon propagator, the
amputated Feynman diagram for the $e^-\to e^-\E^- \E^+$ amplitude
gives the matrix element shown in fig.\ \ref{fig:dHL}.
In the notation we have been using, it evaluates to
\begin {equation}
   \langle e^-,\E^-,\E^+|\delta H|e^-\rangle_{\rm rel}
   =
   \frac{4e^2(x_2 x_3 x_4)^{1/2}}{(x_3{+}x_4)^2} \times \delta_{\rm hel} ,
\end {equation}
where the $x_j$ are the fractions of $p^+$ relative to the original electron;
``$\delta_{\rm hel}$'' is short-hand for conservation of
fermion helicity at each photon vertex; and ``rel'' indicates that
the states have been given conventional relativistic normalization.
In our paper, we use non-relativistic normalization
for states because we describe the medium-averaged evolution
of particles using an equation analogous to the
non-relativistic Schr\"odinger
equation.  To convert normalization, we need to multiply
the matrix element of fig.\ \ref{fig:dHL} by a factor of
$(2E_j)^{-1/2}$ for each external line, giving
\begin {equation}
   \delta H_{e^-\to e^-\ssE^-\ssE^+} =
   \frac{e^2}{\kgamma^2} \times \delta_{\rm hel} \,.
\end {equation}
The sum over final-state helicities in (\ref{eq:II2}) just gives a
factor of 2 for the two possible helicities of the pair-produced
electron $\E^-$.  So
\begin {equation}
   \sum_{\rm hel} |\delta H_{e^-\to e^-\ssE^-\ssE^+}|^2 =
   \frac{2e^4}{\kgamma^4} \,.
\label {eq:dHL2}
\end {equation}

Using (\ref{eq:dHL2}) in (\ref{eq:II2}),
$\Mo = (1{-}x_\gamma)E/x_\gamma \simeq E/x_\gamma$ and
$M_\pr = \yfrakE(1{-}\yfrakE)\kgamma$ from (\ref{eq:M0}) and
(\ref{eq:Mpr}), and
${\cal N}=x_\gamma^2$ from (\ref{eq:calN})
yields
\begin {equation}
  \left[ \frac{d\Gamma}{dx_\gamma\,d\yfrakE} \right]_{\rm(b)} =
  -\frac{\Nf\alpha^2}{\pi^2} \, \frac{4\yfrakE(1{-}\yfrakE)}{x_\gamma} \,
  \Re \int_0^\infty \frac{d(\Delta t)}{\Delta t} \>
    \Omega_\pr \csc(\Omega_\pr\,\Delta t) .
\label {eq:II3}
\end {equation}


\subsubsection{\boldmath$\Delta t$ integration}

The integral in (\ref{eq:II3}) is UV ($\Delta t {\to}0$) divergent.
Like the discussion surrounding (\ref{eq:ZrateHO}), this divergence
may be side-stepped by subtracting the (vanishing) vacuum contribution:%
\footnote{
  One way to do the vacuum-subtracted integral in (\ref{eq:IIint})
  is to first change integration variable to $\tau \equiv i\Omegapr\,\Delta t$,
  giving
  \[
     i\Omegapr \int_0^\infty \frac{d\tau}{\tau}
       \left[ \csch\tau - \frac{1}{\tau} \right]
     =
     i\Omegapr \lim_{\eps\to0^+} \int_0^\infty \frac{\tau^\eps d\tau}{\tau}
       \left[ \csch\tau - \frac{1}{\tau} \right] .
  \]
  Now integrate the two terms separately, following logic similar to
  dimensional regularization to set $\int_0^\infty \tau^{-2+\eps} = 0$.
  For the remaining term, use
  $\csch\tau = 2e^{-\tau}/(1-e^{-2\tau}) = 2(e^{-\tau}+e^{-3\tau}+e^{-5\tau}+\cdots)$
  and integrate term by term to get
  \[
    \int_0^\infty d\tau \> \tau^{-1+\eps} \csch\tau
    = 2 \, \Gamma(\eps) \times (1 + 3^{-\eps} + 5^{-\eps} + \cdots)
    = 2\,\Gamma(\eps) \times (1-2^{-\eps}) \, \zeta(\eps) ,
  \]
  where $\zeta(z)$ is the Riemann zeta function.  Finally, take the limit
  $\eps \to 0$.  (Readers concerned about the justification of these
  steps may check the result numerically.)
}
\begin {equation}
  \int_0^\infty \frac{d(\Delta t)}{\Delta t} \>
    \Omega_\pr \csc(\Omega_\pr\,\Delta t)
  \longrightarrow
  \int_0^\infty \frac{d(\Delta t)}{\Delta t} \>
    \left[ \Omega_\pr \csc(\Omega_\pr\,\Delta t) - \frac{1}{\Delta t} \right]
  =
  -i \Omegapr \ln 2 .
\label {eq:IIint}
\end {equation}
Eq.\ (\ref{eq:II3}) then gives the soft-photon result for this diagram
quoted in (\ref{eq:Bfactorize1}).


\subsection{Diagrams with one longitudinally-polarized photon}
\label{app:I}

From the above discussion of the double-longitudinal diagram of
fig.\ \ref{fig:netrate}b, we can now see why diagrams like
fig.\ \ref{fig:Iexamples}, which have one transverse and one longitudinal
photon, are suppressed in the soft photon
limit.  Consider the evolution of pair-production (real or virtual) in either
diagram.  In the soft-photon limit, this is associated with a
harmonic oscillator propagator
\begin {equation}
  G(\b',\Delta t;\b,0) =
  \frac{\Mpr\Omegapr \csc(\Omegapr\,\Delta t)}{2\pi i}
  \exp\Bigl( i\Mpr\Omegapr \bigl[
    \tfrac12(b^2 + b'^2) \cot(\Omegapr t)
    - \b\cdot\b' \csc(\Omegapr t)
  \bigr] \Bigr)
\label {eq:Gproppair}
\end {equation}
similar to (\ref{eq:Gprop}).  The transverse photon vertex at one end
will come with a factor of transverse momentum, and so a transverse
gradient $\grad$ in $b$-space.  The longitudinal photon vertex at the
other end has no such factor.  At both ends, the transverse
separation between
$\E^-$ and $\E^+$ vanishes, and so the pair production part of
the diagram contains a factor of
\begin {equation}
  \grad_\b G(\b',\Delta t;\b,0) \Bigg|_{\b=\b'=0}
  \qquad \mbox{or} \qquad
  \grad_{\b'} G(\b',\Delta t;\b,0) \Bigg|_{\b=\b'=0} ,
\label{eq:singleI}
\end {equation}
which vanish by parity.  Or one may plug
(\ref{eq:Gproppair}) into (\ref{eq:singleI}) to to see this explicitly.

As a cross check, we have verified this conclusion numerically using
the more general formulas of ref.\ \cite{qedNf}.


\section{Integration}
\label{app:integration}

In this appendix, we show how to perform the integral
\begin {equation}
  I \equiv
  \int_0^\infty dt \> \Omega^2 \csc^2(\Omega t) \,
       \left[ e^{-{\cal G} t} - 1 + {\cal G} t \right]
\label {eq:Idef}
\end {equation}
of (\ref{eq:integral}), where $\Omega$ has complex phase $e^{-i\pi/4}$.
First, write $I = \int u\,dv$ with
$u \equiv e^{-{\cal G}t} - 1 + {\cal G}t$
and $v \equiv \Omega[i-\cot(\Omega t)]$.
Integrate once by parts to get
\begin {equation}
  I =
  -\Omega \int_0^\infty dt \>
   \bigl[ (-{\cal G}e^{-{\cal G} t}+{\cal G}) \bigr]
   \bigl[ i - \cot(\Omega t) \bigr]
  =
  -2i\Omega\, {\cal G} \int_0^\infty dt \>
   \frac{e^{-{\cal G}t} - 1}{e^{2i\Omega t} - 1} \,.
\end {equation}
Next, changing integration variable to $s \equiv e^{-2i\Omega t}$,
\begin {equation}
  I =
  {\cal G} \int_0^1 ds \>
  \frac{ 1 - s^{{\cal G}/2i\Omega} }{ 1-s } \,.
\end {equation}
To obtain a nice form for the answer, it helps to introduce an
unnecessary regulator,
\begin {equation}
  I =
  {\cal G} \lim_{\delta\to0^+} \int_0^1 ds \>
  \frac{ 1 - s^{{\cal G}/2i\Omega} } {(1-s)^{1-\delta} } \,,
\end {equation}
so that we can integrate the two terms in the numerator separately.
That gives
\begin {equation}
  I =
  {\cal G} \lim_{\delta\to0^+}
  \Bigl[
    \tfrac{1}{\delta} - B\bigl(1{+}\tfrac{{\cal G}}{2i\Omega}\, , \delta \bigr)
  \Bigr] ,
\end {equation}
where $B(x,y) = \Gamma(x)\,\Gamma(y)/\Gamma(x{+}y)$ is the
Euler Beta function.  Expanding in $\delta$ then gives the result
\begin {equation}
  I
  = {\cal G} \Bigl[
        \psi\bigl(1{+}\tfrac{\cal G}{2i\Omega}\bigr) + \gammaE
    \Bigr]
\end {equation}
presented in (\ref{eq:integral}).


\end {document}